\documentclass[twocolumn]{aastex61}

\usepackage{morefloats}
\usepackage{epstopdf}
\usepackage{xcolor, graphicx}
\usepackage{amssymb}
\usepackage{amsmath}
\usepackage{longtable}
\usepackage{natbib}
\begin{document}


\title{Magnification Bias of Distant Galaxies in the Hubble Frontier Fields: \\ Testing Wave vs.~Particle Dark Matter Predictions}

\author{Enoch Leung}\affiliation{Department of Physics, The University of Hong Kong, Pokfulam Road, Hong Kong}
\author{Tom Broadhurst}\affiliation{Department of Theoretical Physics, University of the Basque Country UPV/EHU, Bilbao, Spain}\affiliation{IKERBASQUE, Basque Foundation for Science, Bilbao, Spain}
\author{Jeremy Lim}\affiliation{Department of Physics, The University of Hong Kong, Pokfulam Road, Hong Kong}
\author{Jose M.~Diego}\affiliation{IFCA, Instituto de F\'isica de Cantabria (UC-CSIC), Av.~de Los Castros s/n, 39005 Santander, Spain}
\author{Tzihong Chiueh}\affiliation{Department of Physics, National Taiwan University, Taipei 10617, Taiwan}\affiliation{Center for Theoretical Sciences, National Taiwan University, Taipei 10617, Taiwan}
\author{Hsi-Yu Schive}\affiliation{National Center for Supercomputing Applications, Urbana, IL, 61801, USA}
\author{Rogier Windhorst}\affiliation{School of Earth and Space Exploration, Arizona State University, Tempe AZ 85287, USA}

\correspondingauthor{Enoch Leung}
\email{enochleung@connect.hku.hk}


\begin{abstract}

Acting as powerful gravitational lenses, the strong lensing galaxy clusters of the deep {\it Hubble} Frontier Fields (HFF) program permit access to lower-luminosity galaxies lying at higher redshifts than hitherto possible. We analyzed the HFF to measure the volume density of Lyman-break galaxies at $z > 4.75$ by identifying a complete and reliable sample up to $z \simeq 10$. A marked deficit of such galaxies was uncovered in the highly magnified regions of the clusters relative to their outskirts, implying that the magnification of the sky area dominates over additional faint galaxies magnified above the flux limit. This negative magnification bias is consistent with a slow rollover at the faint end of the UV luminosity function, and indicates a preference for Bose-Einstein condensate dark matter with a light boson mass of $m_\mathrm{B} \simeq 10^{-22} \, \mathrm{eV}$ over standard cold dark matter. We emphasize that measuring the magnification bias requires no correction for multiply lensed images (with typically three or more images per source), whereas directly reconstructing the luminosity function will lead to an overestimate unless such images can be exhaustively matched up, especially at the faint end that is accessible only in the strongly lensed regions. In addition, we detected a distinctive downward transition in galaxy number density at $z \gtrsim 8$, which may be linked to the relatively late reionization reported by {\it Planck}. Our results suggests that {\it JWST} will likely peer into an ``abyss'' with essentially no galaxies detected in deep NIR imaging at $z > 10$.

\end{abstract}



\section{Introduction}

The primary goal of the recently completed deep {\it Hubble} Frontier Fields (HFF) program \citep{2017ApJ...837...97L} is to extend our understanding of galaxy formation. Through very deep imaging of the distant universe magnified by the largest known galaxy cluster lenses, the HFF program provides the opportunity to search for the earliest galaxies. To cater to this purpose, a preference  was made for established lenses with available supporting data, such as the remarkable lensing clusters MACS0717 \citep{2009ApJ...707L.102Z} and MACS1149 \citep{2009ApJ...703L.132Z}. In particular, extensive {\it Hubble} photometry had covered four of the HFF clusters in the earlier CLASH\footnote{The Cluster Lensing and Supernova Survey with {\it Hubble} \citep{2012ApJS..199...25P}.} survey, which helped to reconstruct the corresponding 2D magnification maps in advance of the deeper HFF imaging. In addition, 3D spectroscopy targeting the critically lensed regions of three HFF clusters by the MUSE\footnote{The Multi Unit Spectroscopic Explorer \citep{2010SPIE.7735E..08B}.} instrument were also performed to complement the HFF photometric data \citep{2017AandA...600A..90C,2017MNRAS.469.3946L,2018MNRAS.473..663M}, consolidating the identifications of potential multiple image systems. These continuing efforts have led to remarkable discoveries of distant galaxies including a highly magnified one at $z = 9.1$ behind MACS1149 \citep{2018Natur.557..392H}.

Despite the potential of the HFF program to detect galaxies lying at redshifts as high as $z \simeq 11.5$ \citep{2015ApJ...800...84C}, analyses of the HFF data to date have uncovered a surprisingly small number of additional galaxies at such high redshifts, with only one confirmed multiply lensed example at $z\simeq 9.8$ behind A2744 \citep{2014ApJ...793L..12Z}. This scarcity of high-$z$ detections, even with the much greater depth of the HFF program than that of the CLASH survey, raises the issue of whether an effective redshift limit may have been reached that marks the onset of cosmic galaxy formation. Furthermore, only a handful of high-$z$ candidates, all of which are extraordinarily luminous, were identified in wider-field (unlensed) {\it Hubble} surveys at similarly high redshifts based on photometry, and in one case with supporting grism data indicating a redshift of $z = 11.1$ \citep{2016ApJ...819..129O,2016ApJ...830...67B}.

A question related to the epoch of the earliest galaxy formation is the degree to which the galaxy luminosity function extends to low luminosity at high redshifts, as this has a bearing on the nature of dark matter (DM) that governs the growth of structure. For example, a suppressed formation of low-mass DM halos is expected to occur in the Warm Dark Matter (WDM) \citep{2001ApJ...556...93B} or Wave Dark Matter \citep{2014NatPh..10..496S} (also known as Fuzzy Dark Matter; \citealt{2000PhRvL..85.1158H}) models compared with predictions of the standard Cold Dark Matter (CDM) model. Such a suppression can be tracked with star-forming galaxies via the UV luminosity function (LF), which encodes information about the spatial number density of galaxies as a whole (through its normalization), and the relative abundance between galaxies having different UV luminosities (through its shape). Reconstructing the high-$z$ UV LF is thereby crucial for understanding the physics of galaxy formation and evolution in the early universe \citep{2017MNRAS.470..651R}.

In this paper, we study the HFF data principally using the magnification bias method \citep{1995ApJ...438...49B}, with a strong motivation to test competing theories of galaxy formation for which definite predictions have now been made, in particular the Wave Dark Matter model that describes the expected behaviour of bosonic DM, such as light axion-like particles proposed by string theory. This model is of growing interest as one of the most viable interpretations for the observed coldness of dark matter, given the increasingly strict limits on the non-detections of standard Weakly Interacting Massive Particles (WIMPs) \citep{2017NatPh..13..212L}. Dubbed ``$\psi$DM'' to signify the coherent quantum wave property of DM, \citet{2014NatPh..10..496S} showed, with the first high-resolution cosmological simulations in such context, that the large-scale structures of $\psi$DM are statistically indistinguishable from those of standard CDM. On small scales, however, in spite of the close resemblance to the NFW profile \citep{1996ApJ...462..563N} at large radii of individual $\psi$DM halos, they were found to possess distinct solitonic cores which depend only on one free parameter, namely the boson mass. By fitting the soliton profile from $\psi$DM simulations to the large cores of DM-dominated dwarf spheroidal (dSph) galaxies in the Local Group, the DM boson mass was then constrained to be $m_\mathrm{B} \simeq 8.1 \times 10^{-23} \, \mathrm{eV}$ \citep{2014NatPh..10..496S}.

A key prediction of the $\psi$DM model is that galaxy formation is ``delayed'' relative to the standard CDM model, owing to the inherent Jeans scale that forbade galaxies to form at $z \gtrsim 13$ \citep{2014NatPh..10..496S}. The non-discovery of any galaxies at this redshift thus far is some reassurance that the $\psi$DM model is not inconsistent in the way that the conventional WDM scenario was proven to be, where simulations demonstrated the local dSph kpc-scale cores require too large a density of free streaming WDM particles such that the parent halos could not even form in the first place, coined the ``Catch 22'' problem \citep{2012MNRAS.424.1105M}. Recently, the predicted UV LF at $4 \lesssim z \lesssim 10$ in the $\psi$DM context has been determined by \citet{2016ApJ...818...89S} (S16 hereafter) from the simulated halo mass function (MF) using the conditional luminosity function approach \citep{2005ApJ...627L..89C}. We hereby aim to examine these predictions utilizing the unparalleled depth of the strongly lensed HFF data, specifically whether the high-$z$ UV LF exhibits a smooth faint-end rollover, and if so, how it evolved over cosmic history.

This paper is organized as follows. In Section \ref{data_extraction}, we describe the HFF data set from which we used to identify high-$z$ galaxy candidates and the corresponding selection criteria. We then provide in Section \ref{lensing_effects} a comprehensive account of the gravitational lensing effects that are necessary to be considered in order to derive the intrinsic properties of the source galaxies in the cluster fields. Sections \ref{magnification_bias_section} to \ref{sfrd_section} are devoted to the major results, comprising several independent methods that we employed to analyze the UV LF in the HFF fields. We show in Section \ref{lens_model_dependence} the robustness of our results against alternative lens modelling approaches (i.e.~parametric and semi-parametric models) utilized to estimate galaxy magnifications. Lastly, a concise summary and the conclusions of our work can be found in Section \ref{conclusion}. Throughout this research, we adopted the standard cosmological parameters in concordance cosmology for distance determinations\footnote{The reader may wonder whether it is appropriate to use these parameters in the context of the $\psi$DM model. The use of them is indeed justified owing to the fact that the ($\Lambda$)$\psi$DM model has the same asymptotic behaviour as the standard $\Lambda$CDM model in terms of large scale structure formation and cosmic evolution \citep{2014NatPh..10..496S}.}, i.e.~$\Omega_\mathrm{M} = 0.3$, $\Omega_\Lambda = 0.7$, and $H_0 = 70 \, \mathrm{km} \, \mathrm{s}^{-1} \, \mathrm{Mpc}^{-1}$. All magnitudes quoted are in the AB system \citep{1983ApJ...266..713O}.


\section{Data Extraction}\label{data_extraction}

We analyzed all the twelve completed fields of the HFF program, comprising six pairs of cluster field and parallel ``blank'' (or control) field. The galaxy clusters that were observed (in chronological order) are Abell 2744 ($z = 0.308$), MACS J0416.1$-$2403 ($z = 0.396$), MACS J0717.5+3745 ($z = 0.545$), MACS J1149.5+2223 ($z = 0.543$), Abell S1063 (or RXC J2248.7$-$4431; $z = 0.348$), and Abell 370 ($z = 0.375$); hereafter referred to as A2744, MACS0416, MACS0717, MACS1149, AS1063, and A370 respectively. Selected from the catalogs published by \citet{1958ApJS....3..211A}, \citet{1989ApJS...70....1A}, \citet{2007ApJ...661L..33E}, and \citet{2012MNRAS.420.2120M}, these galaxy clusters are among the most powerful gravitational lenses known, making them the most ideal clusters for the purpose of our work. The accompanying control fields (i.e.~the parallel fields) are located $6\arcmin$ away from the centres of their cluster counterparts \citep{2015ApJ...800...84C}, making the effect of lensing by foreground cluster members negligible. Seven {\it HST} bandpass filters spanning from optical (with {\it ACS/WFC}) to near-IR (with {\it WFC3/IR}) were employed in the observations, with a total of up to 140 {\it HST} orbits devoted to each pair of cluster and parallel fields. The 5$\sigma$ point-source limiting AB magnitude reached for each cluster and its accompanying parallel field is $\sim$28.6\,--\,29.1 (see Table \ref{HFF_filters}; \citealt{2017ApJ...837...97L}).

\begin{deluxetable}{c c c}[tp!]
\tablecaption{Observation depths of HFF filters\label{HFF_filters}}
\tablehead{\colhead{Filter} & \colhead{{\it HST} orbits} & \colhead{$5 \sigma$ point-source $m_\mathrm{AB}$ limit\tablenotemark{*}}}
\startdata
F435W ($B_{435}$) & 18 & 28.8 \\
F606W ($V_{606}$) & 10 & 28.8 \\
F814W ($I_{814}$) & 42 & 29.1 \\
F105W ($Y_{105}$) & 24 & 28.9 \\
F125W ($J_{125}$) & 12 & 28.6 \\
F140W ($JH_{140}$) & 10 & 28.6 \\
F160W ($H_{160}$) & 24 & 28.7 \\
\enddata
\tablenotetext{*}{Quoted values are averaged over the entire fields of view.}
\end{deluxetable}

\subsection{Photometry and Redshifts}\label{photometry_section}

HFF photometric catalogs for the twelve cluster and parallel fields, produced from the combined deep-imaging data across the seven {\it HST} optical and NIR filters, were adopted as the primary pool from which we identified potential high-$z$ galaxy candidates. Here we provide a brief summary of the methodology behind the catalog preparation. Readers interested in more information are referred to Section 5 of \citet{2015ApJ...800...84C} (C15 hereafter) for a detailed documentation. (We used the C15 catalogs for all but one of the HFF clusters, i.e.~A370, and its accompanying parallel field, where the corresponding C15 catalogs were only partially complete in the near-IR bands. In this case, we substituted the photometric catalogs with those constructed by \citealt{2018ApJS..235...14S}.)

Automated detection of sources (mostly at 5$\sigma$ levels) was performed in each field using SExtractor (version 2.8.6; \citealt{1996A&AS..117..393B}) based on a detection image constructed from a weighted sum of the HFF images \citep{2017AAS...23031613K}. After that, isophotal apertures enclosing the detected sources were created in each HFF filter image, within which photometry was done to measure their fluxes, AB magnitudes, and the associated uncertainties, where corrections for galactic extinction were simultaneously applied following the extinction law of \citet{2011ApJ...737..103S}.

To determine the photometric redshifts of individual galaxy candidates, the Bayesian Photometric Redshifts (BPZ) \citep{2000ApJ...536..571B,2004ApJS..150....1B,2006AJ....132..926C}) method was employed, where the observed colours $\{C\}$ and the multiband apparent magnitudes $\{m\}$ of a given source were compared against spectral energy distribution (SED) templates of galaxies belonging to different morphological types. The photometric redshift of such a source was then estimated by maximizing the Bayesian probability $p(z|\{C\},\{m\})$ for which it is located at redshift $z$, given the observed colours $\{C\}$ and apparent magnitudes $\{m\}$ measured in the HFF filter set, and assuming that the source falls into a particular morphological type with a prior containing information about the redshift distribution of galaxies belonging to such type \citep{2000ApJ...536..571B}.

\subsection{Data Completeness}\label{data_completeness}

A crucial factor to be taken into account before deriving galaxy number density estimates is the completeness of the data sample, in the sense that it unbiasedly reflects the underlying galaxy population of interest by properly addressing selection effects such as the Malmquist bias. In contrast to the frequent use of Monte Carlo simulations where artificial sources with different absolute magnitudes are placed in the real data and tested whether they can be recovered with the desired S/N ratios (e.g.~\citealt{2015ApJ...803...34B}), we devised an independent empirically-motivated approach that takes into account both the limited detection sensitivity of HST as well as the angular variation in sky intensity. This empirical approach is particularly important for cluster lensing fields, where the sky intensity and hence the effective detection threshold can vary dramatically especially in the vicinity of bright cluster members.

\paragraph{Detection Limit}

\begin{figure}[tp!]
\centering
\includegraphics[width=85mm]{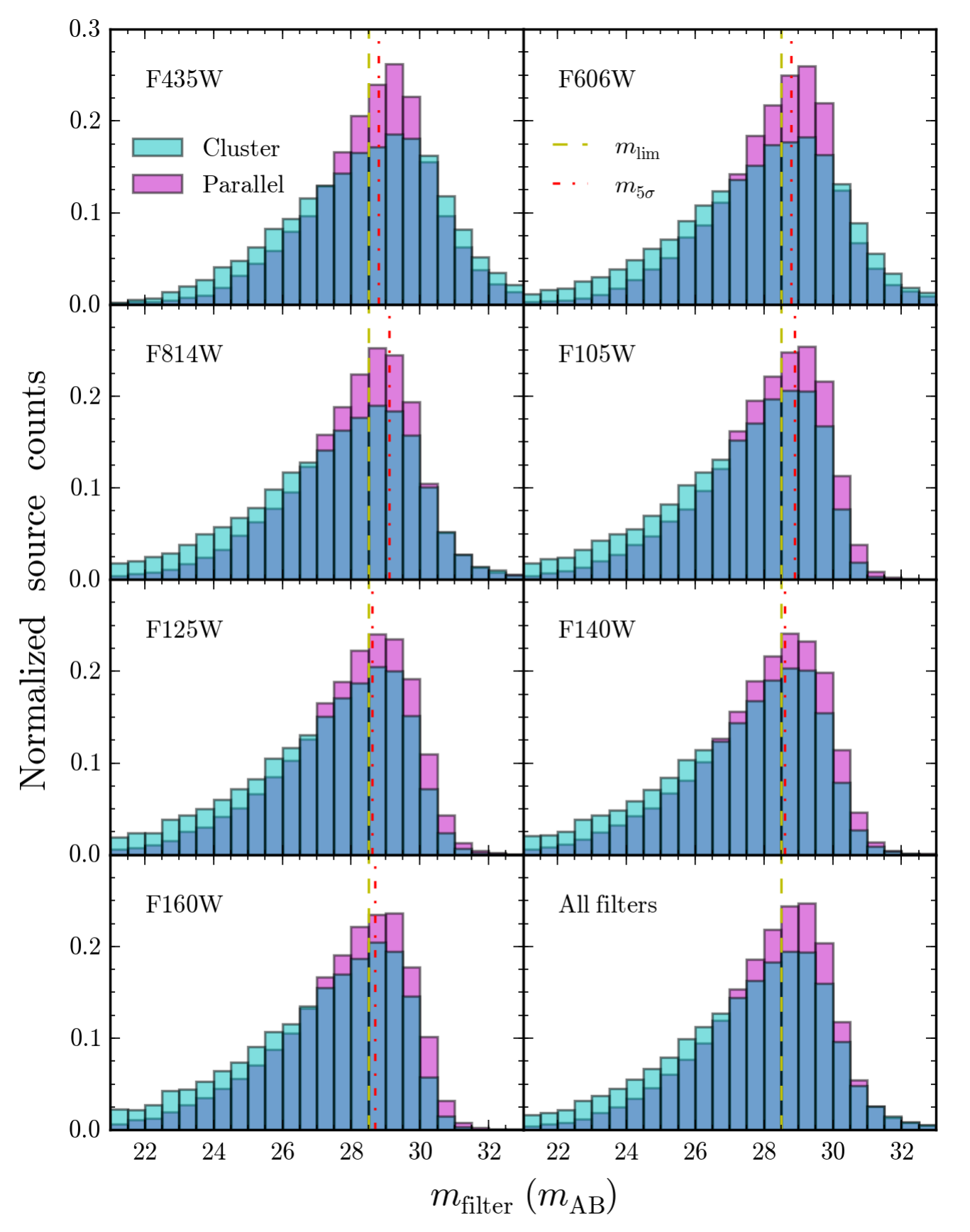}
\caption{\label{multi_fld_mAB_dist}Normalized source counts (i.e.~area under each histogram is equal to unity) in the ten HFF C15 catalogs as a function of the apparent magnitude $m_\mathrm{filter}$ in individual HFF filters (labelled at the upper left corner of each panel). Cyan and magenta bars denote the source counts in the cluster and parallel fields respectively. Yellow dashed lines show the apparent magnitude limit $m_\mathrm{lim} = 28.5$ that we set for all filters in subsequent analysis, whereas the red dash-dot lines show the expected 5$\sigma$ point-source limiting magnitudes for the corresponding HFF filters. The bottom right panel shows the normalized total source counts irrespective of filter.}
\end{figure}

The expected 5$\sigma$ limiting magnitude of the HFF fields varies from filter to filter (and also in position, which we address later) as shown in Table \ref{HFF_filters}. Therefore, we extracted the exhaustive set of sources (which include cluster members in the cluster fields) detected in the HFF cluster-parallel field pairs from the C15 catalogs, and present in Figure \ref{multi_fld_mAB_dist} the distributions of their apparent magnitudes measured in the seven HFF filters respectively. The number of source counts detected in each filter rises as we progress towards fainter magnitudes and turns over at $m_\mathrm{AB} \sim 28$\,--\,29 as expected for a magnitude-limited survey, with this trend seen in both the cluster and parallel fields.

Note that at $m_\mathrm{AB} \lesssim 27$, the relative fraction of sources is higher in the cluster fields than the parallel fields, because of the extra contribution from cluster members as well as some intrinsically luminous background sources which are further lensed (brightened up) by the clusters.  Conversely, the relative source fraction at fainter apparent magnitudes is lower in the cluster fields than the parallel fields.\footnote{In the two optical filters, F435W and F606W, there is a slight excess of the relative source fraction fainter than $m_\mathrm{AB} \sim 30$ in the cluster fields compared to the parallel fields. We traced this excess to a number of image artefacts present in the non-overlapping regions between different exposures in these two filters, which were misidentified as real sources when performing automated photometry in the cluster fields. These artefacts do not pose any threat to our identification of high-$z$ galaxy candidates since such non-overlapping regions were excluded from the sky area of interest within which we carried out our galaxy selections} as discussed in Section \ref{galaxy_candidate_selection}.

To ensure completeness, we adopted a conservative apparent magnitude limit that is well within the expected 5$\sigma$ point-source limits for all the filters. For simplicity, we set a globally constant apparent magnitude limit of $m_\mathrm{lim} = 28.5$, regardless of filter and field type (i.e.~cluster or parallel fields) in subsequent analyses of the HFF data.

\paragraph{Sky Background Variation}

We now consider the complication that the effective detection threshold is not entirely uniform over a given finite region in the image plane, especially in the neighbourhood of bright cluster members. The pervading diffuse or intracluster light boosts the intensity of the sky background, masking out faint background objects that would otherwise be detectable.

\citet{2016ApJ...819..114K} addressed this masking effect by dividing each cluster field into a coarse collection of grid cells (see their Figure 1) and measured the 5$\sigma$ limiting magnitudes in each grid cell.  From the F160W images, they found limiting magnitudes from as bright as $m_{5\sigma} \simeq 28.2$ near the cluster cores to as dim as $m_{5\sigma} \sim 29$ in the outskirts. Here we adopt a more stringent criterion by introducing ``exclusion regions'', within which the sky background is considered to be too bright and furthermore changes so rapidly that no sensible magnitude limit can be defined to guarantee data completeness.

\begin{figure*}[tp!]
\centering
\includegraphics[width=\textwidth]{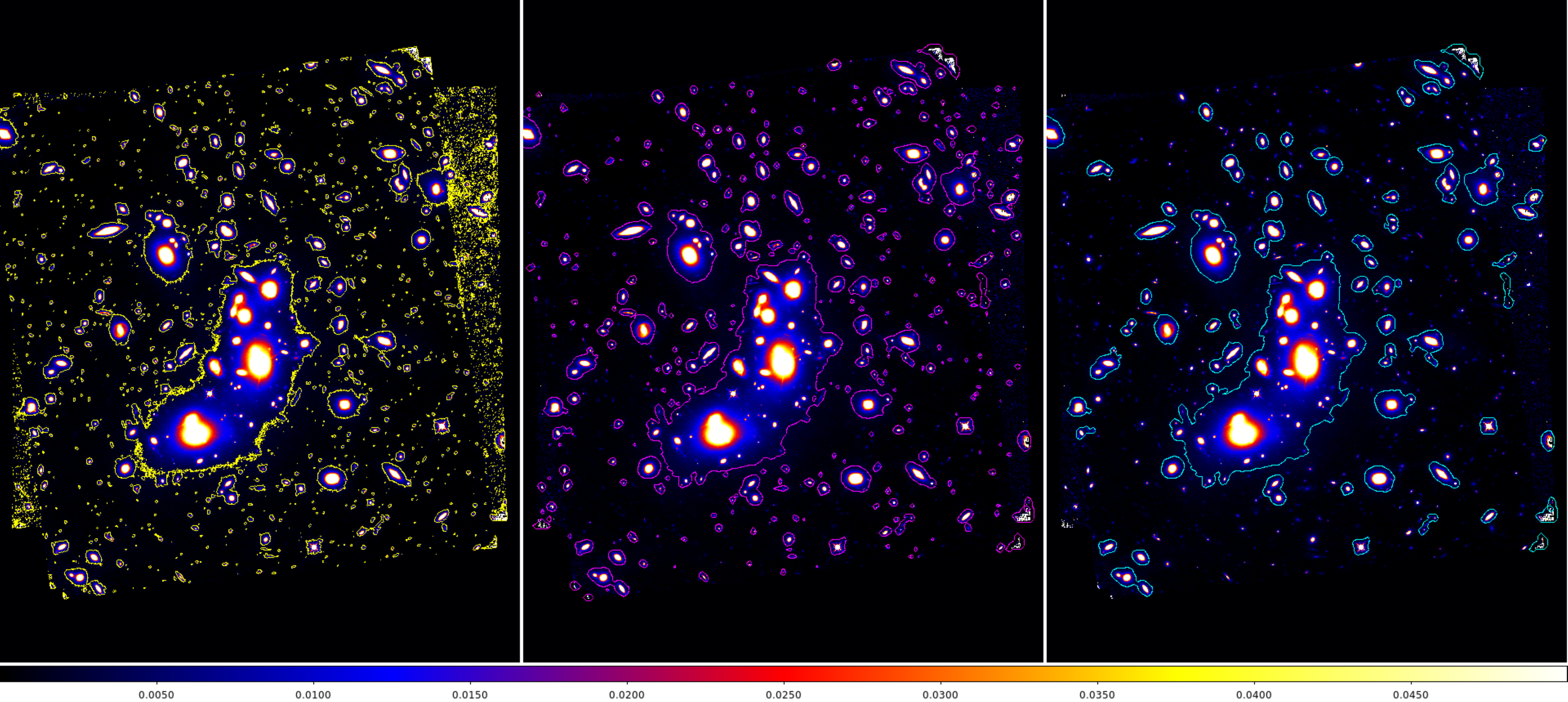}
\caption{\label{a2744_clt_contour_comparison}Illustration of the effects of different contour selection methods on the resultant set of isophotes used to define our exclusion regions, which are overlaid on identical F160W images of the A2744 cluster field (displayed in pseudocolour). {\it Left panel:} contour smoothing scale of $0.12\arcsec$. {\it Middle panel:} contour smoothing scale of $0.96\arcsec$. {\it Right panel:} same as the middle panel but with contours enclosing each a solid angle of less than 0.09 square arcseconds removed.}
\end{figure*}

We chose to work with the $0.03\arcsec \, \mathrm{pix}^{-1}$ F160W drizzled images (v1.0 data products; \citealt{2017AAS...23031613K}) to set up the exclusion regions. The reason for using an NIR filter as our reference is simply because the diffuse or intracluster light, contributed predominantly by giant elliptical cluster members, is most easily detected in the near-IR. A first attempt was made by tracing out single-level isophotal contours on the F160W images, evaluated at a smoothness scale of four image pixels ($0.12\arcsec$), but this resulted in typically $\sim$$10^4$ contours generated largely by statistical fluctuations due to Poisson noise as shown in the the left panel of Figure \ref{a2744_clt_contour_comparison} for the A2744 cluster field. To prevent such a large number of non-physical isophotes, we increased the smoothness of the contours by a factor of eight, whereby the number of isophotes dropped dramatically to $\sim$$10^2$ for each cluster fields as demonstrated in the middle panel of Figure \ref{a2744_clt_contour_comparison}. Many relatively large isolated ``islands'' can be seen in this panel enclosing individual bright cluster members and foreground stars (as well as image artefacts lying around the edges and corners of separate exposures). From visual inspection, we empirically selected only those islands with areas larger than 100 square pixels (0.09 square arcseconds) for masking, within which the detection of faint galaxies is compromised as shown in the right panel of Figure 2. Failure to properly handle these regions would result in biased estimates of the overall galaxy number density in the cluster fields. On the other hand, smaller islands encircling fainter objects, as can be seen in the middle panel of Figure 2, are retained as these regions may contain galaxies of interest that just happen to lie above the selected brightness level for exclusion. For consistency, we also defined the exclusion regions for the six parallel fields in the same manner.

\begin{figure*}[tp!]
\centering
\includegraphics[width=\textwidth]{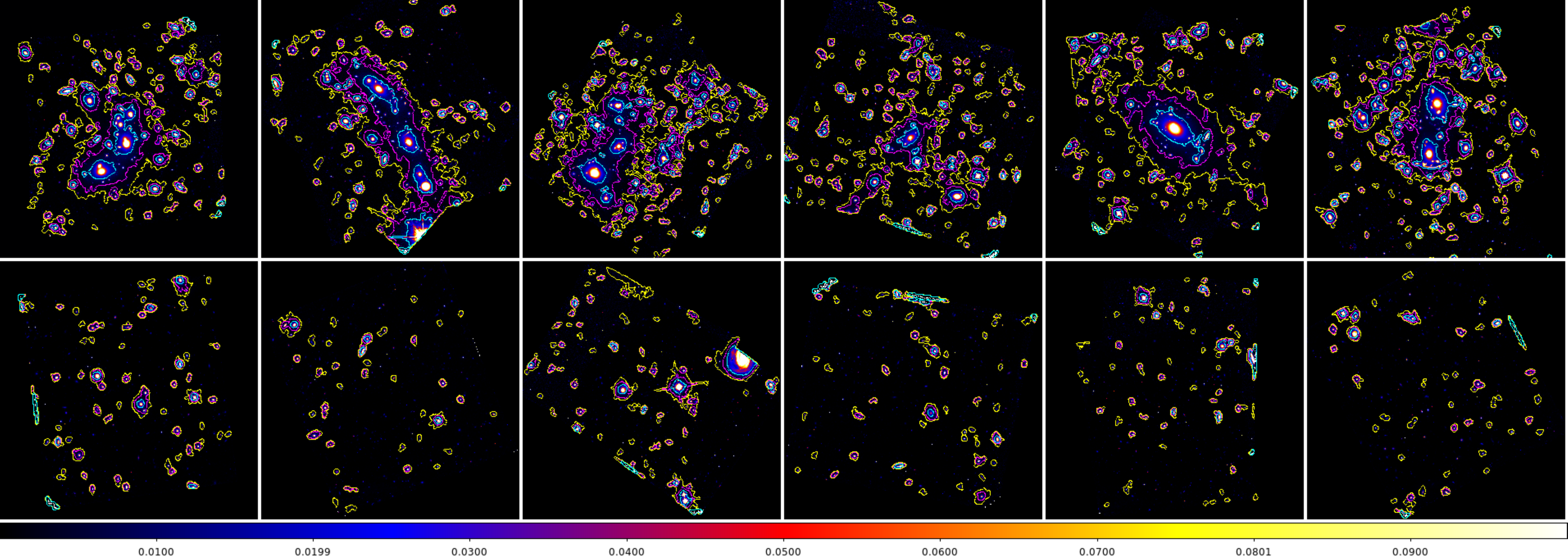}
\caption{\label{multi_fld_contour_comparison}Three different sets of (cleaned) isophotes with brightness levels corresponding to $0.01 \,\mathrm{e}^- \, \mathrm{s}^{-1}$ ({\it cyan}), $0.0025 \, \mathrm{e}^- \, \mathrm{s}^{-1}$ ({\it magenta}), and $0.001 \, \mathrm{e}^- \, \mathrm{s}^{-1}$ ({\it yellow}) respectively, overlaid on all the twelve fields being analyzed in our work. {\it Upper row:} cluster fields. {\it Lower row:} parallel fields. {\it Left to right: }A2744, MACS0416, MACS0717, MACS1149, AS1063, and A370.}
\end{figure*}

\begin{figure}[tp!]
\centering
\includegraphics[width=85mm]{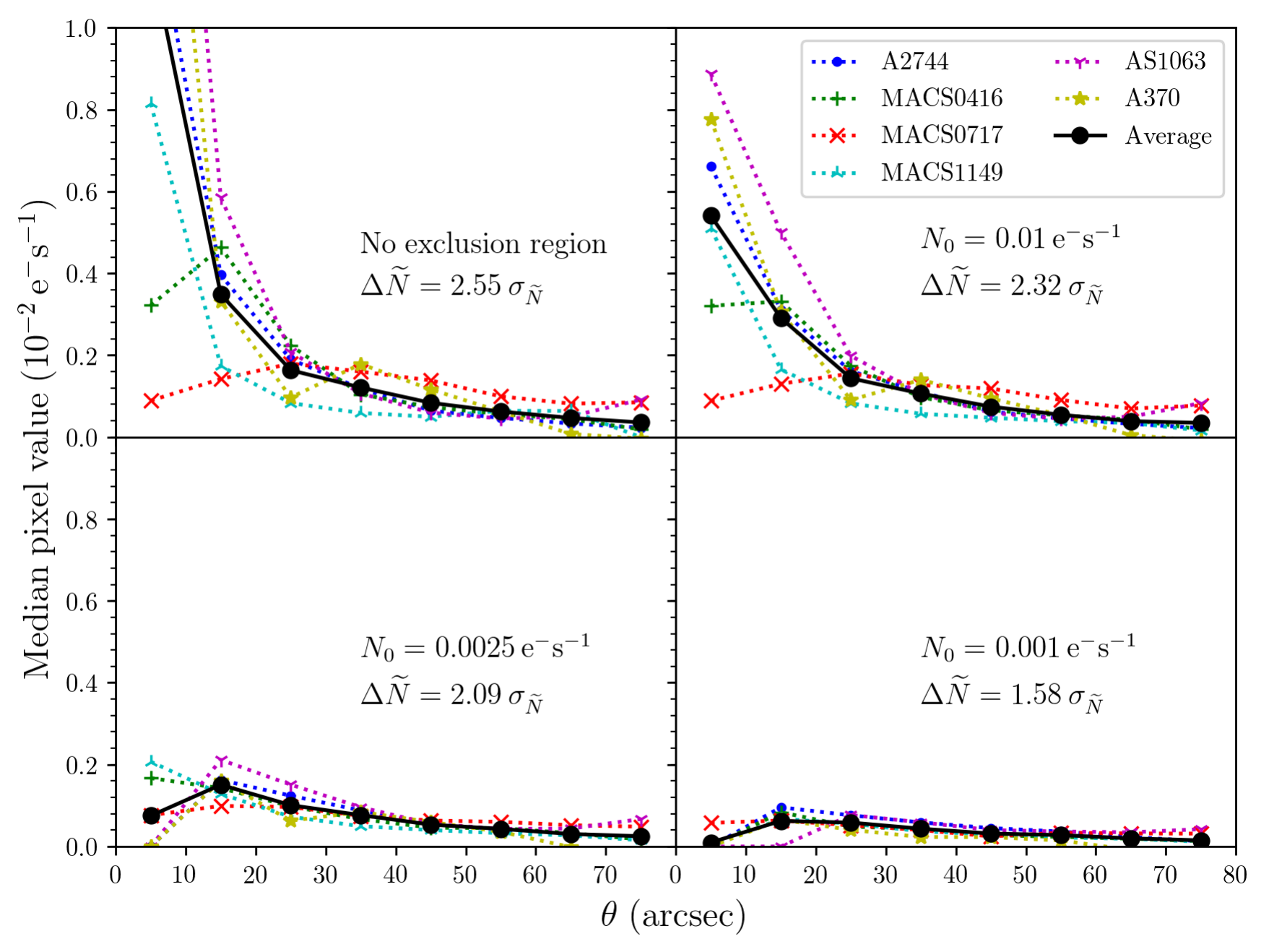}
\caption{\label{multi_clt_sky_brightness_dist_median}Sky intensity as a function of clustercentric angular radius for various choices of the isophotal level ($N_0$) as labelled in the panels, including the case where no exclusion region is defined ({\it upper left}). Black solid lines represent the median pixel values averaged among the six cluster fields at every radial bin. We provide the maximum deviation $\Delta \widetilde{N} \equiv \mathrm{Max}(\widetilde{N}) - \langle \widetilde{N} \rangle$ in units of $\sigma_{\widetilde{N}}$, the standard deviation of the (averaged) median pixel values at all radii, corresponding to each isophotal level. The median sky intensities for individual cluster fields are also shown (in differently coloured dotted lines) as reference.}
\end{figure}

\begin{figure}[tp!]
\centering
\includegraphics[width=85mm]{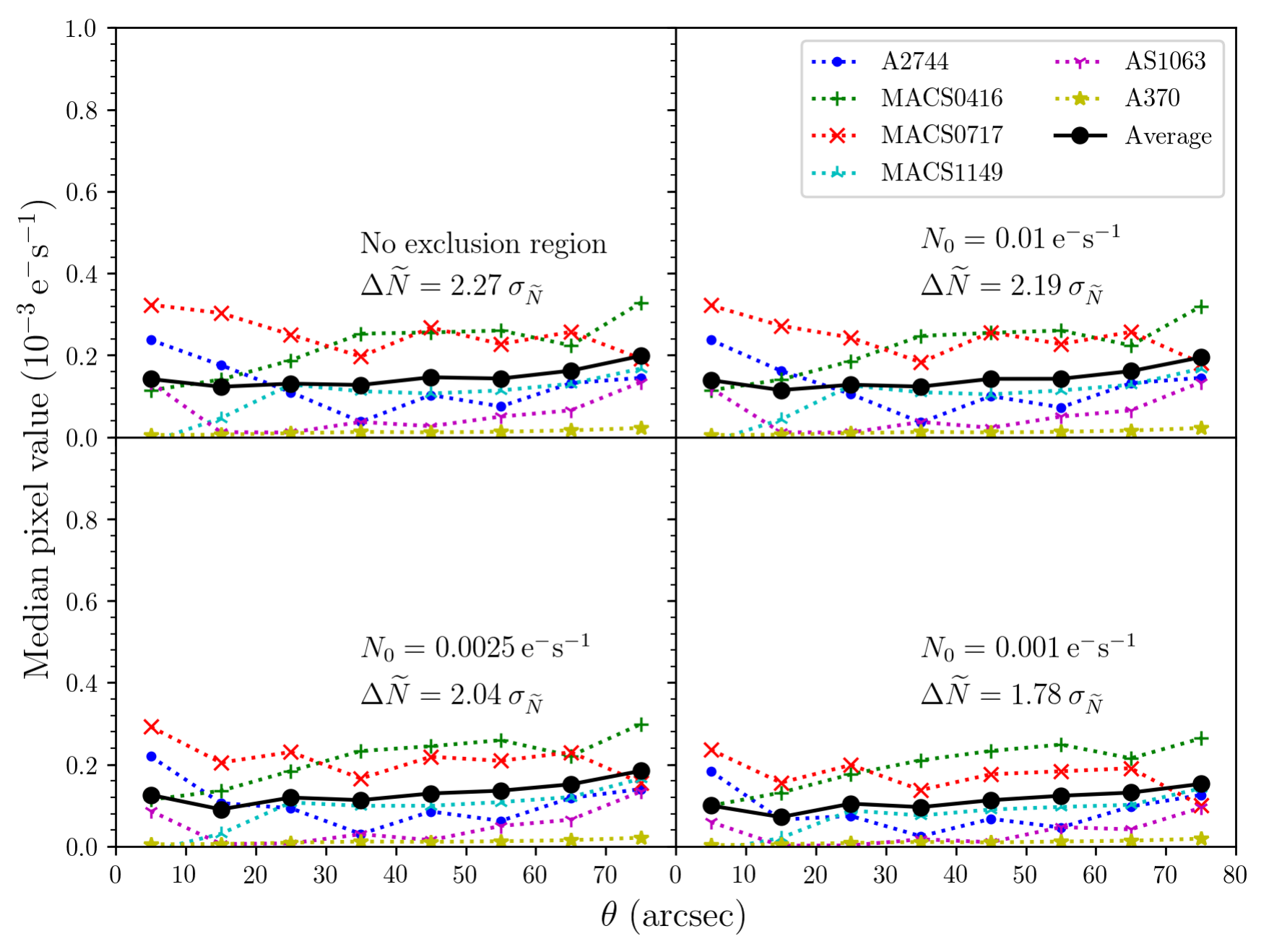}
\caption{\label{multi_par_sky_brightness_dist_median}Same as Figure \ref{multi_clt_sky_brightness_dist_median} for the parallel fields, where the ``centre'' of each such control field was chosen to be approximately the centre of the field of view. Note that the vertical scale is one dex lower than that of Figure \ref{multi_clt_sky_brightness_dist_median}, primarily because there is only little brightening of the sky background, if any, by diffuse light when compared to the cluster fields.}
\end{figure}

As illustrated in Figure \ref{multi_fld_contour_comparison}, the total solid angle enclosed by the isophotes is dependent on the value of the brightness level chosen to evaluate the contours. The lower the isophotal level, the larger the fraction of the field excluded from consideration. Figure \ref{multi_clt_sky_brightness_dist_median} shows how the resultant sky intensity outside of the exclusion regions depends on the choice of the isophotal level, where we plot the median image pixel values, neglecting the exclusion regions, at different clustercentric radii\footnote{The centre of a given cluster is defined here as the centre of the field of view spanned by the corresponding WSLAP+ lens model (see Section \ref{lensing_effects}).}. The median pixel value is an appropriate measure of the sky intensity because it is not sensitive to the extreme values originating from individual galaxies. We can see that when no exclusion regions are defined or when they are defined at an isophotal level of $0.01 \,\mathrm{e}^- \, \mathrm{s}^{-1}$, the sky intensity rises steeply towards the cluster core for four (A2744, MACS1149, AS1063, and A370) of the six cluster fields. The sky intensity in the remaining two cluster fields (MACS0416 and MACS0717) also rises towards the cluster centres, although not as dramatically due to the offset between the brightest cluster galaxies (BCGs) and the cluster centres. On the other hand, this inward rising trend becomes much more modest when we impose more stringent (i.e.~lower) isophotal levels. The degree of smoothness of the sky background can be quantified by comparing the maximum deviation of the median pixel value from the mean (averaged over all radii), $\Delta \widetilde{N} \equiv \mathrm{max}(\widetilde{N}) - \langle \widetilde{N} \rangle$, against the rms fluctuation $\sigma_{\widetilde{N}}$ of the median pixel values over all radii. The typical amplitude of this maximum deviation in the parallel control fields, which are presumably free of diffuse light contamination, is $\sim$$2\sigma_{\widetilde{N}}$ as can be seen in Figure \ref{multi_par_sky_brightness_dist_median}. For reference, a similar level of relative sky intensity variation, $\Delta \widetilde{N} / \sigma_{\widetilde{N}}$, can be achieved in the cluster fields when the isophotal level $N_0$ is equal to $\simeq$$0.0025 \, \mathrm{e}^- \, \mathrm{s}^{-1}$ (or equivalently $\simeq$$24.8 \, \mathrm{mag} \, \mathrm{arcsec}^{-2}$), which corresponds closely enough to a minimum in the sky intensity fluctuation while preserving a reasonably large sky area. Consequently, we adopted the exclusion regions (with $N_0 \simeq 0.0025 \, \mathrm{e}^- \, \mathrm{s}^{-1}$) thus defined in the rest of this work.

We emphasize that the dark sky area of interest, as defined above, excludes regions with background intensity above an empirically determined, conservative threshold. In addition, we limited the selection of objects in these regions to those lying well above the instrumental flux limit. This careful approach is the key to deriving credible constraints on the faint-end slope of the UV LF that are free from sizeable model-dependent ``completeness corrections'' present in most other works, where fake sources are scattered everywhere to see what fraction of them are not recovered, relying heavily on assumptions on the distributions of source shapes, sizes, and light profiles etc. Confidence in this standard approach is questionable as the upward corrections applied at the limiting luminosities from such source-recovery simulations very often exceed by several times the observed numbers of detected sources. It will also become evident in subsequent sections that because of the exceptionally high magnification supplied by foreground galaxy clusters, we can still probe comparably low, if not considerably lower, luminosities down the faint end of the UV LF despite using our conservative magnitude limit. Setting this limit ensures that data completeness is not a compromising issue in our work, compared with other blank-field studies that attempt to correct for spatially varying, profile-dependent detection limits in the photometrically incomplete regions using fake source-recovery simulations.

\subsection{Selection of High-\texorpdfstring{$z$}{z} Galaxy Candidates}\label{galaxy_candidate_selection}

We selected high-$z$ galaxy candidates from the twelve {\it Hubble} Frontier Fields based on the criteria described below. These criteria maximize data completeness beyond a selected redshift and above a selected brightness (apparent magnitude), while minimizing lower-z contaminants.

\paragraph{Photometric Redshift Threshold}

\begin{figure}[tp!]
\centering
\includegraphics[width=85mm]{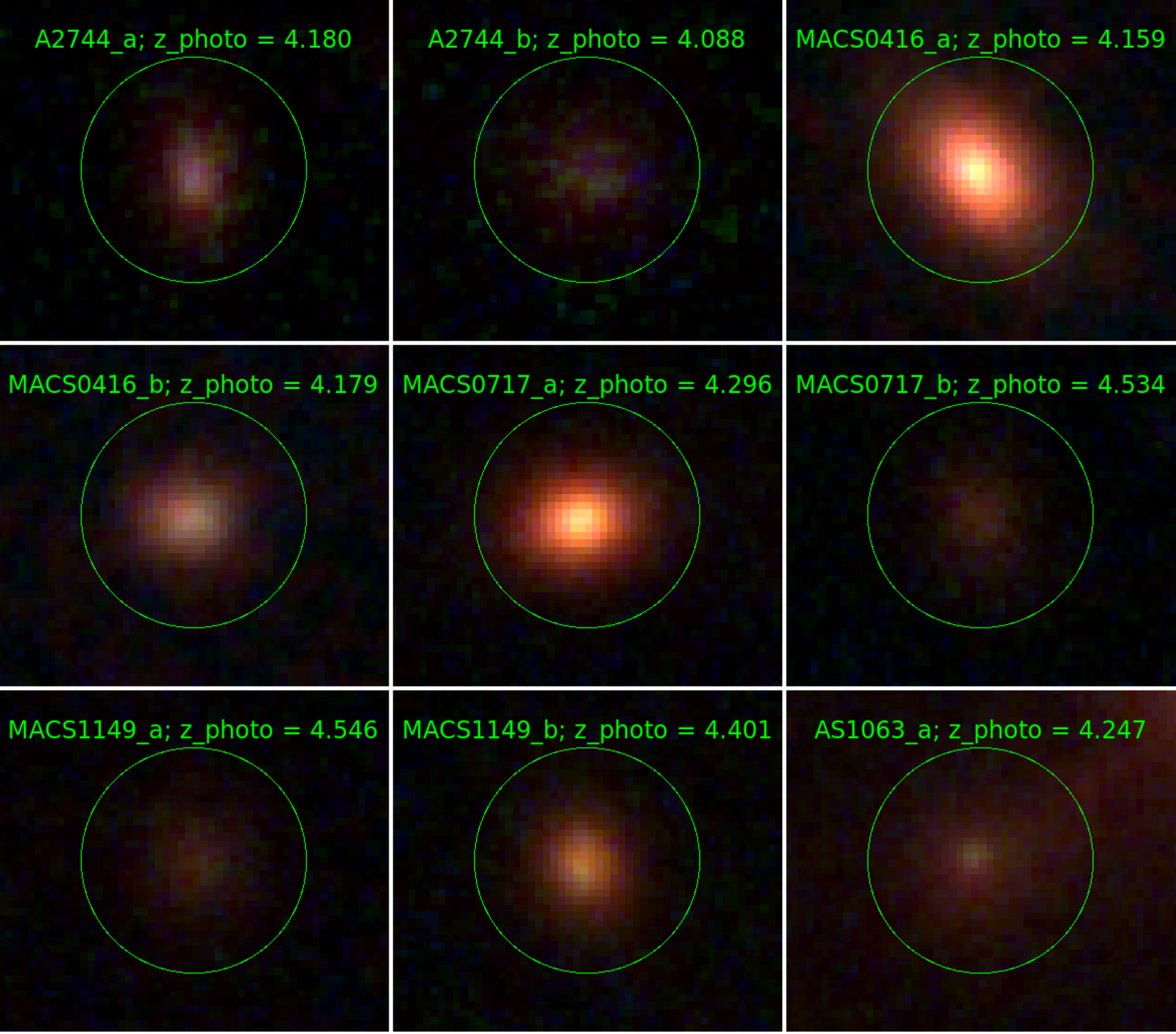}
\caption{\label{multi_clt_mem_misid}Selected sample of low-$z$ cluster members which were misidentified as $z  \sim 4$\,--\,4.5 dropout galaxies in the C15 photometric catalogs. The host cluster and the estimated photo-$z$ are labelled on top of each circled cluster member. All the panels are displayed in the same brightness and angular scales, illustrating the broad ranges of apparent magnitudes, colours, and sizes that they span. These NIR-weighted closeups were constructed from the multi-band HFF drizzled images by stacking F105W, F125W, and F140W as red, F814W as green, and F606W as blue. The angular scale of each green circle is $0.9\arcsec$.}
\end{figure}

The key feature in the spectral energy distributions (SEDs) of galaxies (constructed from the apparent magnitudes measured in the seven HFF bands) that enables them to be photometrically identified as high-$z$ objects is the Lyman break, which corresponds to an abrupt drop in the rest-frame UV continuum flux shortwards of the Lyman limit at 912\,\AA. For $z  \sim 4-$5 galaxies, the Lyman break is redshifted to $\sim$4560\,--\,5472\,\AA, hence falling between the F435W and F606W filters. A drop in flux between these two filters, however, is reminiscent of a similar behavior in the SEDs of low-$z$ early-type E/S0 galaxies, which constitute a significant proportion of the detected sources in the cluster fields. As a consequence, an inherent problem in applying the photometric dropout technique to crowded cluster fields is the inevitable inclusion of low-$z$ cluster members (and less severely other foreground galaxies) misclassified as high-$z$ dropout galaxies (and vice versa). Examples of such cases in several target cluster fields are shown in Figure \ref{multi_clt_mem_misid}. All these galaxies share round, featureless morphologies with colours similar to well-resolved cluster members in the vicinity. This degeneracy is gradually lifted for galaxies at higher redshifts where the Lyman break is redshifted to longer wavelengths, making these galaxies appear significantly redder so as to be easily distinguished from typical cluster members.

\begin{figure}[tp!]
\centering
\includegraphics[width=85mm]{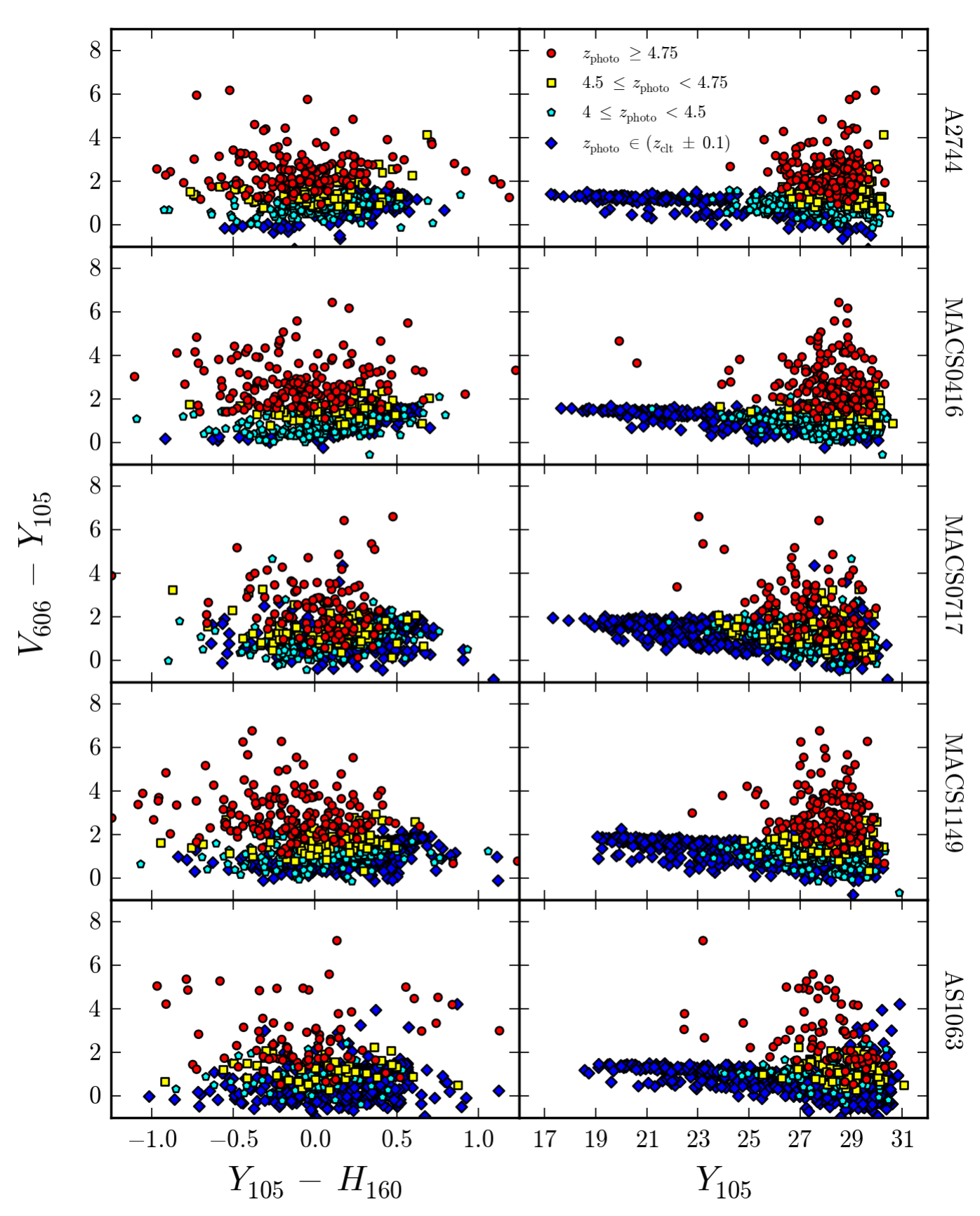}
\caption{\label{multi_clt_color_color_mag_diagram}Colour-colour ({\it left column}) and colour-magnitude ({\it right column}) diagrams for the first five completed cluster fields (labelled at the right vertical axis of each row) compiled using the C15 catalogs. Data points are colour-coded according to the redshift intervals that the sources belong to. The code scheme is defined as follows: $z_\mathrm{clt} - 0.1 < z_\mathrm{photo} < z_\mathrm{clt} + 0.1$ ({\it blue diamonds}), $4 \leq z_\mathrm{photo} < 4.5$ ({\it cyan pentagons}), $4.5 \leq z_\mathrm{photo} < 4.75$ ({\it yellow squares}), and $z_\mathrm{photo} \geq 4.75$ ({\it red circles}), where $z_\mathrm{clt}$ is the respective cluster redshift and $z_\mathrm{photo}$ is the estimated photo-$z$ of the source. Note that the same colour $V_{606} - Y_{105}$ is plotted as the vertical axes of both columns. This particular combination of filters is used to provide the clearest visual distinction between various galaxy populations.}
\end{figure}

In our work, it is important to identify a suitable lower redshift bound such that there is only little, if any, contamination of the high-$z$ galaxy sample from the low-$z$ cluster members and other foreground galaxies. We identified a suitable redshift threshold by plotting $V_{606} - Y_{105}$ against $Y_{105} - H_{160}$ as shown in Figure \ref{multi_clt_color_color_mag_diagram} (left panel). This figure also shows the $V_{606} - Y_{105}$ versus $Y_{105}$ diagram (right panel), where the red sequence of cluster members can be clearly seen (blue diamonds). All the identified galaxies in the C15 catalogs are separated, according to their corresponding photometric redshift estimates, into ``low-$z$'' ($z_\mathrm{photo} \in (z_\mathrm{clt} \pm 0.1)$) and ``high-$z$'' ($z_\mathrm{photo} \geq 4$) galaxies in this figure.\footnote{Quotation marks are being used to emphasize that the classification of galaxies here is solely based on their Bayesian photometric redshift estimates, which do not necessarily reflect their true redshifts.} It is obvious that galaxies at higher (photometric) redshifts tend to be redder in $V_{606} - Y_{105}$ but bluer in $Y_{105} - H_{160}$ (i.e.~migration towards the upper left in the colour-colour plane). Nevertheless, the $4 \leq z_\mathrm{photo} < 4.75$ ``dropout'' galaxies ({\it cyan pentagons} and {\it yellow squares}) are substantially blended into the population of ``low-$z$ cluster members'' ({\it blue diamonds}) in this colour-colour diagram. The same is also true in the colour-magnitude space diagram, demonstrating a scatter in the estimated photo-$z$'s owing to both random and systematic errors. Selecting galaxies within the redshift interval $4 \leq z  < 4.75$ based on their estimated photo-$z$'s would therefore result in a significant loss of actual dropouts as well as the undesired inclusion of cluster members. This misclassification between dropouts and cluster members is more severe in cases where the clusters are at slightly higher redshifts (e.g.~MACS0717 and MACS1149). Fortunately, the population of $z_\mathrm{photo} \geq 4.75$ galaxies ({\it red circles}) is, in general, quite well detached from the red sequence ({\it blue diamonds}) for most of the cluster fields. We therefore limited our analysis to only those galaxies lying beyond this ``critical'' redshift for the cluster fields, and for consistency also the parallel fields, in our work, i.e.~the first of our selection criteria is:
\begin{equation*}\tag{\uppercase\expandafter{\romannumeral 1\relax}}\label{photo-z_constraint}
z_\mathrm{photo} \geq 4.75.
\end{equation*}

Even with this conservative selection criterion, there may still be cluster members misidentified as high-$z$ dropouts, especially for the somewhat higher-redshift MACS0717 and MACS1149 clusters. We therefore verified the identity of every selected high-$z$ galaxy candidate by careful visual inspection, and manually removed all the obvious mimics resembling cluster members like those shown in Figure 6.  The number of such misclassified cases removed constitute about $10\%$ of the $z_\mathrm{photo} \geq 4.75$ galaxy sample, with many being clearly different in morphology (e.g.~more roundish and less compact) than the majority of such galaxy candidates.

\paragraph{Apparent Magnitude Threshold}

As mentioned in Section \ref{data_completeness}, to ensure data completeness, we strictly selected galaxies only if their apparent magnitudes are brighter than the imposed apparent magnitude limit of $m_\mathrm{lim} = 28.5$. In addition, as we would like to accurately estimate the rest-frame UV luminosities of our galaxy candidates, we imposed this apparent magnitude threshold as measured in the filter with an effective wavelength that, when scaled back to the source redshift as $\lambda_\mathrm{rest} = \lambda_\mathrm{eff} / (1 + z)$, is closest to the relevant rest-frame UV wavelength (1500\,\AA). In doing so, we had to be aware that the Ly$\alpha$ forest along the sightline towards the source, which attenuates the observed spectrum from $(1 + z) \times 1216 \, \mathrm{\AA}$ shortwards, can significantly overlap with the passband of the filter and thus lead to an underestimate of its intrinsic brightness.

The working principle used to choose the appropriate filter for a given source redshift interval, $z_\mathrm{int} \equiv [\mathrm{min}(z_\mathrm{int}), \mathrm{max}(z_\mathrm{int}))$, is described schematically as follows. We have a set of {\it HST} bandpass filters available for use from the HFF observations, each of which has an associated effective wavelength, $\lambda_\mathrm{eff}$, and a minimum (transmission) wavelength, $\lambda_\mathrm{min}$. Among the set of seven optical to near-IR filters, we further select a subset of filters in which their wavelength ranges of transmission do not overlap appreciably with the Ly$\alpha$ forest at the maximum source redshift in a given redshift bin interval (i.e.~$f = \{\mathrm{Filter} \in F : \lambda_\mathrm{min} \gtrsim \lambda_{\mathrm{Ly}\alpha\mathrm{,obs}}(\mathrm{max}(z_\mathrm{int}))\}$ where $F = \{\mathrm{F435W}, \dots, \mathrm{F160W}\}$ and $\lambda_{\mathrm{Ly}\alpha\mathrm{,obs}}(z) = (1 + z) \times 1216 \, \mathrm{\AA}$). We then calculate the absolute difference between the effective wavelength and the rest-frame UV wavelength redshifted to the observer's frame for every filter in this subset (i.e.~$\Delta\lambda = \{|\lambda_\mathrm{eff} - \lambda_\mathrm{UV,obs}(M(z_\mathrm{int}))| \ \forall \ \mathrm{Filter} \in f\}$ where $\lambda_\mathrm{UV,obs}(z) = (1 + z) \times 1500 \, \mathrm{\AA}$). Finally, we choose the filter having the smallest such difference to be the filter we use for identifying galaxy candidates in the source redshift interval (i.e.~$\mathrm{Filter}(z_\mathrm{int}) = \mathrm{Filter} \in f : |\lambda_\mathrm{eff} - \lambda_\mathrm{UV,obs}| = \mathrm{min}(\Delta\lambda)$). We applied the above set of procedures to determine the filters for which the corresponding HFF images were used for identifying galaxy candidates within various redshift intervals as listed in Table \ref{filter_selection}. As a result, the second galaxy selection criterion is
\begin{equation*}\tag{\uppercase\expandafter{\romannumeral 2\relax}}\label{apparent_magnitude_constraint}
m_\mathrm{filter}(z_\mathrm{photo} \in z_\mathrm{int}) \leq m_\mathrm{lim} = 28.5.
\end{equation*}

\begin{deluxetable}{c c}[tp!]
\tablecaption{HFF filters chosen for different source redshift intervals\label{filter_selection}}
\tablehead{\colhead{Source redshift interval} & \colhead{Filter}}
\startdata
$4.75 \leq z < 5$ & F814W \\
$5 \leq z < 6$ & F105W \\
$6 \leq z < 7$ & F105W \\
$7 \leq z < 8$ & F125W \\
$8 \leq z < 9$ & F140W \\
$9 \leq z < 10$ & F160W \\
\enddata
\end{deluxetable}

\begin{deluxetable*}{c c c c c c}[tp!]
\tablecaption{Effective solid angles and number of galaxies selected in various target fields\label{Omega_eff_and_gal_no}}
\tablehead{\colhead{Target cluster} & \colhead{Field type} & \colhead{$\Omega_\mathrm{eff}$} & \colhead{No.~of galaxies} & \colhead{Scaled galaxy count\tablenotemark{a}} & \colhead{Surface number density\tablenotemark{b}} \\
\colhead{} & \colhead{} & \colhead{($\mathrm{arcmin}^2$)} & \colhead{} & \colhead{} & \colhead{($\mathrm{arcmin}^{-2}$)}
}
\startdata
A2744 & cluster & 3.53 & \phm{c}67\tablenotemark{c} & 72.1 & 18.1 \\
A2744 & parallel & 4.61 & \phm{c}48\tablenotemark{c} & 40.3 & 10.4 \\
MACS0416 & cluster & 3.45 & \phm{c}40\phm{c} & 44.0 & 9.9 \\
MACS0416 & parallel & 2.37 & \phm{c}74\phm{c} & 120.9 & 31.2 \\
MACS0717 & cluster & 4.67 & \phm{c}34\phm{c} & 27.7 & 6.2 \\
MACS0717 & parallel & 4.64 & \phm{c}91\phm{c} & 75.9 & 19.6 \\
MACS1149 & cluster & 3.72 & \phm{c}62\phm{c} & 63.3 & 15.3 \\
MACS1149 & parallel & 4.12 & \phm{c}112\phm{c} & 105.2 & 27.2 \\
AS1063 & cluster & 3.70 & \phm{c}20\phm{c} & 20.5 & 4.9 \\
AS1063 & parallel & 2.97 & \phm{c}51\phm{c} & 66.5 & 17.2 \\
A370 & cluster & 3.72 & \phm{c}37\phm{c} & 37.8 & 9.1 \\
A370 & parallel & 4.52 & \phm{c}43\phm{c} & 36.8 & 9.5 \\
Cluster fields & total & 22.79 & \phm{c}260\phm{c} & \nodata & 10.4 \\
Parallel fields & total & 23.23 & \phm{c}419\phm{c} & \nodata & 18.0 \\
\enddata
\tablenotetext{a}{For the six cluster fields, the scaled galaxy count, $\hat{N}_i$, for the {\it i}-th cluster field is defined as the number of galaxies, $N_i$, in this field scaled by the ratio between the mean effective solid angle of all the cluster fields, $\bar{\Omega}_\mathrm{eff} = \sum_{i=1}^6 \Omega_{\mathrm{eff},i} / 6$, and its effective solid angle, $\Omega_{\mathrm{eff},i}$, i.e.~$\hat{N}_i \equiv N_i (\bar{\Omega}_\mathrm{eff} / \Omega_{\mathrm{eff},i})$. The same formula applies for the parallel fields except that $\bar{\Omega}_\mathrm{eff}$ is replaced with the mean effective solid angle of all the six parallel fields. The scaled galaxy count is defined such that it reflects the expected number of galaxies observed in a given cluster (parallel) field assuming that all the cluster (parallel) fields have the same effective solid angle.}
\tablenotetext{b}{The (image-plane) surface number density of galaxies in a given field is defined here as the number of observed source galaxies (i.e.~preserving only one galaxy count for each multiple image system identified in the cluster fields; see Section \ref{multiply_lensed_images_section} and Appendix \ref{multiple_image_info}) divided by the effective solid angle.}
\tablenotetext{c}{A2744 is the only case where we identified more galaxy candidates in the cluster field than in the parallel field, which can be partially attributed to the high-$z$ overdensity discovered in the A2744 cluster field by \citet{2014ApJ...795...93Z}.}
\end{deluxetable*}

\paragraph{Field Boundaries}

Because the final HFF images were constructed by stacking multiple single-exposure images having slightly different pointing centres and orientations, the detection threshold is shallower in regions without complete coverage. An example can be seen in Figure \ref{a2744_clt_contour_comparison}, which also shows image artefacts at the edges of individual exposures that render the surrounding pixels unusable. Therefore, the useable field for a given filter is limited to that covered in all exposures. To be uniform across filters, we further restricted the useable field to that common in all the filters. These restrictions ensured that the most complete (and deepest) set of multi-band photometric data possible was utilized to deduce the photo-$z$'s of galaxies.

In addition to the aforementioned field boundary constraints, we needed to obtain the local magnification at the observed position of each selected galaxy in the cluster fields so as to infer its intrinsic (i.e.~unlensed) UV luminosity. This places an extra requirement on the selection of galaxies in a given cluster field to be within the field of view, $\boldsymbol{\Omega}_\mathrm{lens}$, covered by the corresponding lens model being used.

Combining these two constraints with the completeness condition in Section \ref{data_completeness} that the galaxies have to be situated outside of the exclusion regions $\boldsymbol{\Omega}_\mathrm{exclusion}$, the last galaxy selection criterion enforces a requirement on the angular position, $\boldsymbol{\theta}_\mathrm{gal}$, of a given galaxy that
\begin{equation*}\tag{\uppercase\expandafter{\romannumeral 3\relax}}\label{angular_position_constraint}
\boldsymbol{\theta}_\mathrm{gal} \in \boldsymbol{\Omega}_\mathrm{eff} \equiv (\boldsymbol{\Omega}_\mathrm{data} \cap \boldsymbol{\Omega}_\mathrm{lens}) \setminus \boldsymbol{\Omega}_\mathrm{exclusion}.
\end{equation*}

\paragraph{Spurious Objects}

Our high-$z$ galaxy candidates satisfy the composite criterion (\ref{photo-z_constraint}) $\land$ (\ref{apparent_magnitude_constraint}) $\land$ (\ref{angular_position_constraint}). Although the criteria imposed also mitigate against possible mis-selection of spurious objects, some contaminants other than cluster members, such as foreground stars and bad pixels (most likely induced by cosmic ray hits), remain in the pool of selected galaxy candidates. We therefore inspected the selected fields carefully to remove all such remaining contaminants. In this way, we identified 679 galaxy candidates altogether from a total sky area of $\simeq$$46.0 \, \mathrm{arcmin}^2$ contributed by all of the six HFF cluster fields and the accompanying six HFF parallel fields, as summarized in Table \ref{Omega_eff_and_gal_no}. Detailed information of each of the selected galaxy candidates is provided in Appendix \ref{galaxy_info}.

\paragraph{Cosmic Variance}

\begin{deluxetable*}{c c c c}[tp!]
\tablecaption{Sample variance in galaxy counts in the HFF cluster fields and parallel fields\label{fld_galaxy_count_variance}}
\tablehead{\colhead{} & \colhead{Mean (scaled) galaxy count} & \colhead{Standard deviation} & \colhead{Poisson noise} \\
\colhead{} & \colhead{$\langle \hat{N} \rangle$} & \colhead{$\sigma_{\hat{N}}$} & \colhead{$\sqrt{\langle \hat{N} \rangle}$}
}
\startdata
Cluster fields & 44.2 & 20.1 & 6.7 \\
Parallel fields & 74.3 & 33.9 & 8.6 \\
\enddata
\end{deluxetable*}

\begin{figure}[tp!]
\centering
\includegraphics[width=85mm]{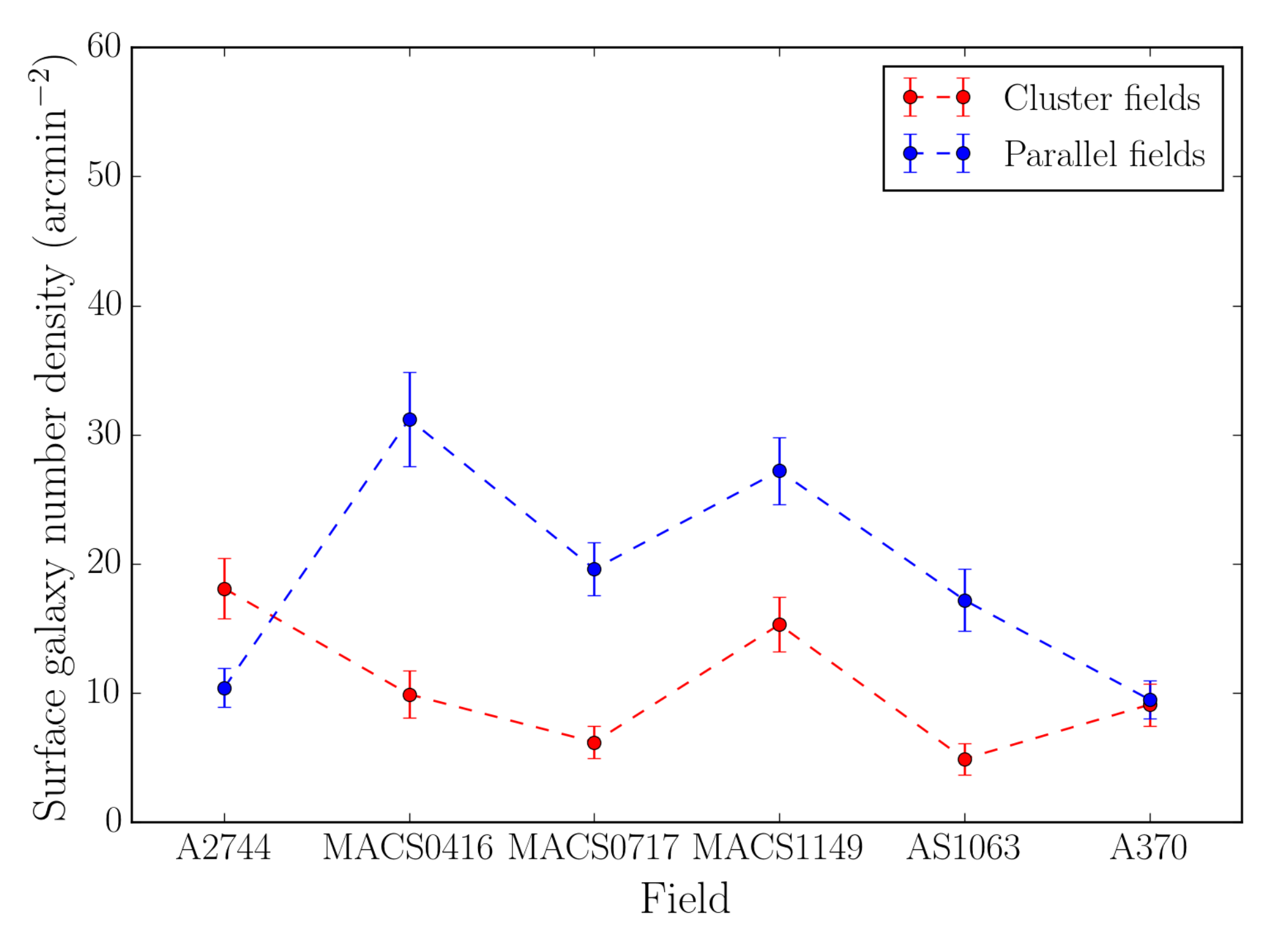}
\caption{\label{clt_par_fld_gal_density_comparison}Comparison of the image-plane surface number densities of galaxies between each of the six HFF cluster fields ({\it red circles}) and the accompanying parallel fields ({\it blue circles}) located at $6\arcmin$ away. Error bars are Poisson errors evaluated as $\sqrt{N}/\Omega_\mathrm{eff}$. Dashed lines are used to join the data points belonging either to the cluster or parallel fields for visual clarity.}
\end{figure}

We list in Table \ref{fld_galaxy_count_variance} the mean galaxy counts (scaled to have the same effective solid angle in every field; see caption (a) of Table \ref{Omega_eff_and_gal_no} for the precise definition) in the cluster and parallel fields respectively, and compare the standard deviations in these mean galaxy counts, $\langle \hat{N} \rangle$, with the errors expected purely from Poisson noise, i.e.~$\langle \hat{N} \rangle^{1/2}$. It can be seen that rather than being solely described by Poisson noise, the fluctuations in galaxy counts between either the HFF cluster or parallel fields are dominated by cosmic variance which outweighs the former contribution by more than a factor of 2. In addition, Figure \ref{clt_par_fld_gal_density_comparison} shows a comparison of the image-plane surface number densities of galaxies between the six pairs of cluster and parallel fields. Except for the case of A2744 where \citet{2014ApJ...795...93Z} identified a high-$z$ overdensity in the cluster field (and perhaps for A370 where the two data points are comparable), the ``trend'' of the remaining data points indicates that the galaxy number densities in the cluster and parallel fields are clearly not independent, and is caused by galaxy clustering on a physical scale of $\simeq$$2.2 \, \mathrm{Mpc}$ at $z \simeq 5$, corresponding to an angular separation of $6\arcmin$ between each cluster-parallel field pair. (The general offset between the pairs of data points is a direct consequence of the negative magnification bias in the cluster fields as will be discussed in detail in Section \ref{magnification_bias_section}.) Hence, the results here reflect that clustering bias in the various HFF target fields is not negligible, thereby it is crucial for us to analyze the ensemble of fields collectively instead of individually so as to mitigate the effect of cosmic variance.


\section{Lensing Effects}\label{lensing_effects}

Background galaxies in the cluster fields are magnified by the foreground galaxy clusters in size and hence also flux (while preserving surface brightness) by the same factor at all wavelengths. This (position- and redshift-dependent) magnification also results in a reduction in the actual source-plane area probed, albeit providing an enhanced (intrinsic) flux detection limit. Knowledge of the local magnification factor is based on a rigorously constructed lens model, which describes the (DM-dominated) mass distribution of a galaxy cluster. The overall cluster lensing potential is contributed primarily by the cluster-scale DM halo, but with local perturbations contributed by individual cluster members. In unrelaxed clusters, which constitute the majority of the HFF clusters, the cluster-scale DM may not necessarily follow that of the luminous galaxies. For our work, there is an additional problem, namely the paucity of cluster members in the cluster outskirts available to serve as reliable tracers of the underlying cluster-scale DM distribution, where there is nevertheless a substantial number of identified background galaxies. We therefore used free-form lens models, which do not assume specific parametric forms to describe the cluster-scale DM component, to compute magnification estimates. The lens models used are those published by \citet{2014ApJ...797...98L} for A2744, \citet{2015MNRAS.447.3130D} for MACS0416, \citet{2015MNRAS.451.3920D} for MACS0717, \citet{2016MNRAS.456..356D} for MACS1149, \citet{2016MNRAS.459.3447D} for AS1063, and \citet{2018MNRAS.473.4279D} for A370. All these lens models were constructed using the semi-parametric lens modelling package WSLAP+ (Weak and Strong Lensing Analysis Package plus member galaxies) \citep{2005MNRAS.360..477D, 2014MNRAS.437.2642S} as introduced below.  We note that a number of lens models developed with WSLAP+ have been subjected to rigorous internal consistency as well as predictive tests. The latter includes the orientation of extended lensed images and the relative fluxes of individual sets of multiply lensed images (for example that made for A2744 by \citet{2014ApJ...797...98L}), and also the correct prediction for the location and time of reappearance of the multi{}ply lensed supernova, SN Refsdal, in MACS1149 \citep{2016MNRAS.456..356D}.

\subsection{WSLAP+}\label{wslap_section}

WSLAP+ is a free-form method used to model gravitational lenses. The mass in the lens plane is modelled as a combination of a large-scale component and a compact component. The large-scale component is a superposition of Gaussian functions loca{}ted at a distribution of grid points that can be regular or adaptive, whereas the compact component accounts for the mass (baryons and DM) associated with the member galaxies. Usually the mass distribution is assumed to follow the distribution of light for the compact component, where the member galaxies are selected from elliptical-type galaxies in the red sequence and/or from a redshift catalog. A detailed description of the code and the various improvements implemented in different versions of the code can be found in \citet{2005MNRAS.360..477D,2007MNRAS.375..958D,2016MNRAS.456..356D} and \citet{2014MNRAS.437.2642S}. 

The inputs of the reconstruction can either be strong or weak lensing (shear) measurements, or a combination of both. For strong lensing data, the inputs are the pixel positions of multiply lensed galaxies (not just the centroids). In the case of featureless elongated arcs near the critical curves, the entire arc is mapped and included as a constraint. However, if the arclets have individual resolved features, they can be incorporated as semi-independent constraints but with the additional condition that they need to coincide in the source plane. Incorporating this information acts as an anchor to constrain the range of possible solutions and reduce the risk of bias due to the minimization being carried out in the source plane. WSLAP+, as its name suggests, is also compatible with weak lensing measurements if they are available for use. Mass reconstruction can therefore be performed on a much wider angular scale. For our work, where we focus only on the highly magnified and hence the inner regions of the cluster fields, strong lensing data alone are sufficient to enable relatively accurate lens models to be produced.

\subsection{Magnifications}\label{magnification_section}

The magnification factor $\mu(\boldsymbol{\theta},z)$ at a given observed position $\boldsymbol{\theta}$ and source redshift $z$ is derived from the model deflection field $\boldsymbol{\hat{\alpha}}(\boldsymbol{\theta})$ via the following relation,
\begin{align}\label{magnification_formula}
\begin{split}
\mu(\boldsymbol{\theta},z) & = \frac{\delta\theta^2}{\delta\beta^2} = \mathrm{det} \Bigg[\frac{\partial\boldsymbol{\theta}}{\partial\boldsymbol{\beta}(\boldsymbol{\theta},z)}\Bigg] \\
& = \mathrm{det} \Bigg[\delta_{ij} - \frac{\partial\alpha_i(\boldsymbol{\theta},z)}{\partial\theta_j}\Bigg]^{-1},
\end{split}
\end{align}
where $\boldsymbol{\beta}(\boldsymbol{\theta},z) = \boldsymbol{\theta} - \boldsymbol{\alpha}(\boldsymbol{\theta},z)$ is the source position, $\boldsymbol{\alpha}(\boldsymbol{\theta},z) = (D_\mathrm{ls}/D_\mathrm{s}) \boldsymbol{\hat{\alpha}}(\boldsymbol{\theta})$ is the reduced deflection angle, $D_\mathrm{ls}$ and $D_\mathrm{s}$ are the angular diameter distances from the lens to the source and from the observer to the source, respectively. Due to a parity flip (corresponding to a change in the orientation of the lensed image from radial to transverse), a change of sign in the magnification factor occurs whenever a lensed image crosses a critical curve. Because we are interested only in the magnitude of the magnification, henceforth we implicitly define $\mu \equiv |\mu|$ in the rest of the paper. To mitigate unrealistic overestimates of the magnification factors near the critical curves resulted from the lack of predictive power of the lens models in these regions, we also enforced an upper limit of $\mu_\mathrm{lim} = 100$ to the magnification factor whenever necessary, such as when we were estimating the intrinsic source-plane volume.

The vast majority of the selected galaxy candidates in the cluster fields do not suffer from extreme shear so as to be appreciably stretched. Such galaxies have half-light radii of typically $\lesssim$$0.5\arcsec$, comparable to the pixel scales of the lens models used. We therefore assumed that the magnitude of the magnification factor varies ``slowly'' compared to the sizes, $\delta\theta_\mathrm{gal}$, of the observed galaxies, i.e.~$(\partial\mu/\partial\theta)\delta\theta_\mathrm{gal} \ll 1$. This assumption is valid in low-magnification regions where the magnification factor is close to 1. Near the critical curves where $\mu \gtrsim 10$, however, the aforementioned condition is generally not satisfied. In such situations, provided that the angular variation in the magnification factor roughly follows $(\partial\mu/\partial\theta)\delta\theta_\mathrm{gal} \sim 0.01\mu^2(\delta\theta_\mathrm{gal}/1\arcsec) \sim 1$, the error in the estimated magnification factor is dominated by the uncertainty in the model deflection field rather than the apparent image size (see Appendix \ref{galaxy_info}). Thus, given the resolutions of the lens models, it suffices to approximate the magnification of an observed galaxy to be constant over the entire image and be equal to the value computed at its centroid.

\begin{figure}[tp!]
\centering
\includegraphics[width=85mm]{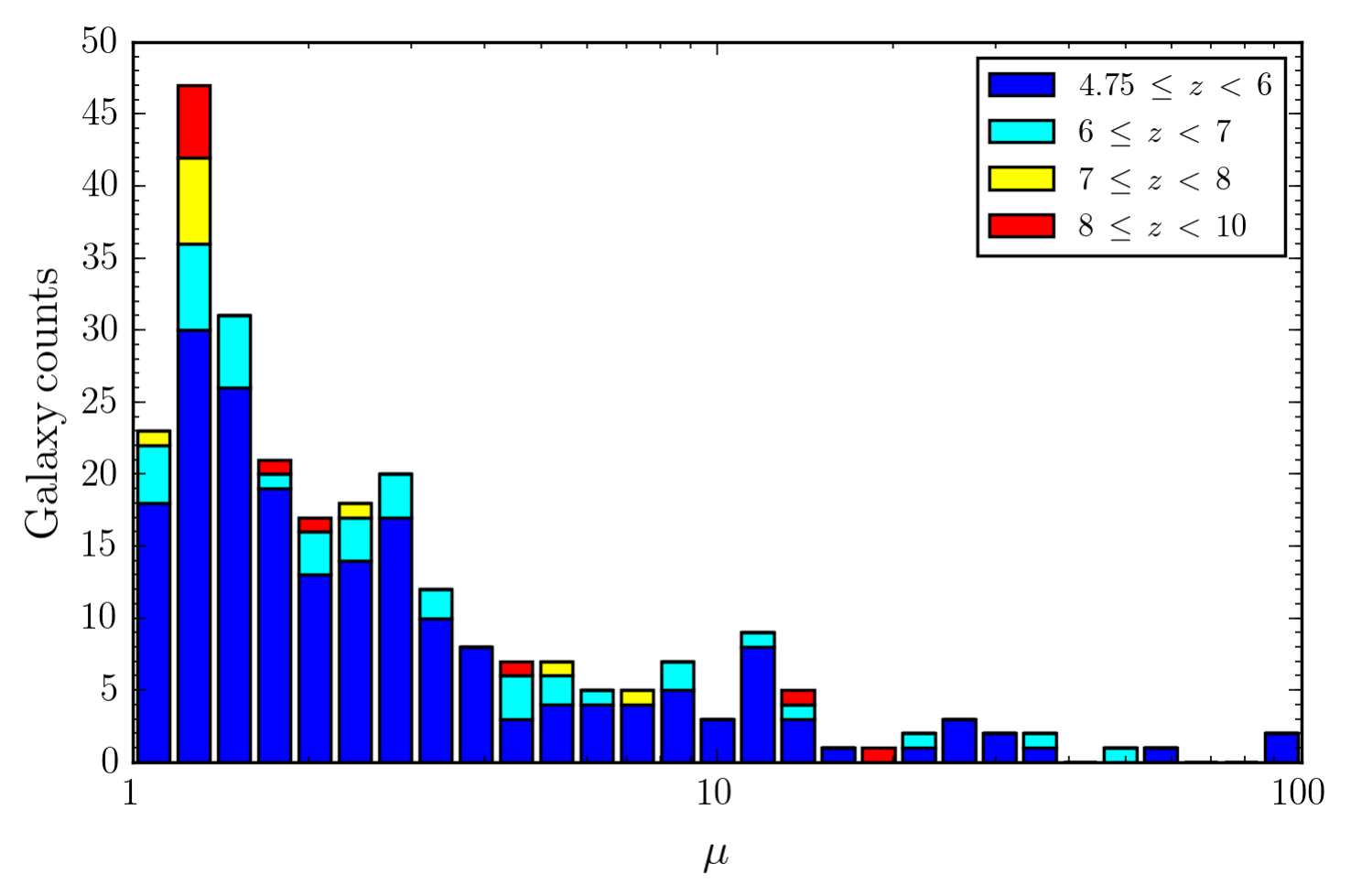}
\caption{\label{multi_clt_z_binned_magnification_distribution}Distribution of the magnification factors $\mu$ of selected galaxy candidates in the cluster fields. The proportions of the galaxy counts belonging to various source redshift intervals are represented in different colours as follows, $4.75 \leq z < 6$ ({\it blue}), $6 \leq z < 7$ ({\it cyan}), $7 \leq z < 8$ ({\it yellow}), and $8 \leq z < 10$ ({\it red}). Note that the horizontal axis is in logarithmic scale.}
\end{figure}

In Figure \ref{multi_clt_z_binned_magnification_distribution}, we plot the distribution of magnification factors for all the 260 galaxy candidates selected from the cluster fields as derived from the free-form lens models through Equation (\ref{magnification_formula}). We can see that the majority of galaxy candidates in the cluster fields have magnification factors $\mu < 10$, implying that they are located relatively far from the critical curves. The relative paucity of galaxy candidates near critical curves is due in part to our choice of exclusion regions (Section \ref{data_completeness}), which mask out large areas of extremely magnified regions around the critical curves close to the cD galaxies. Furthermore, the number of highly magnified galaxies is expected to be relatively low owing to the empirical relation $A(>\!\mu) \propto \mu^{-2}$, where $A(>\!\mu)$ is the area in the source plane with a corresponding (image-plane) magnification factor higher than $\mu$, implying that the probability for a given source galaxy to be magnified by, say a factor of 100, should be 100 times lower than that of being magnified by a factor of 10 \citep{1992grle.book.....S,2017arXiv170610281D}.

The associated parallel fields likely experience a minute lensing effect from the clusters in the neighbouring cluster fields. To account for this effect, we assumed a 5\% fiducial magnification (i.e.~$\mu = 1.05$) for all the selected galaxy candidates in the parallel fields.  As we shall see, this small correction has little effect on the results.

\subsection{UV Luminosities}

We derive in this section an operational formula for determining the UV luminosities of our galaxy candidates. First, we consider the effect of magnification on the observed flux of a lensed galaxy. As the apparent solid angle subtended by the galaxy is magnified by a factor $\mu(\boldsymbol{\theta},z)$, its unlensed and lensed spectral flux densities (per unit frequency), denoted as $f_{\nu,\mathrm{unlensed}}$ and $f_{\nu,\mathrm{lensed}}$ respectively, are related according to
\begin{equation}
f_{\nu,\mathrm{unlensed}}(\nu_{\mathrm{eff},R}) = \frac{f_{\nu,\mathrm{lensed}}(\nu_{\mathrm{eff},R})}{\mu(\boldsymbol{\theta},z)},
\end{equation}
as measured through a filter with an effective frequency $\nu_{\mathrm{eff},R} \equiv c / \lambda_{\mathrm{eff},R}$ and a bandpass response $R(\nu)$. Hence, in terms of apparent magnitudes, which we label as $m_R$, the corresponding unlensed and lensed quantities satisfy the relation
\begin{align}
m_{R,\mathrm{unlensed}} & = m_R - 2.5 \ \log_{10} \Bigg(\frac{f_{\nu,\mathrm{unlensed}}(\nu_{\mathrm{eff},R})}{f_{\nu,\mathrm{lensed}}(\nu_{\mathrm{eff},R})}\Bigg) \nonumber\\
& = m_R + 2.5 \ \log_{10} \big(\mu(\boldsymbol{\theta},z)\big).
\end{align}
For an equivalent measurement made now in the rest frame of the galaxy using a general filter having a bandpass response $Q(\nu)$, the absolute magnitude of the galaxy is given by (e.g.~\citealt{2002astro.ph.10394H})
\begin{equation}
M_Q = m_{R,\mathrm{unlensed}} - DM(z) - K_{QR}(z),
\end{equation}
where $DM(z) \equiv 5 \ \log_{10} (D_\mathrm{L}(z) / 10 \, \mathrm{pc})$ is the distance modulus, with $D_\mathrm{L}(z)$ being the luminosity distance at redshift $z$. $K_{QR}(z)$ is the $K$ correction, which originates from the redshifting of the bandwidth due to cosmological expansion, given by
\begin{align}\label{k_correction_general_formula}
\begin{split}
K_{QR}(z) = -2.5 \ \log_{10} \Bigg[(1 + z) \ & \frac{\int f_\nu(\nu_\mathrm{o})R(\nu_\mathrm{o})\frac{d\nu_\mathrm{o}}{\nu_\mathrm{o}}}{\int g_\nu^R(\nu_\mathrm{o})R(\nu_\mathrm{o})\frac{d\nu_\mathrm{o}}{\nu_\mathrm{o}}} \\
\times \ & \frac{\int g_\nu^Q(\nu_\mathrm{e})Q(\nu_\mathrm{e})\frac{d\nu_\mathrm{e}}{\nu_\mathrm{e}}}{\int f_\nu(\frac{\nu_\mathrm{e}}{1 + z})Q(\nu_\mathrm{e})\frac{d\nu_\mathrm{e}}{\nu_\mathrm{e}}}\Bigg]
\end{split}
\end{align}
where $\nu_\mathrm{o}$ and $\nu_\mathrm{e}$ are the observed and emitted frequencies respectively such that $\nu_\mathrm{e} = (1+z) \, \nu_\mathrm{o}$, and $g_\nu^R(\nu) = g_\nu^Q(\nu) = 3631 \, \mathrm{Jy}$ is the zero-point spectral flux density for AB magnitudes. To simplify the above equation, we can further choose a rest-frame bandpass filter designed such that $Q(\nu_\mathrm{e}) = R(\nu_\mathrm{o})$ and the effective frequency $ \nu_{\mathrm{eff},Q} = (1 + z) \, \nu_{\mathrm{eff},R}$, then Equation (\ref{k_correction_general_formula}) will be reduced to
\begin{align}
K_{QR}(z) & = -2.5 \ \log_{10} \Bigg[(1 + z) \, \frac{\int f_\nu(\nu_\mathrm{o})R(\nu_\mathrm{o})\frac{d\nu_\mathrm{o}}{\nu_\mathrm{o}}}{\int f_\nu(\frac{\nu_\mathrm{e}}{1 + z})Q(\nu_\mathrm{e})\frac{d\nu_\mathrm{e}}{\nu_\mathrm{e}}}\Bigg] \nonumber\\
& = -2.5 \ \log_{10} (1 + z).
\end{align}

Having obtained the rest-frame spectral flux density in terms of the measurable quantities, we can proceed to extrapolate it from the effective wavelength $\lambda_{\mathrm{eff},R}$ of the given filter to the rest-frame UV wavelength at 1500\,\AA. Assuming that the UV continua of high-$z$ galaxies obey a power-law relation $f_\lambda \propto \lambda^\beta$ (e.g.~\citealt{1999ApJ...521...64M}),\footnote{Unfortunately due to notational conventions in the literature, the symbol $\beta$ is both used here to refer to the UV continuum slope and used in Sections \ref{wslap_section} and \ref{magnification_section} to denote the source position, but its usage should be clear to readers from the relevant context.} we can determine the rest-frame UV spectral flux density from
\begin{align}
f_\lambda(\lambda_\mathrm{UV}) & = f_\lambda(\lambda_{\mathrm{eff},Q}) \Bigg(\frac{\lambda_\mathrm{UV}}{\lambda_{\mathrm{eff},Q}}\Bigg)^\beta \nonumber\\
\implies f_\nu(\nu_\mathrm{UV}) & = f_\nu(\nu_{\mathrm{eff},Q}) \Bigg[(1 + z) \, \frac{\lambda_\mathrm{UV}}{\lambda_{\mathrm{eff},R}}\Bigg]^{\beta + 2}
\end{align}
where we have used the relation $|f_\nu d\nu| = |f_\lambda d\lambda|$ to get the second line, $\nu_\mathrm{UV} \equiv c / \lambda_\mathrm{UV}$, $\lambda_\mathrm{UV} = 1500 \, \mathrm{\AA}$, and $\lambda_{\mathrm{eff},Q} = \lambda_{\mathrm{eff},R} / (1 + z)$. Therefore, combining the above equations, the rest-frame UV absolute magnitude can be computed with
\begin{align}\label{M_UV_formula}
& \ M_\mathrm{UV}(m_\mathrm{filter},\lambda_\mathrm{eff},\boldsymbol{\theta},z) \nonumber\\
= & \ M_Q - 2.5 \ \log_{10} \Bigg(\frac{f_\nu(\nu_\mathrm{UV})}{f_\nu(\nu_{\mathrm{eff},Q})}\Bigg) \nonumber\\
= & \ m_\mathrm{filter} + 2.5 \ \log_{10} \Bigg[\frac{\mu}{(1 + z)^{\beta + 1}} \bigg(\frac{\lambda_\mathrm{eff}}{\lambda_\mathrm{UV}}\bigg)^{\beta + 2} \bigg(\frac{10 \, \mathrm{pc}}{D_\mathrm{L}}\bigg)^2\Bigg]
\end{align}
where the magnification factor $\mu$ is evaluated at the observed position $\boldsymbol{\theta}$ and redshift $z$ of the galaxy using the lens model according to Equation (\ref{magnification_formula}), and the luminosity distance $D_\mathrm{L}$ is also calculated at the source redshift. We have also replaced $m_R$ with $m_\mathrm{filter}$, and $\lambda_{\mathrm{eff},R}$ with $\lambda_\mathrm{eff}$, to stress that this formula is applicable to arbitrary bandpass filters used for our measurements. Finally, the UV luminosity is defined by
\begin{align}\label{L_UV_formula}
\begin{split}
L_\mathrm{UV} \equiv & \ 4 \, \pi \, (10 \, \mathrm{pc})^2 \, 10^{-0.4 \, M_\mathrm{UV}} \\
& \ \times (3.631 \times 10^{-20} \, \mathrm{erg \, s^{-1} \, cm^{-2} \, Hz^{-1}})
\end{split}
\end{align}
where $3.631 \times 10^{-20} \, \mathrm{erg \, s^{-1} \, cm^{-2} \, Hz^{-1}}$ is the zero-point spectral flux density for the AB magnitude system.

\begin{figure}[tp!]
\centering
\includegraphics[width=85mm]{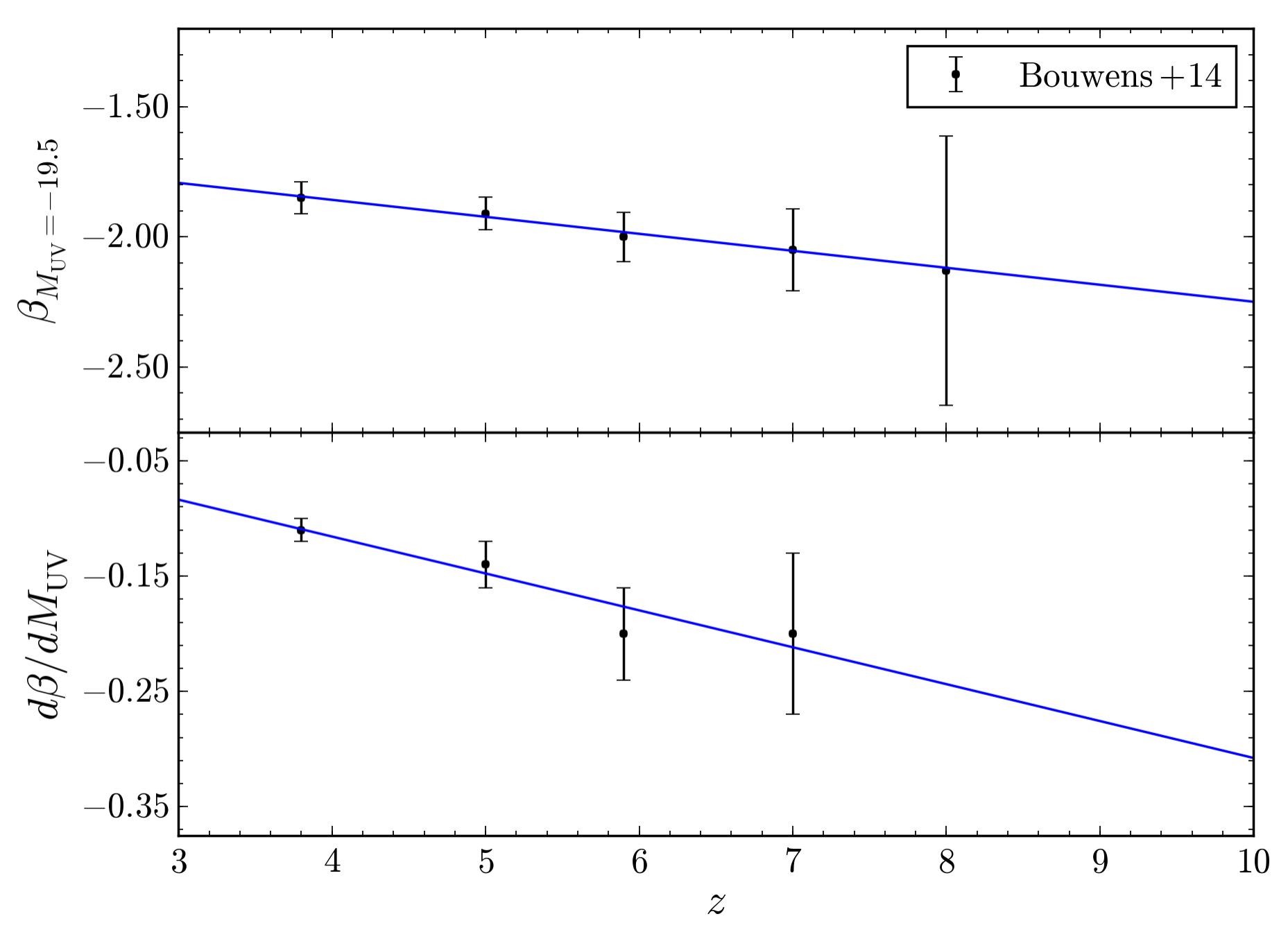}
\caption{\label{Bouwens_beta_fitting}Best-fit intercept (at $M_\mathrm{UV} = -19.5$) and slope of the UV continuum slope $\beta(z,M_\mathrm{UV})$ using the data from \citet{2014ApJ...793..115B}.}
\end{figure}

There remains a minor complication, which is that the UV continuum slope $\beta$ depends mildly on redshift as well as the UV luminosity of a galaxy. We performed chi-squared fitting on the data presented by \citet{2014ApJ...793..115B} with measurements of the colours of over 4000 galaxies at $z \sim 4-$8 to obtain an analytic expression for $\beta$ as follows:
\begin{align}\label{beta_formula}
\begin{split}
\beta(z,M_\mathrm{UV}) = & \ (-0.065 \, z - 1.597) \\
& \ + (-0.032 \, z + 0.012)(M_\mathrm{UV} + 19.5),
\end{split}
\end{align}
where we assumed both the intercept and slope depend linearly on $z$. The best-fit colour-luminosity relationship is plotted against redshift in Figure \ref{Bouwens_beta_fitting}, where $\beta$ becomes larger with decreasing redshift and for brighter galaxies. We supplied for each galaxy candidate an initial guess for $\beta$ to arrive at a rough estimate of $M_\mathrm{UV}$, which was substituted back to Equation (\ref{beta_formula}) to refine the value of $\beta$, then we repeated the calculation again. After several iterations, $M_\mathrm{UV}$ converged to provide a reasonably accurate estimate of its value.

\subsection{Multiply Lensed Images}\label{multiply_lensed_images_section}

In the strongly lensed cluster fields, multiply lensed images of the same source galaxy can typically appear three times, and sometimes up to five times, depending on individual multiple image configurations and how many of them get magnified above the flux limit. If we do not correct for this multiplicity, the LF will be overestimated by a factor of $\sim$3 at the faint end where we typically have a few galaxies only with such low luminosities.\footnote{At this point, we alert the reader that identifying and correcting for multiply lensed images are not required for the magnification bias test that we are going to describe in Section \ref{magnification_bias_section}, helping to make this test the most robust result of our work.}

To address this problem, we first performed a coarse search within the pool of selected galaxy candidates, in which for each galaxy candidate $i$ with delensed source position $\boldsymbol{\beta}_i$ and estimated photometric redshift $z_{\mathrm{photo,}i}$, we identified all potential counterimages $j$ whose delensed source positions $\boldsymbol{\beta}_j$ and estimated photometric redshifts $z_{\mathrm{photo,}j}$ satisfy the tolerance constraints $|\boldsymbol{\beta}_i - \boldsymbol{\beta}_j| < 5\arcsec$ and $|z_{\mathrm{photo,}i} - z_{\mathrm{photo,}j}| < 0.1[1 + (z_{\mathrm{photo,}i} + z_{\mathrm{photo,}j}) / 2]$. The former constraint was chosen because the typical rms scatter of the delensed positions of multiply lensed images is a few arcseconds \citep{2014ApJ...797...98L,2015MNRAS.451.3920D,2016MNRAS.459.3447D,2016MNRAS.456..356D}. For the latter constraint, we assumed that the error in photo-$z$ scales roughly as $(1 + z)$, and to allow a tolerance of 10\% deviation from such relation. We then linked up such groups sharing one or more common galaxy members to form ``clusters'' of potential multiple images, in a way essentially similar to the Friends-of-Friends algorithm commonly used for identifying galaxy clusters.

After narrowing down potential multiple image systems from the original collection of high-$z$ galaxies in each cluster field, we visually examined every candidate within a given system and determined whether any of them could be multiple images of a single source galaxy. The criteria that we employed to confirm the validity of a multiple image system were based on the following considerations. First, we checked whether all the potential multiple images display similar visual colours when inspected on NIR-weighted images specially catered to view high-$z$ galaxies (such as those shown in Figure \ref{multi_clt_mem_misid}). This exercise was repeated by varying the combination of the weightings of individual filters to ensure the robustness of the multiple image candidates. Second, we required the geometrical configuration and image morphologies to be consistent with general expectations in strong lensing. For instance, a pair of closely separated multiple images should be bisected by a critical curve, the presence of which can be checked by generating critical curves from the lens model for a given source redshift. In addition, a counterimage is expected at a large separation on the far side of the same critical curve. This counterimage, however, may not be bright enough to be detected, as is often the case if the highly magnified image pair near the critical curve is already relatively faint. Whether or not such counterimages can be identified of course depends also on whether it lies within a defined exclusion region. Another example of a plausible multiple image configuration is that images near the tangential critical curves should be tangentially stretched, whereas images near the radial critical curves should be radially stretched. The extent of distortion scales positively with the angular size of an image, and so the effect is in general relatively small but still noticeable for our sample of galaxy candidates. Lastly, genuine multiple images should have UV absolute magnitudes, computed using Equation (\ref{M_UV_formula}), that are roughly consistent with each other provided their magnification uncertainties are not too large (which constitute the dominant source of error in estimating the UV luminosities of highly magnified galaxies).

In Appendix \ref{multiple_image_info}, we provide lists of the twenty multiply lensed image systems (sixteen of which are doubly lensed, and the rest being triply lensed) that we thus identified from our pool of high-$z$ galaxy candidates for the six HFF cluster fields.


\section{Results {\uppercase\expandafter{\romannumeral 1\relax}}: Magnification Bias}\label{magnification_bias_section}

As mentioned in Section \ref{lensing_effects}, the advantage of studying the high-$z$ galaxy population in the cluster fields is that we gain extra depth to detect intrinsically faint galaxies, while the drawback of doing so is that the effective sampling volume is reduced. This effect is known as the magnification bias \citep{1995ApJ...438...49B,2005ApJ...619L.143B,2008ApJ...684..177U}, and can be quantified as
\begin{equation}
\Phi_\mathrm{lensed}(<\!m_\mathrm{lim},\mu,z) = \frac{\Phi(<\!M_\mathrm{UV,lim}(\mu),z)}{\mu}
\end{equation}
where $\Phi(<\!M_\mathrm{UV,lim},z) \equiv \int_{-\infty}^{M_\mathrm{UV,lim}}\phi(M_\mathrm{UV},z)dM_\mathrm{UV}$ is the cumulative UV LF, $\phi(M_\mathrm{UV},z)$ is the UV LF, and $m_\mathrm{lim}$, $\mu$, and $M_\mathrm{UV,lim}$ are related through Equation (\ref{M_UV_formula}). If the LF is a simple power law, i.e.~$\phi(L) \propto L^\alpha$, then the expected galaxy number density scales as $\Phi_\mathrm{lensed}(<\!M_\mathrm{lim}) \propto \mu^{-\alpha - 2}$ where $\alpha \neq -1$ \citep{1995ApJ...438...49B}. A value of $\alpha = -2$ therefore results in a magnification-invariant galaxy number density, whilst the smaller volume probed is more than compensated for by the deepened flux limit if $\alpha < -2$ (positive magnification bias) and vice versa (negative magnification bias).

\begin{figure}[tp!]
\centering
\includegraphics[width=85mm]{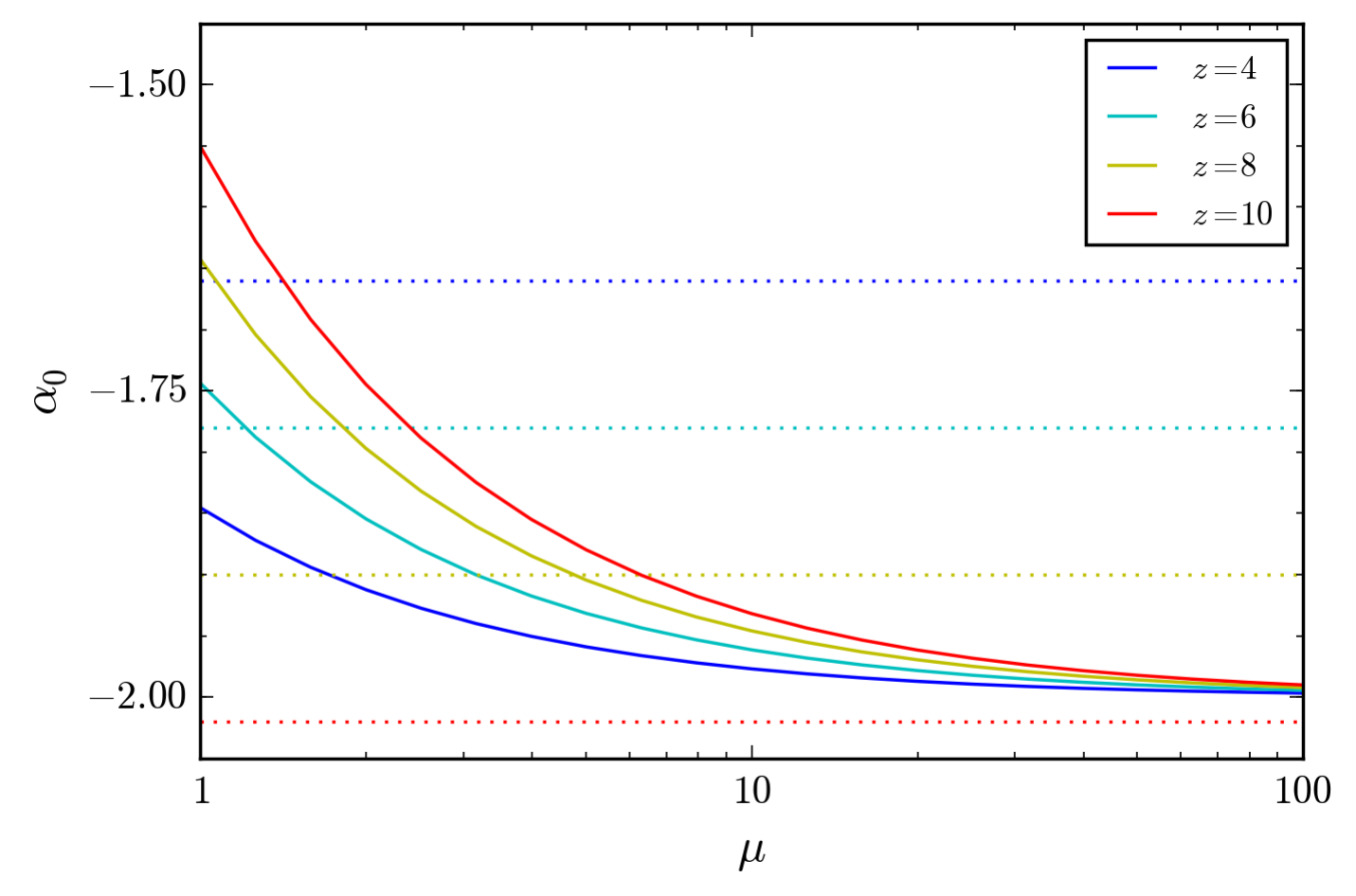}
\caption{\label{critical_schechter_alpha}Critical faint-end slope of the UV LF as a function of the magnification factor (with different values of $x(z) \equiv L_\mathrm{UV,lim}(z,\mu=1)/L_\star(z)$ at various redshifts (coloured solid lines)), assuming that the LF is described by a Schechter function. The best-fit $\alpha$'s at the corresponding redshifts obtained by S16 are also shown as coloured dotted lines for reference. Note that the peak observed galaxy number density occurs at the intersection, if any, between $\alpha(z)$ and $\alpha_0(z)$, whereas the magnification bias is always positive and increases monotonically if $\alpha(z) < -2$.}
\end{figure}

As will become clear in the following description, magnification bias provides an independent, and superior, method for constraining the UV LF, especially at the faint end, compared with conventional direct determinations of UV LFs. We will also show below that the slope of the faint-end UV LF provides the most powerful discriminator between different DM (and/or galaxy formation) models. We shall begin with a more realistic UV LF described by a Schechter function \citep{1976ApJ...203..297S}, which is that expected in the standard CDM model. It is a power-law function with an exponential cutoff at the bright end such that
\begin{equation}\label{CDM_LF}
\phi_\mathrm{CDM}(L_\mathrm{UV}) = \frac{\phi_\star}{L_\star} \Bigg(\frac{L_\mathrm{UV}}{L_\star}\Bigg)^\alpha \mathrm{exp}\Bigg(-\frac{L_\mathrm{UV}}{L_\star}\Bigg),
\end{equation}
where $\phi_\star$ is the normalization constant, $L_\star$ is the characteristic luminosity of the bright-end exponential suppression, and $\alpha$ is the faint-end (logarithmic) slope,\footnote{The faint-end LF slope $\alpha$ mentioned here should be distinguished from the deflection angle $\boldsymbol{\alpha}$ used in Section \ref{magnification_section}.} then we have
\begin{align}\label{CDM_lensed_cumulative_LF}
\Phi_\mathrm{CDM,lensed}(<\!M_\mathrm{UV,lim}) & = \frac{\phi_\star}{\mu} \Gamma\Bigg(\alpha + 1, \frac{L_\mathrm{UV,lim}(\mu)}{L_\star}\Bigg) \nonumber\\
& = \frac{\phi_\star}{\mu} \Gamma\Bigg(\alpha + 1, \frac{L_0}{\mu L_\star}\Bigg)
\end{align}
where $\Gamma(a,z) = \int_z^{\infty} t^{a - 1} \mathrm{e}^{-t} dt$ is the upper incomplete gamma function, $L_0 \equiv L_\mathrm{UV,lim}(\mu=1)$, and $L_\mathrm{UV,lim}$ can be obtained from $M_\mathrm{UV,lim}$ via Equation (\ref{L_UV_formula}). The derivative of $\Phi_\mathrm{CDM,lensed}(<\!M_\mathrm{UV,lim})$ with respect to $\mu$ is given by
\begin{align}
\frac{\partial\Phi_\mathrm{CDM,lensed}}{\partial\mu} = & \ \frac{\phi_\star}{\mu^2} \Bigg[\Bigg(\frac{L_0}{\mu L_\star}\Bigg)^{\alpha + 1}\mathrm{exp}\Bigg(-\frac{L_0}{\mu L_\star}\Bigg) \nonumber\\
& \ - \Gamma\Bigg(\alpha + 1, \frac{L_0}{\mu L_\star}\Bigg)\Bigg],
\end{align}
and so there exists a critical faint-end slope $\alpha_0$ such that the sign of this derivative switches from positive to negative as $\alpha$ increases from below to above $\alpha_0$, indicating that the expected galaxy number density increases (decreases) upon an infinitesimally small increase in $\mu$ if $\alpha$ is smaller (larger) than $\alpha_0$. In other words, $\alpha_0$ corresponds to the maximum possible galaxy number density for a given magnification factor $\mu$. Note that $\alpha_0$ decreases asymptotically to $-2$ as $\mu$ increases, because $(x/\mu)^{\alpha + 1}\mathrm{e}^{-x/\mu} \simeq \mu/x \simeq \Gamma(\alpha + 1, x/\mu)$ as $\mu \to \infty$ and $\alpha \to -2$. This behaviour is consistent with the earlier result for the power-law LF whereby $\alpha = -2$ gives a magnification-invariant galaxy number density. In this case where $\mu$ is sufficiently high, the effective luminosity limit $L_\mathrm{UV,lim}$ becomes so low such that the integrated Schechter function is dominated by its power-law component. Assuming the best-fit redshift dependence of $\alpha$ as obtained by S16, the magnification bias is strictly positive (i.e.~higher galaxy number density than in the absence of lensing) for all $\mu$ beyond a critical redshift of $z \simeq 9.7$. The critical faint-end slope of the LF for different redshifts is plotted against the magnification factor in Figure \ref{critical_schechter_alpha} to illuminate the above discussion.

We now consider deviations of the UV LF from the classical Schechter form, in particular those featuring a slow rollover at the faint end that reflects a suppression of small-scale structure formation. Such a suppression is naturally predicted by alternative DM models such as the Wave Dark Matter ($\psi$DM) \citep{2014NatPh..10..496S} or the Warm Dark Matter (WDM) \citep{2001ApJ...556...93B} models. In the following, we will adopt a parametrization for the $\psi$DM UV LF, proposed by S16, in the form of a modified Schechter function expressed as
\begin{equation}\label{psiDM_LF}
\phi_{\psi\mathrm{DM}}(L_\mathrm{UV}) = \phi_\mathrm{CDM}(L_\mathrm{UV}) \Bigg[1 + \bigg(\frac{L_\mathrm{UV}}{L_\psi}\bigg)^\gamma\Bigg]^{\delta/\gamma},
\end{equation}
where $\phi_\mathrm{CDM}(L_\mathrm{UV})$ was defined earlier in Equation (\ref{CDM_LF}). The second term on the right-hand side of Equation (\ref{psiDM_LF}) represents the suppression in the predicted number density of low-luminosity galaxies, and $L_\psi$ is the characteristic luminosity below which the UV LF approaches asymptotically to
\begin{align}
& \ \phi_{\psi\mathrm{DM}}(L_\mathrm{UV}) \nonumber \\
\simeq & \ \frac{\phi_\star}{L_\star} \Bigg(\frac{L_\mathrm{UV}}{L_\star}\Bigg)^\alpha \Bigg[1 + \bigg(\frac{L_\mathrm{UV}}{L_\psi}\bigg)^\gamma\Bigg]^{\delta/\gamma} & \text{if $L_\mathrm{UV} \ll L_\star$,} \nonumber\\
\simeq & \ \frac{\phi_\star}{L_\star} \Bigg(\frac{L_\mathrm{UV}}{L_\star}\Bigg)^\alpha \Bigg(\frac{L_\mathrm{UV}}{L_\psi}\Bigg)^\delta & \text{if $L_\mathrm{UV} \ll L_\psi$,}
\end{align}
where $L_\psi \ll L_\star$. The best-fit parameters derived from the bright-end UV LF reconstructed by \citet{2015ApJ...803...34B}, assuming a (log-)linear redshift evolution, are given by
\begin{align}\label{LF_parameters}
M_\star & = -20.90 - 0.004 (z - 6), \nonumber\\
\phi_\star & = 5.2 \times 10^{-0.28 (z - 6) - 4} \, \mathrm{Mpc}^{-3}, \nonumber\\
\alpha & = -1.78 - 0.06 (z - 6), \nonumber\\
M_\psi & = -17.44 + 5.19 \ \log_{10} \bigg(\frac{m_{22}}{0.8}\bigg) - 2.71 \ \log_{10} \bigg(\frac{1 + z}{7}\bigg), \nonumber\\
\gamma & = -1.10, \nonumber\\
\delta & = 1.69 + 0.03 (z - 6),
\end{align}
where $M_\star$ and $M_\psi$ are the absolute magnitudes corresponding to the luminosities $L_\star$ and $L_\psi$ respectively, and $m_{22} \equiv m_\mathrm{B} / 10^{-22} \, \mathrm{eV}$ is the dimensionless mass of the DM bosons. An energy scale for these bosons of $\sim$$10^{-22} \, \mathrm{eV}$ is favoured based on considerations for the DM-dominated cores of local dwarf spheroidal galaxies \citep{2014NatPh..10..496S}. Note that the CDM Schechter function can be recovered as $m_{22} \to \infty$ such that $L_\psi \to 0$, implying that the DM particles are so massive and the associated de Broglie wavelength so short that quantum effects are negligible in the formation of DM halos.

To determine the corresponding lensed cumulative UV LF, we first modify the expression for the $\psi$DM UV LF to be a piecewise function defined by
\begin{align}\label{piecewise_psiDM_LF}
& \ \phi'_{\psi\mathrm{DM}}(L_\mathrm{UV}) \nonumber\\
\equiv & \begin{cases}
\displaystyle \frac{\phi_\star}{L_\star} \Bigg(\frac{L_\mathrm{UV}}{L_\star}\Bigg)^\alpha \Bigg[1 + \bigg(\frac{L_\mathrm{UV}}{L_\psi}\bigg)^\gamma\Bigg]^{\delta/\gamma} & \text{if $L_\mathrm{UV} < L_\mathrm{c}$,} \\[3ex]
\displaystyle \frac{\phi_\star}{L_\star} \Bigg(\frac{L_\mathrm{UV}}{L_\star}\Bigg)^\alpha \mathrm{exp}\Bigg(-\frac{L_\mathrm{UV}}{L_\star}\Bigg) & \text{if $L_\mathrm{UV} > L_\mathrm{c}$,}
\end{cases}
\end{align}
where $L_\mathrm{c} \equiv (-\delta L_\star / \gamma L_\psi^\gamma)^{1 / (1 - \gamma)} \sim \sqrt{3 L_\star L_\psi / 2}$. This particular definition of $L_\mathrm{c}$ has been chosen such that $\phi'_{\psi\mathrm{DM}}(L_\mathrm{UV})$ is continuous at $L_\mathrm{UV} = L_\mathrm{c}$ up to the lowest-order expansions in $L_\psi / L_\mathrm{UV}$ and $L_\mathrm{UV} / L_\star$ respectively for the two suppression terms. With this approximate $\psi$DM UV LF at hand, we can deduce that
\begin{align}\label{psiDM_lensed_cumulative_LF}
& \ \Phi_{\psi\mathrm{DM,lensed}}(<\!M_\mathrm{UV,lim}) \nonumber\\
\simeq & \ \Phi'_{\psi\mathrm{DM,lensed}}(<\!M_\mathrm{UV,lim}) \nonumber\\[0.75ex]
= & \begin{cases}
\displaystyle \frac{\phi_\star}{\mu} \Gamma\Bigg(\alpha + 1, \frac{L_0}{\mu L_\star}\Bigg) & \displaystyle \text{if $\mu \leq \frac{L_0}{L_\mathrm{c}}$,} \\[4.5ex]
\displaystyle \frac{\phi_\star}{\mu} \Bigg\{\Gamma\Bigg(\alpha + 1, \frac{L_\mathrm{c}}{L_\star}\Bigg) & \\[3ex]
\displaystyle + \frac{1}{\alpha + 1} \Bigg[\Bigg(\frac{L_\mathrm{c}}{L_\star}\Bigg)^{\alpha + 1}\,_2F_1\Bigg(A, B; & \\[3ex]
\displaystyle A + 1; -\bigg(\frac{L_\mathrm{c}}{L_\psi}\bigg)^\gamma\Bigg) - \Bigg(\frac{L_0}{\mu L_\star}\Bigg)^{\alpha + 1} & \\[3ex]
\displaystyle _2F_1\Bigg(A, B; A + 1; -\bigg(\frac{L_0}{\mu L_\psi}\bigg)^\gamma\Bigg)\Bigg]\Bigg\} & \text{if $\mu > \displaystyle \frac{L_0}{L_\mathrm{c}}$,}
\end{cases}
\end{align}
where $_2F_1(a, b; c; z)$ is the hypergeometric function, $A = (\alpha + 1) / \gamma$, $B = -\delta / \gamma$, and $\alpha \neq -1$ (if $\mu > L_0 / L_\mathrm{c}$). We can also justify that $\Phi_{\psi\mathrm{DM,lensed}}(<\!M_\mathrm{UV,lim}) < \Phi_\mathrm{CDM,lensed}(<\!M_\mathrm{UV,lim})$ for all $\mu$ provided that $\phi_{\psi\mathrm{DM}}(L_\mathrm{UV}) < \phi'_{\psi\mathrm{DM}}(L_\mathrm{UV}) \leq \phi_\mathrm{CDM}(L_\mathrm{UV})$ for all $L_\mathrm{UV}$, meaning that we will always observe a lower cumulative galaxy number density for the $\psi$DM model than for the CDM model irrespective of the magnification factor.

\begin{figure}[tp!]
\centering
\includegraphics[width=85mm]{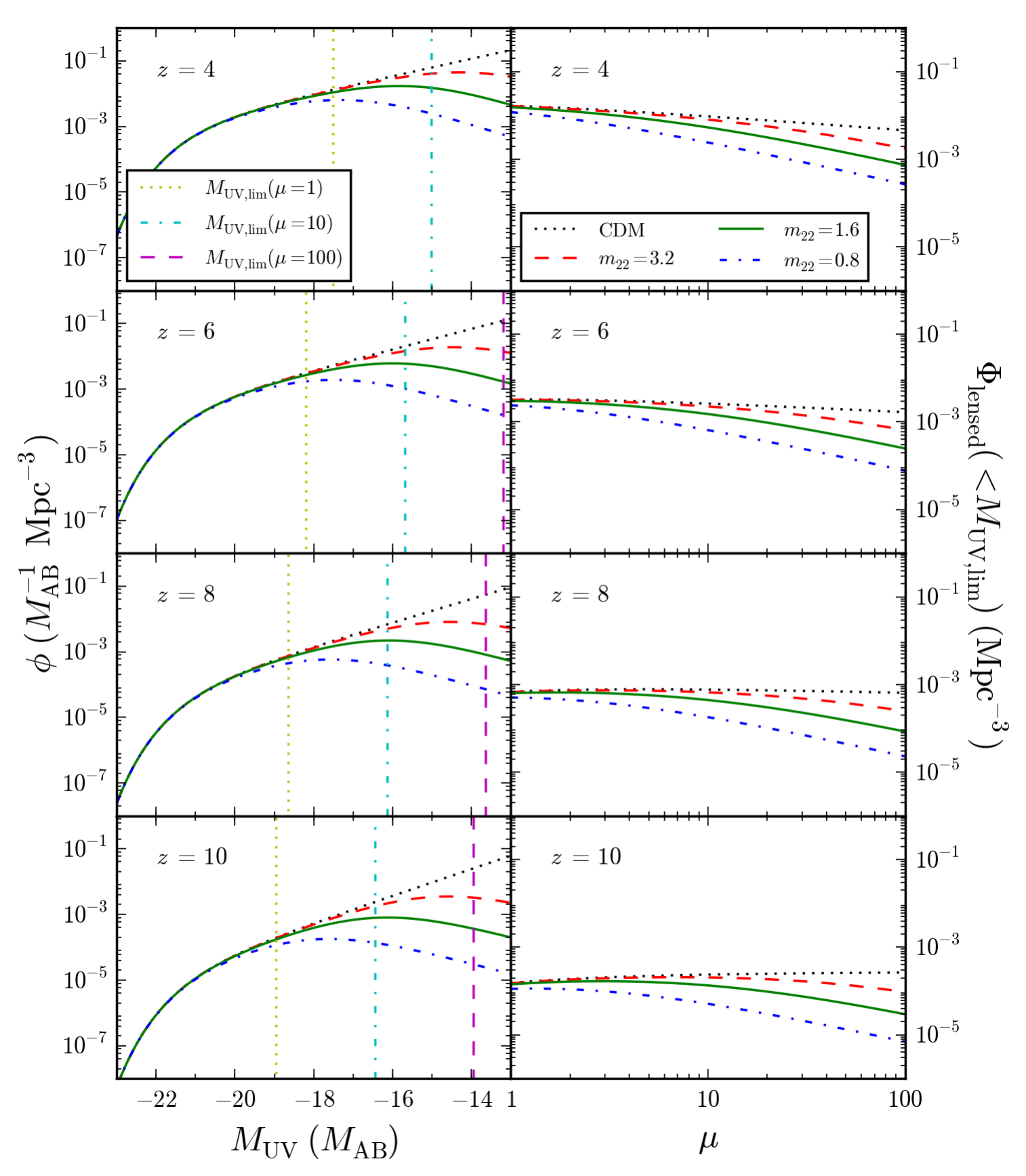}
\caption{\label{LF_mag_bias_comparison}CDM ({\it black dotted lines}) and $\psi$DM UV LFs at $z = 4, 6, 8,$ and 10 are shown in the left column, where the latter is plotted for three different choices of DM boson mass at $m_{22} = 0.8$ ({\it blue dash-dot lines}), 1.6 ({\it green solid lines}), and 3.2 ({\it red dashed lines}). The degree of faint-end suppression increases with decreasing $m_{22}$. In the right column, we plot the corresponding lensed cumulative UV LFs against the magnification factor. Note that the difference in the predicted galaxy number densities between CDM and $\psi$DM grows larger towards higher $\mu$, as evident from the increasing deviation between the CDM and $\psi$DM UV LFs towards brighter $M_\mathrm{UV,lim}(\mu)$ ({\it coloured vertical lines}).}
\end{figure}

In the left column of Figure \ref{LF_mag_bias_comparison}, we plot the CDM UV LF at selected redshifts, along with the $\psi$DM UV LFs for three different values of $m_{22}$ to illustrate the effect of different DM boson masses on the suppression of galaxy formation at the faint end. The corresponding lensed cumulative UV LFs (integrated up to the effective UV absolute magnitude limit) are plotted in the right column as a function of the magnification factor. As can be seen from the right panels, the overall magnification bias is always negative except for the $z = 10$ case in the CDM model, implying that the addition of faint galaxies magnified above the flux limit fails to compensate for the loss in galaxies owing to the diminished sky area. Note that the slope of the magnification bias with respect to $\mu$ changes with redshift, reflecting the cosmic evolution of the UV LF (in $\alpha$ for the CDM model; in $\alpha$, $M_\psi$, and $\delta$ for the $\psi$DM model).

We emphasize at this stage that the specific functional form of the proposed $\psi$DM UV LF as expressed in Equation (\ref{psiDM_LF}) is nothing more than a convenient parameterization choice representing a broader class of UV LFs having different levels of faint-end suppression. While the underlying DM models and the relevant galaxy formation mechanisms may differ from each other, in which case the parameters for the UV LFs may not necessarily resemble those expressed in Equation (\ref{LF_parameters}), the systematic faint-end deficiency in galaxies at all redshifts compared to the Schechter function is nevertheless a shared feature among such class of UV LFs.

The effect of magnification bias is sufficiently discernible in strongly lensed fields that it can be used as a powerful diagnostic tool to constrain the faint-end slope of the UV LF. Its greatest advantage compared with conventional direct LF determinations is that magnification bias is a local effect measured in the image plane rather than the source plane, removing the need to make what are often difficult corrections for all the existing multiply lensed images as required in conventional direct LF determinations to avoid overcounting. This follows from the fact that analyzing multiply lensed regions in the image plane (where multiple images may or may not be present) merely amounts to assigning heavier statistical weights to the respective delensed regions in the source plane without further consequence, for which a given source region of such kind is multiply sampled at generally different (image-plane) magnifications that give rise to different expected galaxy number densities observed in the image plane, thus removing the original degeneracy of galaxies in the source position space by projecting them into the (image-plane) magnification space. A potential drawback of this repeated sampling is that any selection bias introduced by galaxy clustering in the source plane is amplified, although such effect can be alleviated given a large enough survey volume.

There is yet another important advantage of using the magnification bias as a diagnostic of the faint-end LF slope compared with a direct reconstruction of the UV LF. The latter requires conducting photometry, and is therefore subject to photometric errors, especially for faint objects close to the detection threshold. This easily results in a heavily biased faint-end slope, as is often the case where the data points at the far faint-end are inferred from extremely few galaxy counts. By contrast, for the purpose of measuring the magnification bias, all we require is for a source to be detected above a designated flux limit, and so an accurate measure of its brightness is not necessary.\footnote{The magnification bias test is subject to photometric uncertainties only near the detection threshold, where the Eddington bias inevitably comes into play, but it has limited influence on the inferred magnification bias given that we already chose a conservatively bright magnitude limit that avoids appreciable incompleteness.}

To assemble sufficient galaxy counts for comparison with different model predictions in the expected magnification bias, particularly in the high-magnification regime where the lensed galaxy density can be over an order of magnitude lower compared with the unlensed value, we measured the overall surface number density of galaxies having $z_\mathrm{photo} \geq 4.75$ (see Section \ref{galaxy_candidate_selection} for our selection criteria) in all the six HFF cluster fields. We divided these galaxies into ten equally spaced logarithmic bins in $\mu$. Since the uncertainty in the magnification factor derived from the lens models increases with its magnitude, binning in $\log_{10} \, \mu$ reduces the scattering of data points along the magnification-factor axis. To determine the total image-plane solid angle associated with each magnification bin, we summed up the solid angle subtended by all the image-plane pixels (inside the regions concerned, i.e.~$\boldsymbol{\theta}_\mathrm{pix} \in \boldsymbol{\Omega}_\mathrm{eff}$) with estimated magnification factors that fall within the range of the corresponding magnification bin, i.e.~iso-magnification regions. Because the local magnification factor of every pixel varies, albeit mildly, with source redshift (especially at $z \gtrsim 4$ as concerned), we first computed a set of such solid angles at the central redshifts of individual redshift intervals separated by increments of $\Delta z = 0.25$. This selected width in redshift interval is sufficiently small that the magnification factor typically changes very slowly with redshift, i.e.~$(\partial\mu/\partial z)\Delta z \ll 1$. We then took the average of these solid angles as the effective solid angle encompassing all the galaxies residing in a given magnification bin. For the {\it k}-th magnification bin, we computed the surface number density of galaxies according to
\begin{equation}
\Sigma_k(<\!M_\mathrm{UV,lim}) = \sum_{i = 1}^{N_k} \frac{1}{\Omega_{\mathrm{tot},k}} = \frac{N_k}{\Omega_{\mathrm{tot},k}},
\end{equation}
where $N_k$ is the number of observed galaxies belonging to this bin, and
\begin{equation}
\Omega_{\mathrm{tot},k} = \frac{1}{N_z} \sum_{l = 1}^{N_\mathrm{field}} \sum_{m = 1}^{N_z} \sum_{n = 1}^{N_\mathrm{pix}} \Omega_\mathrm{pix},
\end{equation}
where $\Omega_\mathrm{pix}$ is the solid angle subtended by an image-plane pixel, $N_\mathrm{pix}$ is the number of image-plane pixels having magnification factors (when evaluated at the central redshift of the respective redshift interval) within the range specified by the magnification bin, $N_z$ is the number of redshift intervals with identical widths $\Delta z$, and $N_\mathrm{field}$ is the number of cluster fields. We also estimated the error for each magnification bin by summing up the reciprocals of the squared solid angles and then taking the square root of this sum, i.e.
\begin{equation}
\sigma_k = \Bigg(\sum_{i = 1}^{N_k} \frac{1}{\Omega_{\mathrm{tot},k}^2}\Bigg)^{1/2} = \frac{\sqrt{N_k}}{\Omega_{\mathrm{tot},k}},
\end{equation}
which simply reduces to the Poisson noise when $\Omega_{\mathrm{tot},k}$ is taken as a constant determined entirely by the properties of the lenses.

\begin{deluxetable*}{c c c c c c c c c c}
\tablecaption{Galaxy surface number density and magnification bias measured from a joint analysis of the six HFF cluster fields\label{magnification_bias_table}}
\tablehead{\colhead{\phm{a}$\mu$\tablenotemark{a}} & \colhead{No.~of galaxies} & \colhead{\phm{b}$\Sigma(<\!M_\mathrm{UV,lim})$\tablenotemark{b}} & \colhead{\phm{c}$\Sigma(<\!M_\mathrm{UV,lim})/\Sigma_0$\tablenotemark{c}} & \colhead{} & \colhead{\phm{a}$\mu$\tablenotemark{a}} & \colhead{No.~of galaxies} & \colhead{\phm{b}$\Sigma(<\!M_\mathrm{UV,lim})$\tablenotemark{b}} & \colhead{\phm{c}$\Sigma(<\!M_\mathrm{UV,lim})/\Sigma_0$\tablenotemark{c}} \\
\colhead{} & \colhead{} & \colhead{($10^{-3} \, \mathrm{arcsec}^{-2}$)} & \colhead{} & \colhead{} & \colhead{} & \colhead{} & \colhead{($10^{-3} \, \mathrm{arcsec}^{-2}$)} & \colhead{}
}
\startdata
1.26 & 101 & $4.512\pm0.449$ & $1.000\pm0.100$ &  & 12.59 & 15 & $2.673\pm0.690$ & $0.592\pm0.153$ \\
2.00 & 55 & $3.826\pm0.516$ & $0.848\pm0.114$ &  & 19.95 & 5 & $1.355\pm0.606$ & $0.300\pm0.134$ \\
3.16 & 41 & $3.495\pm0.546$ & $0.775\pm0.121$ &  & 31.62 & 5 & $2.008\pm0.898$ & $0.445\pm0.199$ \\
5.01 & 16 & $1.655\pm0.414$ & $0.367\pm0.092$ &  & 50.12 & 2 & $1.188\pm0.840$ & $0.263\pm0.186$ \\
7.94 & 18 & $2.422\pm0.571$ & $0.537\pm0.127$ &  & 79.43 & 2 & $0.677\pm0.479$ & $0.150\pm0.106$ \\
\enddata
\tablenotetext{a}{Magnification bins are equally spaced in $\log_{10} \, \mu$.}
\tablenotetext{b}{Surface number densities are evaluated in the image plane.}
\tablenotetext{c}{$\Sigma_0$ is defined as the unlensed value of $\Sigma(<\!M_\mathrm{UV,lim})$, practically taken as that at the lowest-magnification bin with $\mu \simeq 1$.}
\end{deluxetable*}

\begin{figure}[tp!]
\centering
\includegraphics[width=85mm]{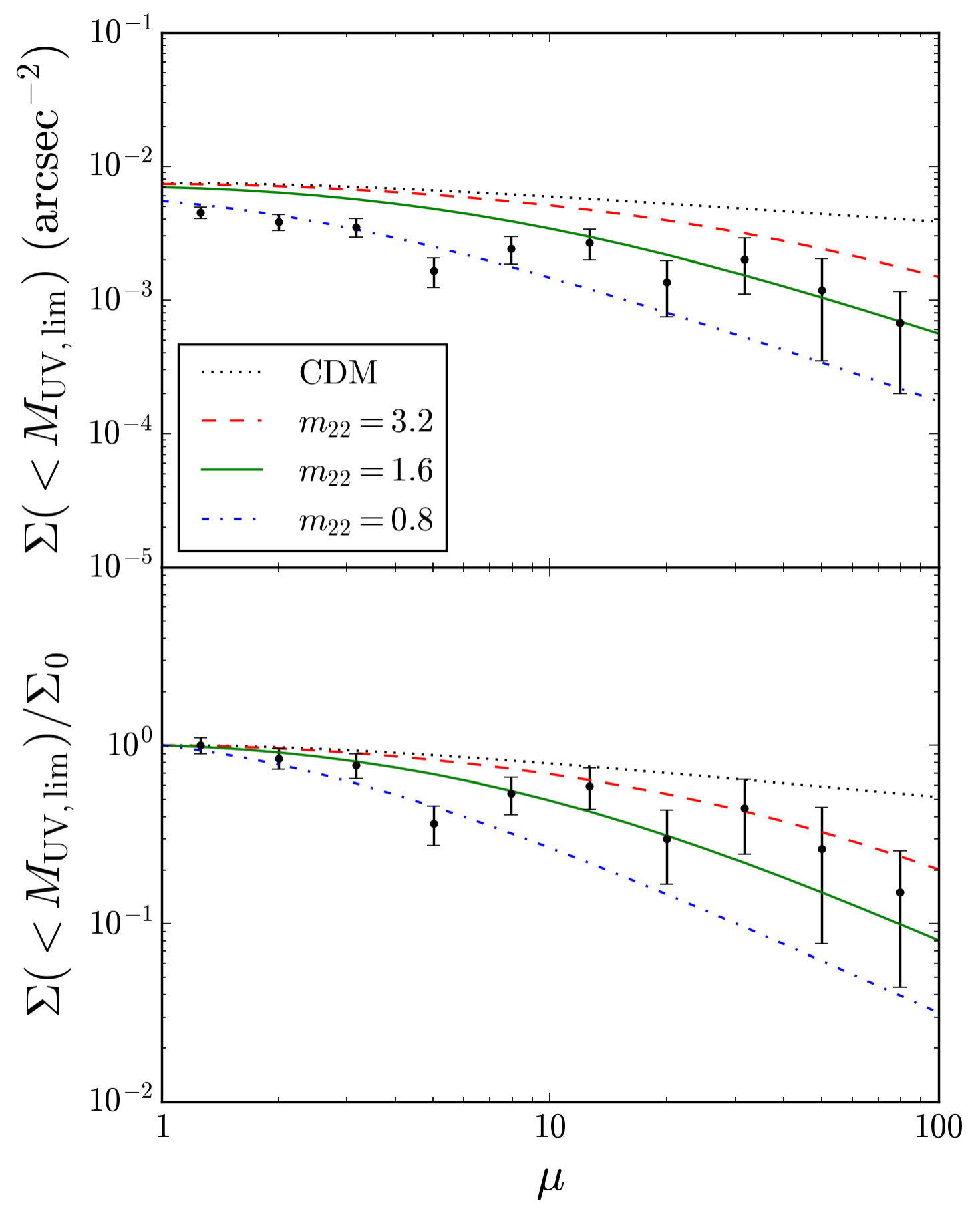}
\caption{\label{magnification_bias}Surface number density of galaxies ({\it upper panel}) and magnification bias ({\it lower panel}) in the six HFF cluster fields, measured as a function of the magnification factor. The predictions for the CDM and $\psi$DM models were computed using the UV LFs specified by Equations (\ref{CDM_LF}), (\ref{psiDM_LF}), and (\ref{LF_parameters}).}
\end{figure}

We present in the upper panel of Figure \ref{magnification_bias} and in Table \ref{magnification_bias_table} the measured surface number density of galaxies against the magnification factor. Also plotted in Figure \ref{magnification_bias} are the various CDM and $\psi$DM model predictions computed according to
\begin{equation}
\Sigma(<\!M_\mathrm{UV,lim}) = \int_{z_\mathrm{min}}^{z_\mathrm{max}} \Phi_\mathrm{lensed}(<\!M_\mathrm{UV,lim},z) \frac{dV_\mathrm{c}(z)}{d\Omega},
\end{equation}
where $z_\mathrm{min}$ and $z_\mathrm{max}$ are the lower and upper redshift bounds of our analysis respectively, and $dV_\mathrm{c}(z)/d\Omega$ is the differential comoving volume per unit solid angle at redshift $z$. As can be clearly seen, our results reveal a constant deficit of galaxy number density at all magnification factors relative to the CDM prediction (assuming a Schechter UV LF), with this deficiency becoming more prominent towards higher magnifications. Just in case our results are affected by cosmic variance, resulting in a systematic offset in the normalization of the galaxy number density, we normalized both the data and the model predictions with respect to their unlensed values, $\Sigma_0 \equiv \Sigma(\mu \simeq 1)$, in the lower panel of Figure \ref{magnification_bias}, so as to restrict our attention only to the effective slope of the magnification bias. Nonetheless, the disagreement between the data and the CDM prediction persists at a $\simeq$$6.4\sigma$ significance level (see Table \ref{magnification_bias_red_chisq_sigma}).

At this point, we remind the reader that the bright-end UV LF is well constrained by existing deep-field surveys to follow a Schechter-like form (e.g.~\citealt{2015ApJ...803...34B}). The tension between our results for the magnification bias and the CDM prediction implies a sub-Schechter behaviour at the faint-end UV LF that can naturally be explained by the existence of a smooth rollover. As can be seen in Figure \ref{magnification_bias} with the corresponding goodness-of-fit statistics listed in Table \ref{magnification_bias_red_chisq_sigma}, our data points are well encapsulated within the range of predictions by the $\psi$DM model for boson masses spanning $0.8 \times 10^{-22} \, \mathrm{eV} \lesssim m_\mathrm{B} \lesssim3.2 \times 10^{-22} \, \mathrm{eV}$.

\begin{deluxetable}{c c c c}[tp!]
\tablecaption{Reduced $\chi^2$ with $\sigma$ values for various model LF fits (assuming fitting parameters given by Equation (\ref{LF_parameters})) on the observed magnification bias\label{magnification_bias_red_chisq_sigma}}
\tablehead{\colhead{Model} & \colhead{$m_{22}$} & \colhead{\phm{a}$\chi^2_\mathrm{red}$\tablenotemark{a}} & \colhead{\phm{b}$\chi^2_\mathrm{red}$\tablenotemark{b}}}
\startdata
CDM &  & $44.93 \ (20.15\sigma)$ & $6.74 \ (6.41\sigma)$ \\
$\psi$DM & 3.2 & $29.30 \ (15.92\sigma)$ & $3.49 \ (3.83\sigma)$ \\
$\psi$DM & 1.6 & $13.29 \ (10.02\sigma)$ & $1.71 \ (1.79\sigma)$ \\
$\psi$DM & 0.8 & \phm{$1$}$1.84 \ (\phm{1}1.97\sigma)$ & $1.68 \ (1.75\sigma)$ \\
\enddata
\tablecomments{Reduced $\chi^2$ values were computed using only the first eight magnification bins, with each consisting of at least five galaxy detections.}
\tablenotetext{a}{Reduced $\chi^2$ computed for the galaxy surface number density $\Sigma(<\!M_\mathrm{UV,lim})$.}
\tablenotetext{b}{Reduced $\chi^2$ computed for the magnification bias $\Sigma(<\!M_\mathrm{UV,lim})/\Sigma_0$.}
\end{deluxetable}

\section{Results {\uppercase\expandafter{\romannumeral 2\relax}}: Clustercentric Radial Density Profile}\label{radial_test_section}

The magnification bias also manifests itself in the cluster fields through the modulated spatial density of galaxies in the image plane. Galaxy clusters (especially if they are dynamically relaxed) can be visualized as approximately spherical/elliptical gravitational lenses where the magnification factor is generally high ($\mu \gtrsim 10$) around the cluster cores and decreases rapidly to unity (i.e.~no magnification) towards the outskirts. The (projected) galaxy number density is therefore expected to change with clustercentric radial distance, where the rate of change depends sensitively on the faint-end slope of the UV LF as demonstrated in the previous discussion on the magnification bias.

\begin{figure}[tp!]
\centering
\includegraphics[width=85mm]{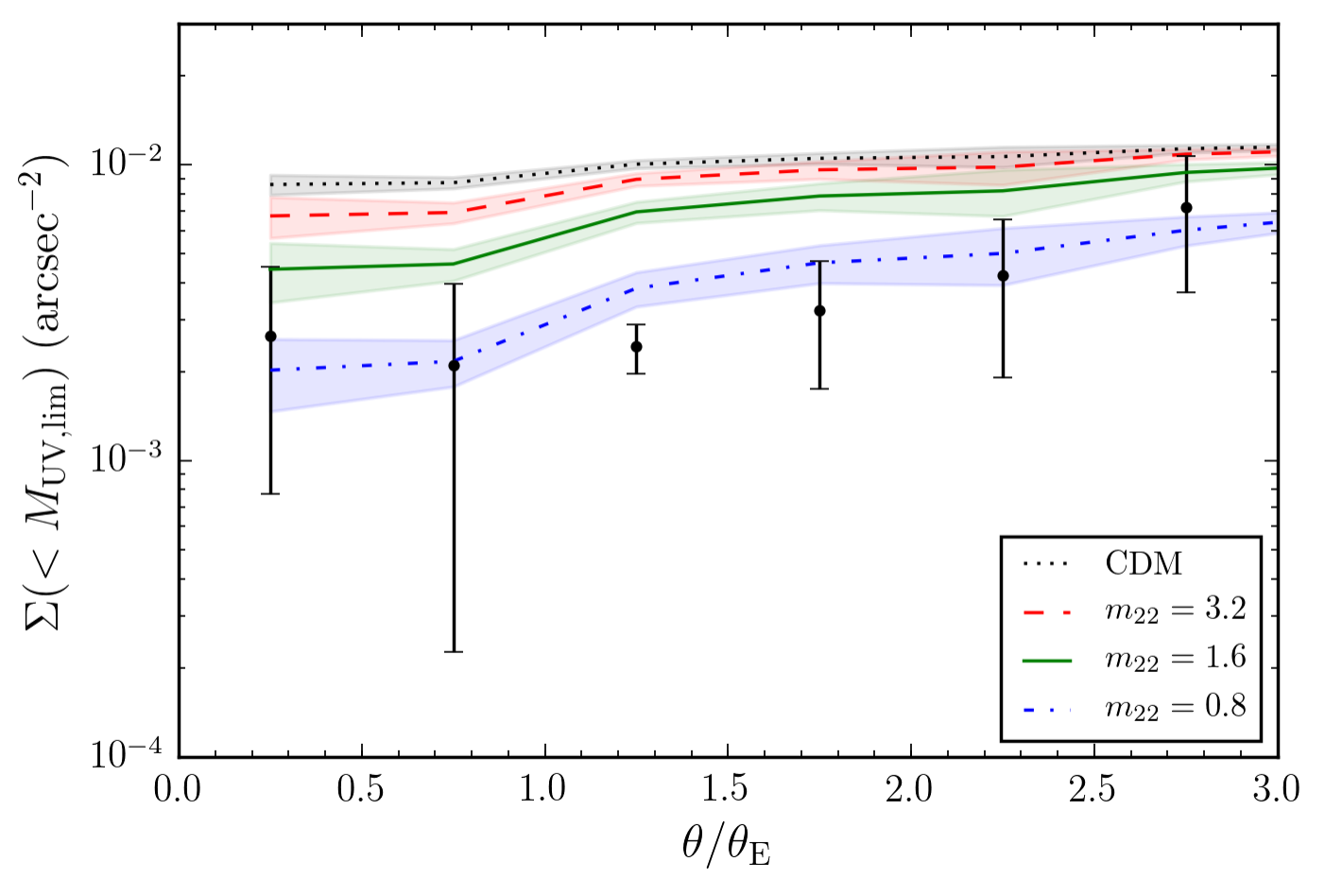}
\caption{\label{multi_clt_n(theta)}Surface number density of galaxies in the six HFF cluster fields as a function of clustercentric angular radius (scaled with effective Einstein radius $\theta_\mathrm{E}$). The data points and the model predictions were obtained by computing the weighted averages between clusters where the respective effective solid angles $\Omega_\mathrm{eff}$ (i.e.~without exclusion regions) were assigned as weights, while the associated errors represent field-to-field sample variance.}
\end{figure}

In Figure \ref{multi_clt_n(theta)}, we plot the measured surface number density of galaxies in different clustercentric radial bins averaged over all the HFF cluster fields, with error bars denoting field-to-field sample variance, which is dominated by cosmic variance in the source plane as evident in Table \ref{fld_galaxy_count_variance}. Because galaxy clusters have different physical sizes, thus leading to different characteristic angular scales for the variation in the (lensed) galaxy number density, we scaled accordingly the clustercentric angular radial position of each selected high-$z$ galaxy in units of the effective Einstein radius $\theta_\mathrm{E}$ (evaluated at a fiducial redshift of $z = 4$) of the corresponding cluster.\footnote{The effective Einstein radius, $\theta_\mathrm{E}$, of a given galaxy cluster is defined by the square root of the total solid angle enclosed by the cluster critical curve, $\Omega_\mathrm{crit}$, divided by a factor of $\pi$, i.e.~$\pi\theta_\mathrm{E}^2 = \Omega_\mathrm{crit}$.} In an ideal lens that is perfectly axisymmetric, we should see a prominent dip in the observed galaxy number density at the Einstein radius where the magnification factor diverges, reflecting the negative magnification bias. In realistic cluster lenses, however, this dip is smeared into a trough within the Einstein radius due to their intrinsic ellipticities and also the presence of irregularly distributed substructures. The predicted galaxy number density therefore gradually climbs up beyond the Einstein radius, and then starts to level off at very large clustercentric radius, owing to the fact that the magnification factor decreases asymptotically to unity in the cluster outskirts. As can be seen in Figure \ref{multi_clt_n(theta)}, both the normalization in the surface number density of galaxies and its rising trend towards the cluster outskirts are in accord with the predictions of the $\psi$DM model for $m_{22} = 0.8$, the same DM model and boson mass preferred by the magnification bias test (see Table \ref{magnification_bias_red_chisq_sigma}). At this point, we emphasize that the clustercentric radial density profile does not provide as robust a discriminator for the relevant boson mass because of the uncertainties and approximations introduced by averaging over different clusters as mentioned earlier in this paragraph.

\begin{figure}[tp!]
\centering
\includegraphics[width=85mm]{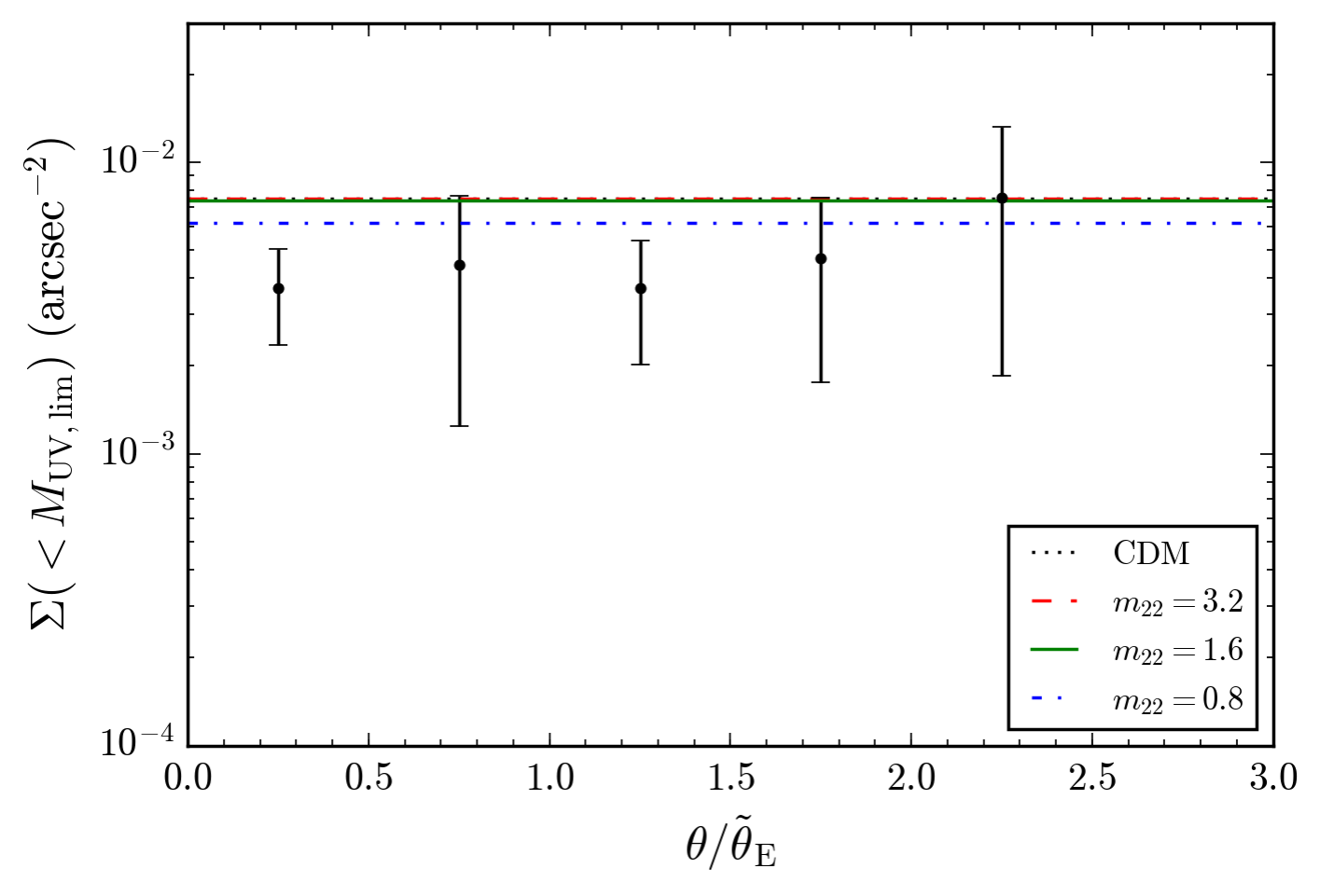}
\caption{\label{multi_par_n(theta)}Same as Figure \ref{multi_clt_n(theta)} for the parallel fields.}
\end{figure}

We also repeated the above exercise with the parallel fields, where we selected the centers of their fields of view as the ``origin'', and then determined the ``radial density profile'' of high-$z$ galaxies. The ``radial positions'' of these galaxies were scaled with the median effective Einstein radii of the six HFF clusters, which is approximately $30\arcsec$. In the absence of magnification induced by foreground galaxy clusters, the radial surface density of background galaxies should remain constant, if, ideally there are no fluctuations due to galaxy clustering in the source plane. The measured data as presented in Figure \ref{multi_par_n(theta)} are consistent with being constant across a broad range of ``radial distances'', which is also reflected in Table \ref{multi_par_goodness_of_fit} where there is no significant tension with the predictions from any of the four DM models (based on the UV LF reconstructed by \citet{2015ApJ...803...34B}) assuming a zero spatial gradient in magnification. Owing to the lack of magnification bias in the very weakly lensed parallel fields, this plot is not as sensitive to the different model predictions as the analogous plot for the cluster fields given the statistical uncertainties, although the data points indicate a slight underdensity relative to the model predictions (especially the standard CDM case) over a ``radius'' covering as large as $\sim$$60\arcsec$.

\begin{deluxetable}{c c c c c}[tp!]
\tablecaption{Goodnesses of fit for various model LF fits on the radial number density profile of galaxies in the parallel fields\label{multi_par_goodness_of_fit}}
\tablehead{\colhead{} & \colhead{CDM} & \colhead{$\psi$DM} & \colhead{$\psi$DM} & \colhead{$\psi$DM} \\
\colhead{$m_{22}$} & \colhead{} & \colhead{3.2} & \colhead{1.6} & \colhead{0.8}}
\startdata
$\chi^2_\mathrm{red}$ & $3.00$ & $3.00$ & $2.84$ & $1.24$ \\
S.D. & $2.56\sigma$ & $2.56\sigma$ & $2.45\sigma$ & $1.06\sigma$ \\
\enddata
\end{deluxetable}

\section{Results {\uppercase\expandafter{\romannumeral 3\relax}}: UV Luminosity Function}\label{LF_section}

The implication of a faint-end rollover in the UV LF from a strongly negative magnification bias can be checked against by directly reconstructing the LF using the inferred UV luminosities of the galaxies. For this purpose, we incorporated extra data from the parallel fields to enhance the statistics at the bright end. We also highlight the perils inherent in such a reconstruction especially in strongly lensed fields towards the end of this section.

\begin{figure}[tp!]
\centering
\includegraphics[width=85mm]{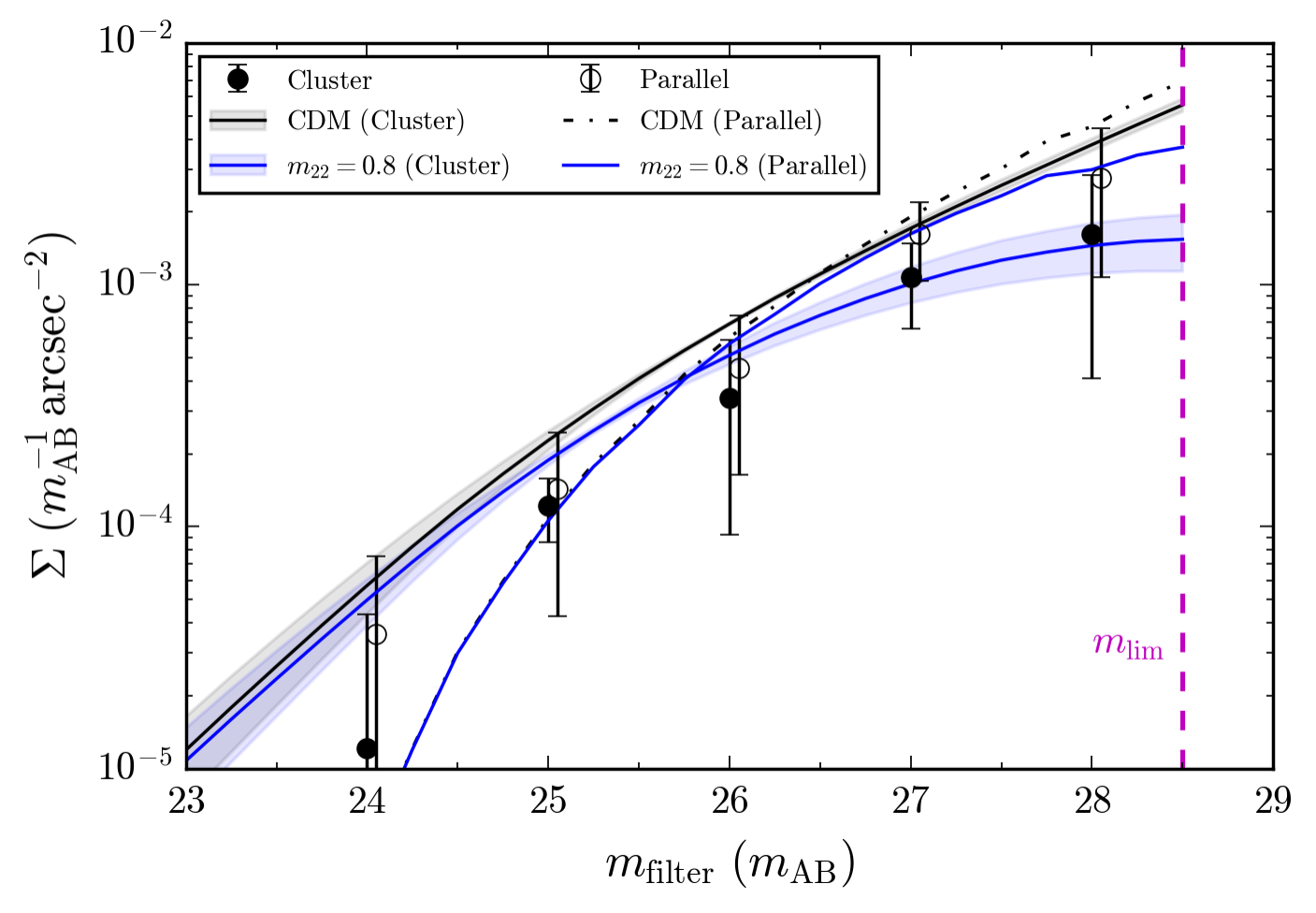}
\caption{\label{multi_fld_n(m)}Illustration of the effect of lensing on the resultant apparent magnitude distribution of galaxies. We plot the observed surface number density of galaxies in the cluster fields ({\it solid data points}) and parallel fields ({\it hollow data points, offset slightly for visual clarity}) respectively, as a function of their apparent magnitudes measured in the HFF filters designated according to their redshifts (see Table \ref{filter_selection}). The error bars reflect field-to-field cosmic variance. Also shown are the predicted apparent magnitude distributions in the cluster fields ({\it solid lines with shaded 1$\sigma$ variance}) and parallel fields ({\it dash-dot lines}) separately computed for the CDM model ({\it black lines}) and the $\psi$DM model with $m_{22} = 0.8$ ({\it blue lines}).}
\end{figure}

It is instructive at this point to first demonstrate the effect of lensing on the expected apparent magnitude distribution of galaxies as observed in the image plane, before we describe how we performed a direct reconstruction of the UV LF, which is intrinsic to the source plane. This effect is illustrated in Figure \ref{multi_fld_n(m)}, where the black and blue dash-dot lines indicate the predicted surface number densities of galaxies as a function of apparent magnitude in the virtually unlensed parallel fields\footnote{A very weak 5\% fiducial magnification in the parallel fields is assumed throughout our work, which is almost negligible for all practical purposes.} for the CDM model and the $\psi$DM model (only with $m_{22} = 0.8$ for simplicity) respectively. These predictions are essentially equivalent to the corresponding model LFs integrated over source redshift, except for an implicit conversion from the rest-frame UV absolute magnitude to the apparent magnitude in the observer's frame. The black and blue solid lines indicate the corresponding predictions for the six HFF cluster fields, accounting for lensing effects based on their respective lens models. Since the apparent magnitude of a given lensed galaxy depends not only on its redshift but also the local magnification factor, the different 2D magnification profiles of the various clusters give rise to slightly different predicted apparent magnitude distributions. In Figure \ref{multi_fld_n(m)}, we depict such 1$\sigma$ field-to-field variance as (colour-coded) shaded regions enclosing the mean values denoted by the corresponding solid lines.

The modification in the predicted apparent magnitude distribution in the presence of lensing is a consequence of three separate factors. First, in the cluster fields, the apparent magnitudes of all the galaxies that would have been detected even without lensing are brightened, so the galaxies are shifted by different extents towards brighter apparent magnitudes depending on the local magnification factor. This effect produces an excess (deficit) in galaxies brighter (fainter) than $m_\mathrm{filter} \sim 26$\,--\,27 as can be seen in Figure \ref{multi_fld_n(m)}. The reason why the predicted excesses for the CDM and $\psi$DM models are comparable is due to the fact that the unboosted (or unlensed) UV luminosity limit, $M_\mathrm{UV,lim}(\mu=1)$, truncates the UV LF at the bright end where the two model LFs are essentially identical (see Figure \ref{LF_mag_bias_comparison}).

Secondly, lensing makes possible the detection of galaxies that would otherwise lie below the detection threshold. The number of galaxies brought above the detection threshold by lensing depends on the faint-end slope of the UV LF, which is different between the CDM and $\psi$DM models. In the CDM model, the approximately magnification-invariant faint-end slope of $\alpha \sim -2$ (see discussion in Section \ref{magnification_bias_section}) leads to a virtually self-similar buildup of galaxies towards fainter apparent magnitudes at the same rate as in the parallel fields where there is no lensing. However, in the $\psi$DM model, the presence of a faint-end turnover in the UV LF supplies too few faint galaxies to compensate for the loss of galaxies being magnified and hence brightened towards lower apparent magnitudes as mentioned above.

Thirdly and lastly, lensing reduces the intrinsic source-plane area probed. As a consequence, considering this factor in isolation, i.e.~assuming a fixed (magnification-independent) effective survey depth, the local surface number density of galaxies is suppressed by a factor of $\mu$. This effect necessitates a proper accounting over the entire image plane with a spatially varying magnification. Moreover, the degree of such suppression is larger towards brighter apparent magnitudes as brighter sources are typically associated with higher magnifications.

To see whether the expected modification in the apparent magnitude distribution due to lensing is reflected in the data, we determined the surface number density of galaxies as a function of their apparent magnitudes in the cluster and parallel fields respectively as shown in Figure \ref{multi_fld_n(m)}. The relatively large uncertainties at the $m_\mathrm{filter} = 24$ magnitude bin imposed by the extremely low galaxy counts (one and three detections respectively for the cluster and parallel fields) prevent us from identifying any statistically significant bright-end excess anticipated for the cluster fields. On the other hand, the data points at the faint end for both the cluster and parallel fields appear to conform better to the predictions of the $\psi$DM model for $m_{22} = 0.8$, and are consistent with the results obtained for the magnification bias test.

We present below the UV LF reconstructed using the classical \citet{1968ApJ...151..393S} ``$1/V_\mathrm{max}$'' method. Adopting the stepwise parameterization of the UV LF by \citet{1988MNRAS.232..431E} in the form
\begin{equation}\label{stepwise_parameterization}
\phi(M_\mathrm{UV}) = \sum_k W(M_\mathrm{UV} - M_k) \phi_k,
\end{equation}
where the index $k$ runs over all the UV absolute magnitude bins, with $M_k$ being the mean UV absolute magnitude of the {\it k}-th bin having a width of $\Delta M_k$, and
\begin{equation}\label{W_formula}
W(x) = 
\begin{cases}
1 & \text{if $-\Delta M_k / 2 \leq x \leq \Delta M_k / 2$,} \\
0 & \text{otherwise,}
\end{cases}
\end{equation}
the contribution of each galaxy was estimated by taking the reciprocal of the maximum (comoving) volume $V_\mathrm{max}$ inside which a hypothetical source possessing the same UV luminosity could be observed in the magnitude-limited survey, i.e.
\begin{equation}
\phi_k = \sum_{i = 1}^{N_k} \frac{1}{V_{\mathrm{max},i}},
\end{equation}
where $N_k$ is the number of galaxies belonging to the {\it k}-th magnitude bin.

\begin{figure}[tp!]
\centering
\includegraphics[width=85mm]{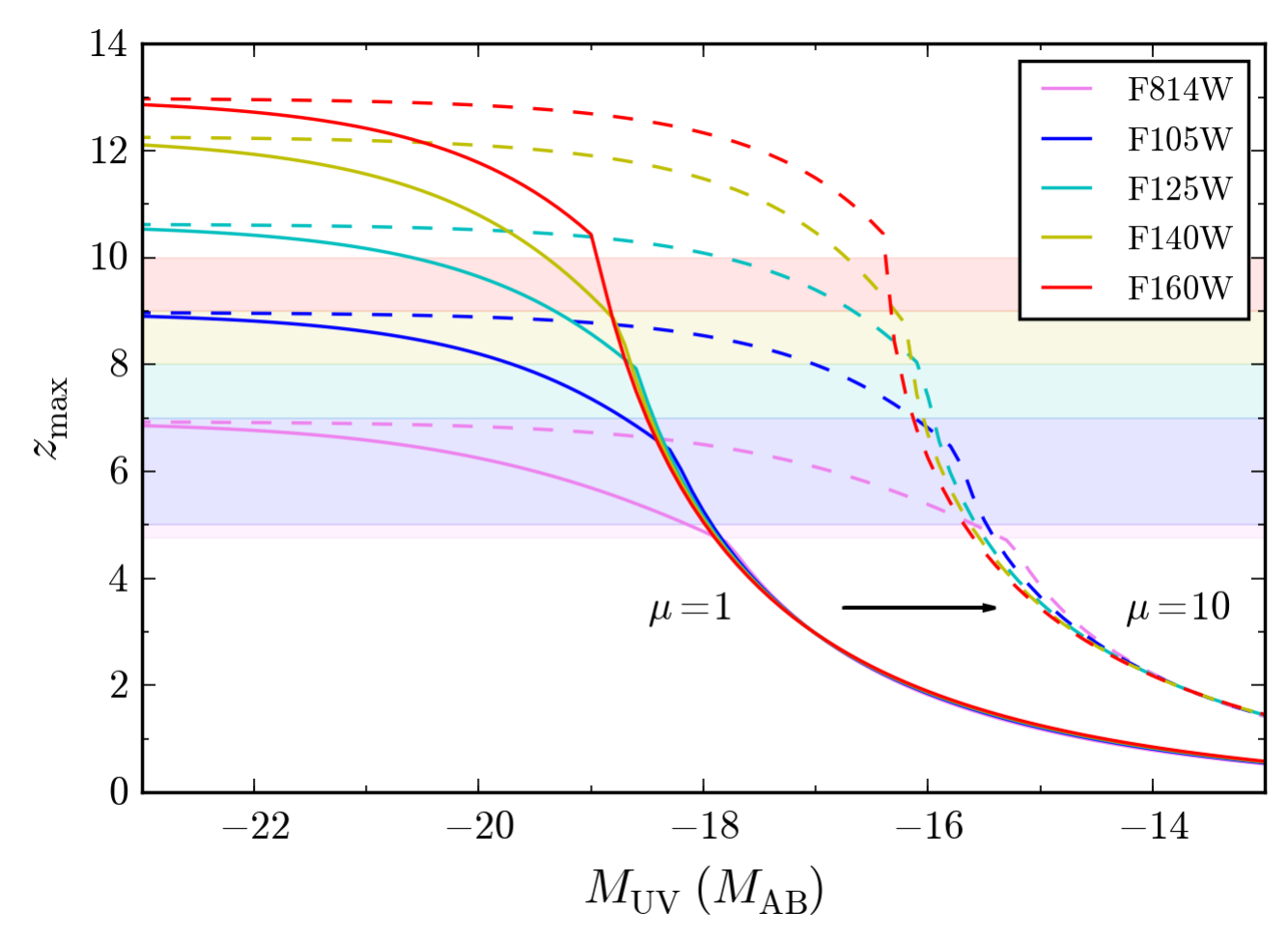}
\caption{\label{z_max_all_filters}Schematic diagram illustrating the maximum detectable redshift of a hypothetical source with UV absolute magnitude $M_\mathrm{UV}$ using a selection of HFF filters, and the impact of lensing magnification on the effective observation depth. The solid lines denote the maximum detectable redshifts without any magnification, whereas the dashed lines denote the maximum detectable redshifts with a magnification factor $\mu = 10$, which pushes the UV luminosity of the faintest detectable galaxy by 2.5 magnitudes faintwards. The shaded regions in different colours depict the redshift ranges, within which the selection of our target galaxies was based on their apparent magnitudes observed through the corresponding HFF filters.}
\end{figure}

The determination of $V_{\mathrm{max},i}$ for the {\it i}-th galaxy was first transformed into the problem of finding the maximum detectable redshift, $z_{\mathrm{max},i}$, that a hypothetical source having the same UV absolute magnitude, $M_{\mathrm{UV},i}$, could have just been detectable, solved through inverting Equation (\ref{M_UV_formula}). Note that the detection of a galaxy in a given filter will be hindered if there is an overlap between its passband and the Ly$\alpha$ forest, resulting in a portion of the light in this filter being absorbed. Taking this factor into account in our computation of $z_\mathrm{max}$, its dependence on the UV absolute magnitude of a galaxy is illustrated in Figure \ref{z_max_all_filters} for an apparent magnitude limit of $m_\mathrm{lim} = 28.5$. The sharp break with increasing brightness in a given coloured curve, which represents a particular HFF filter choice, coincides with the characteristic redshift where the Ly$\alpha$ forest starts to migrate into the passband of that filter, thus attenuating the amount of light reaching the observer and rendering galaxies beyond this redshift increasingly difficult to be detected. The value that $z_\mathrm{max}$ flattens off to is the limiting redshift where the Ly$\alpha$ forest completely overlaps with the passband of that filter, making any galaxy lying at a yet higher redshift impossible to be detected. In the presence of lensing, the effective survey sensitivity is enhanced, so that the UV absolute magnitude of the faintest detectable galaxy at a given redshift is shifted faintwards by $2.5 \, \log_{10} \, \mu$. The redshift range of galaxies selected in each filter, as specified in Section \ref{galaxy_candidate_selection}, is indicated by the shaded region (of the same colour) for the respective filter. As can be seen in Figure \ref{z_max_all_filters} (with Figure \ref{LF_mag_bias_comparison} as reference), the redshift range for each filter virtually lies below the corresponding $z_\mathrm{max}$ curve within the effective UV absolute magnitude limit, i.e.~$\mathrm{max}(z_\mathrm{int}) \lesssim z_\mathrm{max} \ \forall \ M_\mathrm{UV} < M_\mathrm{UV,lim}(\mu)$.

Nonetheless, before we could proceed to calculate $V_{\mathrm{max},i}$ for galaxies observed in the cluster fields, we had to deal with the problem of overcounting associated multiply lensed regions in the source plane. To address this issue, we constructed for each cluster field a grid in the source plane with a higher resolution than the model deflection field. We also derived the (reduced) deflection angle $\boldsymbol{\alpha}(\boldsymbol{\theta}_\mathrm{pix})$ and the magnification factor $\mu(\boldsymbol{\theta}_\mathrm{pix})$ for each image-plane pixel at the same resolution as the model deflection field. The centroid coordinate $\boldsymbol{\theta}_\mathrm{pix}$ of every such pixel was then delensed via the lens equation $\boldsymbol{\beta}(\boldsymbol{\theta}_\mathrm{pix}) = \boldsymbol{\theta}_\mathrm{pix} - \boldsymbol{\alpha}(\boldsymbol{\theta}_\mathrm{pix})$ to the source plane, and we labelled all those fine source-plane grid cells lying within the positional range $\Big[\beta_x(\boldsymbol{\theta}_\mathrm{pix}) \pm l_\mathrm{pix} / 2 \sqrt{\mu(\boldsymbol{\theta}_\mathrm{pix})}, \beta_y(\boldsymbol{\theta}_\mathrm{pix}) \pm l_\mathrm{pix} / 2 \sqrt{\mu(\boldsymbol{\theta}_\mathrm{pix})}\Big]$ with the value $\mu(\boldsymbol{\theta}_\mathrm{pix})$, where $l_\mathrm{pix}$ is the side length of an image-plane pixel. By doing so, we in effect ``painted'' a box on the source-plane grid, centring at the delensed source-plane position $\boldsymbol{\beta}(\boldsymbol{\theta}_\mathrm{pix})$ and having an area $1/\mu(\boldsymbol{\theta}_\mathrm{pix})$ times that of the original image-plane pixel, with a shaded ``grey colour'' proportional to $\mu(\boldsymbol{\theta}_\mathrm{pix})$. This ``painting job'' was iterated over all the image-plane pixels lying in the galaxy selection regions. Whenever we encountered a source-plane grid cell that had already been ``painted'', we always relabelled the cell (if necessary) with the largest value of $\mu$ thus found for that cell, so that we eventually ended up with a grid in the source place having at each grid cell the highest magnification factors among the possibly several multiply lensed image-plane pixels. In this way, we maximized the observation depth at every grid cell in the source plane, and at the same time avoid the overcounting of multiply lensed regions in the image plane by working in the nondegenerate source plane.

After correcting for both the multiply lensed images and the multiply lensed regions in the image plane, which account for the multiplicity in galaxy counts and volume estimates respectively, we then computed the maximum survey volume where the {\it i}-th source galaxy could be detected using
\begin{equation}\label{V_max_formula}
V_{\mathrm{max},i} = \sum_{m = 1}^{N_\mathrm{field}} \sum_{n = 1}^{N_\mathrm{pix}} \int_0^{z_{\mathrm{max},i}(\mu_n,M_{\mathrm{UV},i})} \frac{dV_\mathrm{c}(z)}{d\Omega} \Delta \Omega_n,
\end{equation}
where $N_\mathrm{field}$ is the number of relevant target fields, the index {\it n} runs over all the $N_\mathrm{pix}$ pixels (or grid cells) in a given field, and $\Delta \Omega_n$ is the solid angle subtended by the {\it n}-th pixel. It should be noted that $z_{\mathrm{max},i}$ depends not only on the UV absolute magnitude of the source galaxy, but also varies across pixels depending on the magnification factor.

\begin{figure*}[tp!]
\centering
\gridline{\fig{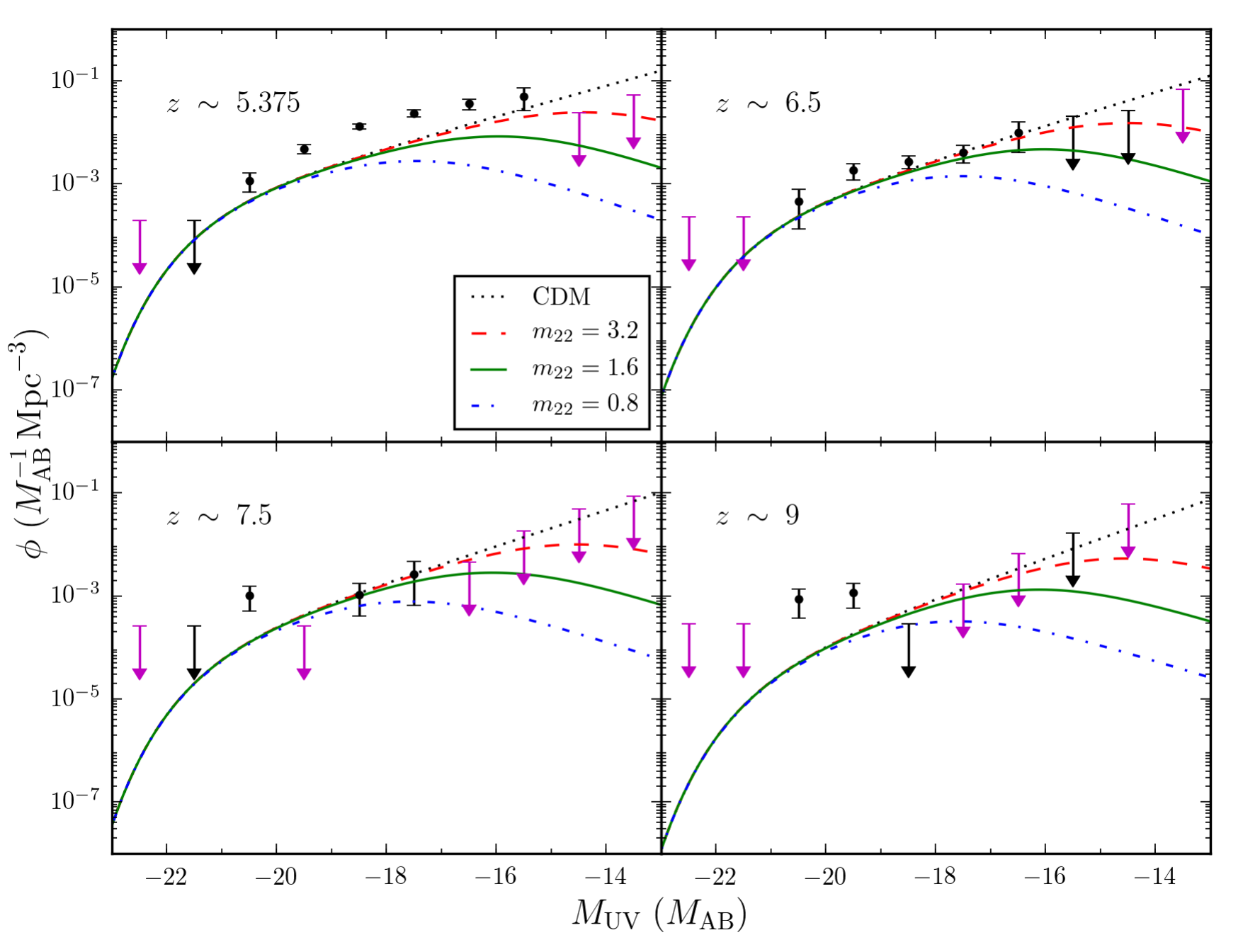}{85mm}{(a) Cluster fields only.} \fig{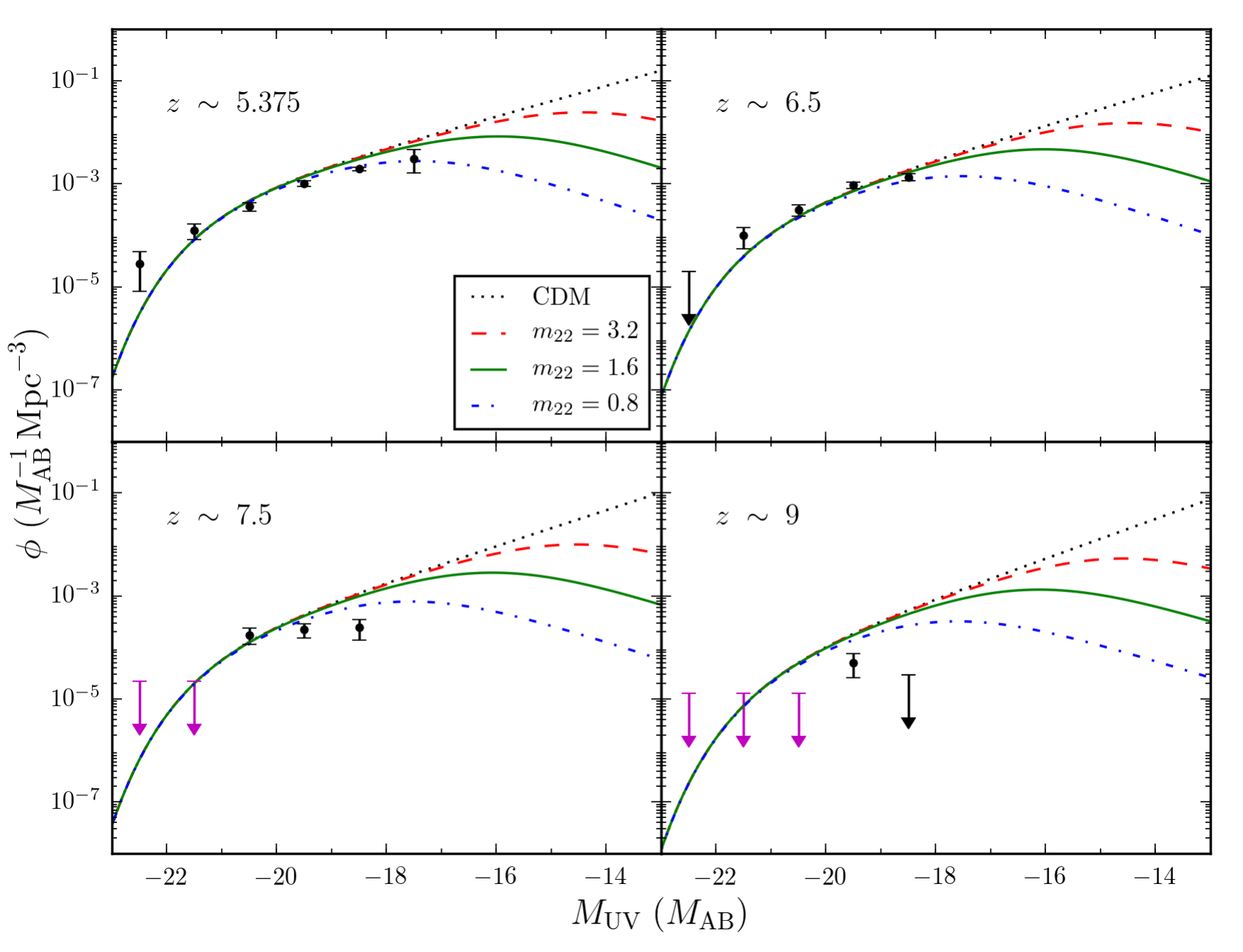}{85mm}{(b) Parallel fields only.}}
\gridline{\fig{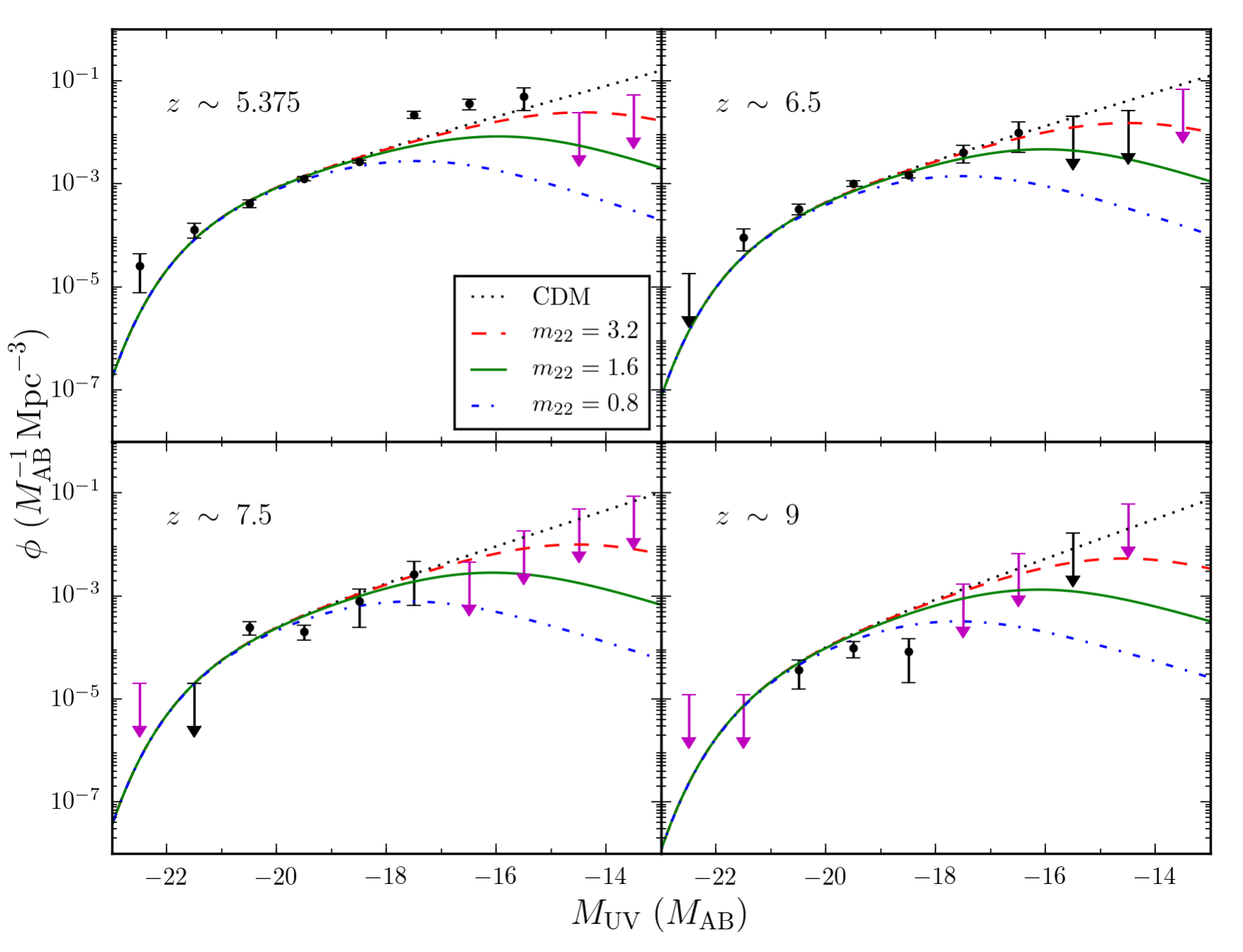}{85mm}{(c) Cluster plus parallel fields.}}
\caption{\label{UV_LF_inv_V_max}Stepwise UV luminosity functions reconstructed using the \citet{1968ApJ...151..393S} ``$1/V_\mathrm{max}$'' estimator at $z \sim 5.375$, 6.5, 7.5, and 9 respectively for the (a) cluster fields, (b) parallel fields, and (c) cluster plus parallel fields (also tabulated in Tables \ref{clt_inv_V_max_LF}-\ref{UV_LF_inv_V_max_table}). Magnitude bins with single galaxy detections are displayed as black upper limits, while those with null detections are displayed as magenta upper limits, computed simply as $V_\mathrm{max}^{-1}(M_k)$. Also shown are the predicted UV LFs according to the standard CDM model ({\it black dotted lines}; Equation (\ref{CDM_LF})) as well as the $\psi$DM model with different DM boson masses ({\it coloured lines}; Equation (\ref{psiDM_LF})), assuming fitting parameter values (including the normalization) given by Equation (\ref{LF_parameters}).}
\end{figure*}

We show in Figure \ref{UV_LF_inv_V_max} the UV LFs reconstructed independently for the six cluster fields (panel a) and the six parallel fields (panel b), as well as that jointly reconstructed using both the cluster and parallel fields (panel c). We divided our analysis into four separate redshift intervals with their mean redshifts labelled in the individual subpanels of Figure \ref{UV_LF_inv_V_max}. For each redshift interval, the integration limits in Equation (\ref{V_max_formula}) were also replaced with the corresponding redshift bound when determining $V_\mathrm{max}$. In addition, we followed \citet{1989ApJ...338...13C} to estimate the associated errors through
\begin{equation}
\sigma_k = \Bigg(\sum_{i = 1}^{N_k} \frac{1}{V_{\mathrm{max},i}^2}\Bigg)^{1/2}.
\end{equation}

Owing to the limited sample size within the relatively small survey volume, there are magnitude bins at certain redshift intervals having only single or even null galaxy detections. For these magnitude bins, we inferred the upper limits for the UV LF as $\phi_k \leq 1/V_\mathrm{max}(M_i)$ for single galaxy detections, and $\phi_k < 1/V_\mathrm{max}(M_k)$ for null detections. The faint-end slope of the UV LF is therefore likely to be flatter than the visual impression given by the chain of upper limits as shown in Figure \ref{UV_LF_inv_V_max}.

To examine in a more robust manner whether there exists a flattening in the faint-end slope, we employed the stepwise maximum likelihood (SWML) method introduced by \citet{1988MNRAS.232..431E} to independently reconstruct the UV LF. The advantage of this alternative approach over the ``$1/V_\mathrm{max}$'' method is that the shape of the LF (i.e. the ratios between various $\phi_k$'s) does not require any information regarding the relative volumes between different magnitude bins. It is therefore not sensitive to the large-scale 2D magnification profiles predicted by the lens models, but only requires knowledge of the magnifications of individual lensed galaxies. This method is also robust against field-to-field variance in the normalization of the UV LF \citep{2015ApJ...803...34B} that can potentially be introduced by cosmic variance or inaccuracies in the lens models (especially near the critical curves), as evident in the normalization offset between the cluster and parallel fields LF reconstructions shown in Figure \ref{UV_LF_inv_V_max}. Consequently, the SWML method should be able to provide a more accurate measure of the faint-end LF slope by avoiding the highly uncertain volume estimates of the multiply lensed and highly magnified regions in the cluster fields, together with the limited field samples resulting in slightly varying normalizations of the UV LF.

\begin{figure*}[tp!]
\centering
\includegraphics[width=127.5mm]{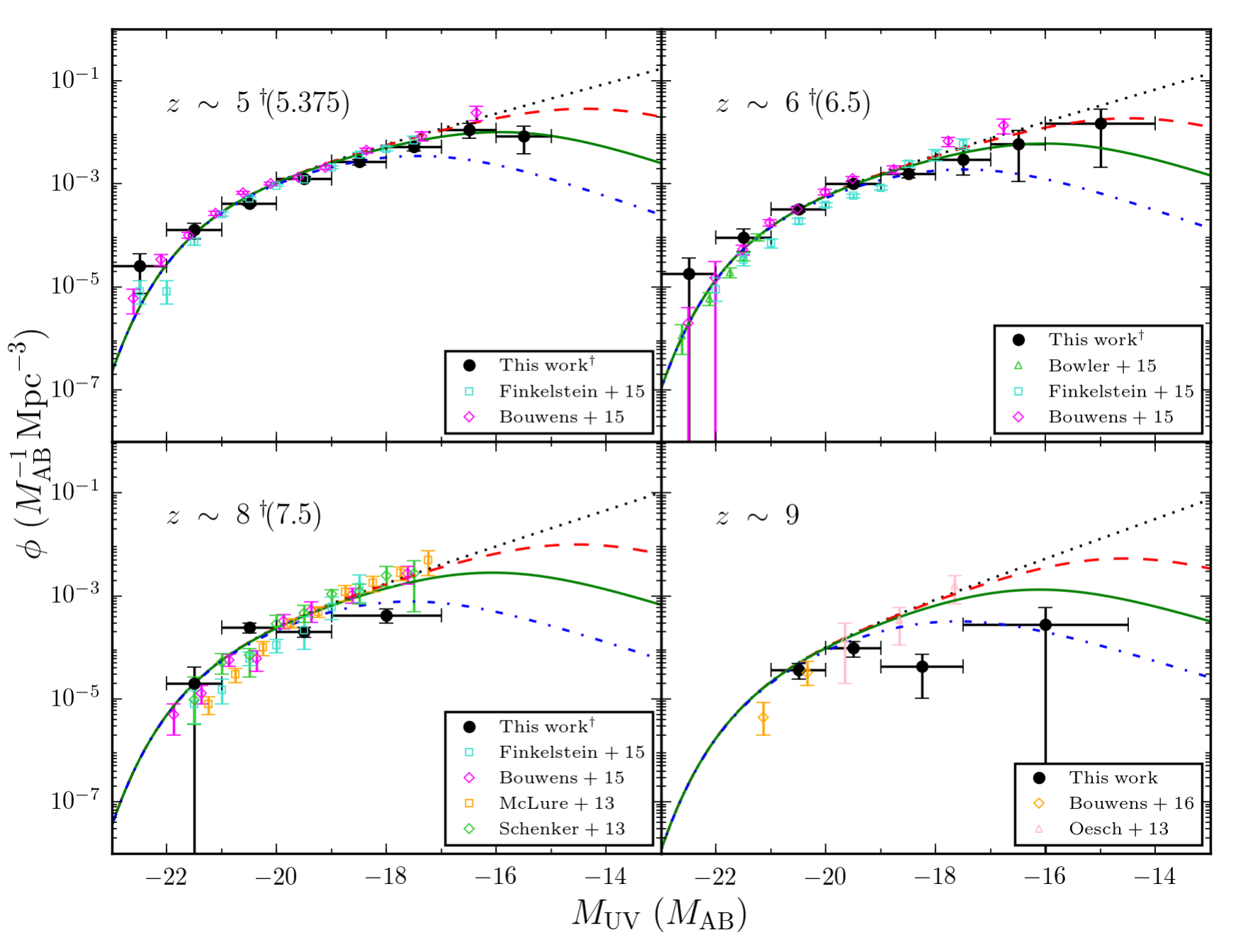}
\caption{\label{UV_LF_multi_works}Stepwise maximum likelihood UV luminosity functions at $z \sim 5.375$, 6.5, 7.5, and 9 respectively, constructed using data from all the twelve HFF fields analyzed in this work ({\it filled black circles}). For comparison, we plot as well the UV LFs determined by previous studies from multiple blank-field surveys at $z \sim 5, 6, 8$, and 9 respectively, including \citet{2016ApJ...830...67B} ({\it orange diamonds}; CANDELS), \citet{2015MNRAS.452.1817B} ({\it limegreen triangles}; UltraVISTA/COSMOS+UDS/SXDS), \citet{2015ApJ...810...71F} ({\it turquoise squares}; CANDELS+HUDF+HFF A2744 PAR+HFF MACS0416 PAR), \citet{2015ApJ...803...34B} ({\it magenta diamonds}; CANDELS+HUDF09+HUDF12+ERS+BoRG/HIPPIES), \citet{2013ApJ...773...75O} ({\it pink triangles}; HUDF09+HUDF12+CANDELS), \citet{2013MNRAS.432.2696M} ({\it orange squares}; HUDF12+HUDF09+ERS+CANDELS+BoRG), and \citet{2013ApJ...768..196S} ({\it limegreen diamonds}; HUDF+ERS+CANDELS-Deep+BoRG/HIPPIES). Also shown are the UV LFs predicted by the standard CDM model ({\it black dotted lines}), and the $\psi$DM model with boson masses $m_\mathrm{B} = 3.2 \times 10^{-22} \, \mathrm{eV}$ ({\it red dashed lines}), $1.6 \times 10^{-22} \, \mathrm{eV}$ ({\it green solid lines}), and $0.8 \times 10^{-22} \, \mathrm{eV}$ ({\it blue dash-dot lines}).}
\end{figure*}

Retaining the stepwise parameterization expressed in Equations (\ref{stepwise_parameterization}) and (\ref{W_formula}), the likelihood $\mathcal{L}$ can be written as
\begin{align}
\mathcal{L} = & \ \prod_k \prod_{i = 1}^{N_k} \frac{\phi(M_{\mathrm{UV},i})}{\int_{-\infty}^{M_\mathrm{max}(z_i,\mu_i)}\phi(M_\mathrm{UV}) dM_\mathrm{UV}} \nonumber\\
= & \ \prod_i \frac{\sum_k W(M_{\mathrm{UV},i} - M_k) \phi_k}{\sum_j \phi_j \Delta M_k H(M_j - M_\mathrm{max}(z_i,\mu_i))},
\end{align}
where the index $i$ in the second line is redefined for simplicity to run over the source galaxies in all magnitude bins. The faintest observable absolute magnitude $M_\mathrm{max}(z_i,\mu_i)$ corresponding to the {\it i}-th source galaxy can be obtained from Equation (\ref{M_UV_formula}) by substituting $m_\mathrm{filter}$ with $m_\mathrm{lim}$, and so we have
\begin{equation}
H(x) = 
\begin{cases}
\displaystyle 1 & \text{if $x < -\Delta M_k / 2$,} \\[1.5ex]
\displaystyle -\frac{x}{\Delta M_k} + \frac{1}{2} & \text{if $-\Delta M_k / 2 \leq x \leq \Delta M_k / 2$,} \\[1.5ex]
\displaystyle 0 & \text{if $x > \Delta M_k / 2$.}
\end{cases}
\end{equation}
We fixed the normalization of the SWML UV LF by imposing an extra constraint,
\begin{equation}
g \equiv \frac{\phi_\mathrm{SWML}(-20.50) \ \phi_\mathrm{SWML}(-19.50)}{\phi_{V_\mathrm{max}^{-1}}(-20.50) \ \phi_{V_\mathrm{max}^{-1}}(-19.50)} - 1 = 0,
\end{equation}
into the log-likelihood function in the form of a Lagrange multiplier $\log \, \mathcal{L} \to \log \, \mathcal{L} + \lambda g$ with the requirement that $\lambda = 0$ (so the shape of the LF is unaffected). The normalization constraint was chosen such that the values of $(\log_{10} \, \phi(-20.50) + \log_{10} \, \phi(-19.50)) / 2$ are identical between the UV LFs reconstructed respectively using the SWML and ``$1/V_\mathrm{max}$'' methods, for which the latter can be easily verified from Figure \ref{UV_LF_inv_V_max}(c) to agree reasonably well with the bright-end normalization in each redshift bin as derived from the much larger data set of \citet{2015ApJ...803...34B} having $\sim$$1000 \, \mathrm{arcmin}^2$ blank-field sky coverage. The errors in the various $\phi_k$'s were extracted from the diagonal elements of the covariance matrix $\mathrm{cov}(\phi_k)$, given by the inverse of the Fisher information matrix $\mathbf{I}(\phi_k)$.

\begin{deluxetable*}{c c c c c c c c c c c c c c c}
\tablecaption{Stepwise maximum likelihood UV LF in the twelve HFF fields at $z \sim 5.375$, $z \sim 6.5$, $z \sim 7.5$, and $z \sim 9$\label{UV_LF_SWML_table}}
\tablehead{\colhead{$M_\mathrm{UV}$} & \colhead{} & \colhead{No.~of galaxies} & \colhead{} & \colhead{$\phi$} & \colhead{} & \colhead{} & \colhead{} & \colhead{} & \colhead{} & \colhead{$M_\mathrm{UV}$} & \colhead{} & \colhead{No.~of galaxies} & \colhead{} & \colhead{$\phi$} \\
\colhead{($M_\mathrm{AB}$)} & \colhead{} & \colhead{} & \colhead{} & \colhead{($M_\mathrm{AB}^{-1} \, \mathrm{Mpc}^{-3}$)} & \colhead{} & \colhead{} & \colhead{} & \colhead{} & \colhead{} & \colhead{($M_\mathrm{AB}$)} & \colhead{}  & \colhead{} & \colhead{}& \colhead{($M_\mathrm{AB}^{-1} \, \mathrm{Mpc}^{-3}$)}
}
\startdata
\multicolumn{5}{c}{$z \sim 5.375$} & & & & & & \multicolumn{5}{c}{$z \sim 6.5$} \\
\hline
-22.50 & & 2 & & $0.000026 \pm 0.000019$ & & & & & & -22.50 & & 1 & & $0.000018 \pm 0.000018$ \\
-21.50 & & 10 & & $0.000130 \pm 0.000043$ & & & & & & -21.50 & & 5 & & $0.000091 \pm 0.000042$ \\
-20.50 & & 32 & & $0.000415 \pm 0.000042$ & & & & & & -20.50 & & 18 & & $0.000327 \pm 0.000044$ \\
-19.50 & & 98 & & $0.001274 \pm 0.000130$ & & & & & & -19.50 & & 56 & & $0.001017 \pm 0.000138$ \\
-18.50 & & 202 & & $0.002667 \pm 0.000330$ & & & & & & -18.50 & & 65 & & $0.001587 \pm 0.000294$ \\
-17.50 & & 78 & & $0.005262 \pm 0.000870$ & & & & & & -17.50 & & 7 & & $0.003019 \pm 0.001552$ \\
-16.50 & & 21 & & $0.011220 \pm 0.003534$ & & & & & & -16.50 & & 3 & & $0.005929 \pm 0.004814$ \\
-15.50 & & 5 & & $0.008545 \pm 0.004677$ & & & & & & -15.00 & & 2 & & $0.015050 \pm 0.012922$ \\
-14.00 & & 0 & & $<\!0.038092$ & & & & & & -13.50 & & 0 & & $<\!0.067898$ \\
\hline
\multicolumn{5}{c}{$z \sim 7.5$} & & & & & & \multicolumn{5}{c}{$z \sim 9$} \\
\hline
-22.50 & & 0 & & $<\!0.000020$ & & & & & & -22.00 & & 0 & & $<\!0.000012$ \\
-21.50 & & 1 & & $0.000020_{-0.000020}^{+0.000021}$ & & & & & & -20.50 & & 3 & & $0.000037 \pm 0.000012$ \\
-20.50 & & 12 & & $0.000243 \pm 0.000052$ & & & & & & -19.50 & & 8 & & $0.000098 \pm 0.000033$ \\
-19.50 & & 10 & & $0.000202 \pm 0.000043$ & & & & & & -18.25 & & 2 & & $0.000042 \pm 0.000032$ \\
-18.00 & & 13 & & $0.000426 \pm 0.000129$ & & & & & & -16.00 & & 1 & & $0.000280_{-0.000280}^{+0.000320}$ \\
-16.00 & & 0 & & $<\!0.011354$ & & & & & & \nodata & & \nodata & & \nodata \\
-14.50 & & 0 & & $<\!0.048022$ & & & & & & \nodata & & \nodata & & \nodata \\
\enddata
\tablecomments{Upper limits for null-detection bins were estimated using the ``$1/V_\mathrm{max}$'' method assuming $\phi_k < 1 / V_\mathrm{max}(M_k)$.}
\end{deluxetable*}

\begin{deluxetable*}{c c c c c c c}
\tablecaption{Reduced $\chi^2$ with $\sigma$ values for various model LFs on the SWML UV LFs\label{SWML_LF_chisq_sigma}}
\tablehead{\colhead{Model} & \colhead{$m_{22}$} & \colhead{$z \sim 5.375$} & \colhead{$z \sim 6.5$} & \colhead{$z \sim 7.5$} & \colhead{$z \sim 9$} & \colhead{Overall}}
\startdata
CDM &  & $3.23 \ (3.57\sigma)$ & $1.48 \ (1.45\sigma)$ & $15.19 \ (10.33\sigma)$ & $90.58 \ (24.51\sigma)$ & $35.49 \ (26.75\sigma)$ \\
$\psi$DM & 3.2 & $1.68 \ (1.76\sigma)$ & $1.31 \ (1.21\sigma)$ & $13.71 \ (\phm{1}9.71\sigma)$ & $65.61 \ (20.69\sigma)$ & $26.58 \ (22.73\sigma)$ \\
$\psi$DM & 1.6 & $0.66 \ (0.31\sigma)$ & $1.40 \ (1.34\sigma)$ & $\phm{1}9.46 \ (\phm{1}7.69\sigma)$ & $37.24 \ (15.26\sigma)$ & $15.87 \ (16.80\sigma)$ \\
$\psi$DM & 0.8 & $2.24 \ (2.47\sigma)$ & $2.50 \ (2.67\sigma)$ & $\phm{1}2.46 \ (\phm{1}2.63\sigma)$ & $\phm{1}8.89 \ (\phm{1}6.56\sigma)$ & $\phm{1}5.62 \ (\phm{1}8.32\sigma)$ \\
\enddata
\end{deluxetable*}

Utilizing all the available data from the twelve HFF fields, we present the SWML estimates of the UV LF at $z \sim 5.375$,  6.5,  7.5, and 9 respectively in Figure \ref{UV_LF_multi_works} and Table \ref{UV_LF_SWML_table}. A relatively shallow faint-end slope is significantly preferred over a steeply rising power law for the UV LF at all the redshift intervals plotted except for that centred at $z \sim 6.5$ (in which the statistical uncertainty of the faintest data point is so large as to not favour nor disfavour a shallower faint-end slope than that of the CDM model), as reflected in the goodness-of-fit statistics for different model LFs compiled in Table \ref{SWML_LF_chisq_sigma}. The directly reconstructed UV LF therefore supports our results from the magnification bias and clustercentric radial density profiles in Sections \ref{magnification_bias_section} and \ref{radial_test_section} respectively for a flattening, if not rollover, in the UV LF at the faint end.

Note that we do not have any galaxy detection fainter than $M_\mathrm{UV} \sim -17$ in the $z \sim 7.5$ redshift interval, the absence of which is marginally consistent with a Schechter-like LF as discussed previously for the ``1/$V_\mathrm{max}$'' reconstruction (see Figure \ref{UV_LF_inv_V_max}). Our faintest data point at $M_\mathrm{UV} = -18$ is well within the conservative apparent magnitude limit at $m_\mathrm{lim} = 28.5$ (corresponding to an effective detection limit at $M_\mathrm{UV,lim} \sim -14$), and hence should be relatively robust against photometric and also magnification uncertainties of individual galaxy candidates. The lack of such low-luminosity galaxies in the HFF cluster fields is in tension with previous blank-field studies at similar UV absolute magnitudes (see Figure \ref{UV_LF_multi_works}). The same is true for the substantially steeper faint-end LF slope compared to ours as derived by \citet{2017ApJ...835..113L} using imaging data from the first two HFF clusters, A2744 and MACS0416, on which \citet{2017ApJ...836...61M} relied to conclude a lower limit on the $\psi$DM boson mass to be $m_\mathrm{B} \geq 8 \times 10^{-22} \, \mathrm{eV}$, whereas our lower galaxy number density measured from the complete HFF data set revises significantly downwards the limit on the boson mass and is fully consistent with constraints from local dSph galaxy kpc-scale cores. Here we stress again that our galaxy selection method is deliberately conservative so that unlike many other works, we do not rely on large upward model-dependent corrections to the faint end of the LF from sources that could barely be detected in single passbands. Without lensing, these desperately faint sources fall into the photometrically incomplete regime, subject to large systematic uncertainties that might not be well understood, and is therefore likely to be the primary cause for their tension with our results.

\section{Results {\uppercase\expandafter{\romannumeral 4\relax}}: Evolutions of the Cumulative UV LF and the Cosmic Star Formation Rate Density}\label{sfrd_section}

\begin{figure}[tp!]
\centering
\includegraphics[width=85mm]{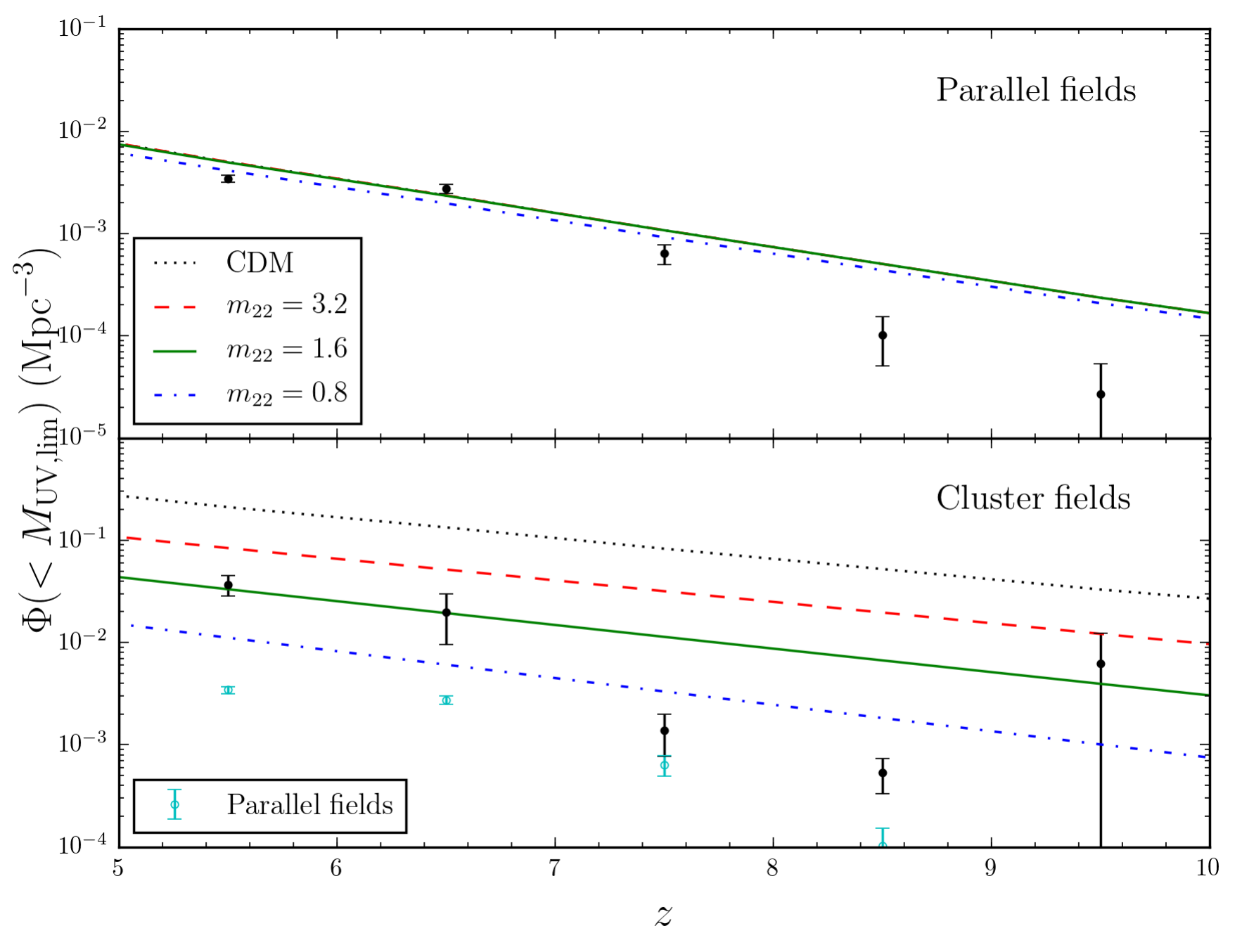}
\caption{\label{multi_fld_cumulative_UV_LF}Cumulative UV LFs, i.e.~cumulative galaxy number densities, reconstructed independently for the parallel fields ({\it upper panel}) and the cluster fields ({\it lower panel}) at $z \sim 5$\,--\,10, taking into consideration the $\sim$$5 \, \mathrm{mag}$ extra observing depth ($\sim$$1 \, \mathrm{dex}$ in $\Phi(<\!M_\mathrm{UV,lim})$) at a given redshift for the latter case. Also overlaid in the lower panel are the data points from the parallel fields ({\it cyan circles}) for direct comparison.}
\end{figure}

A consistent result obtained across independent tests of the UV LF described in Sections \ref{magnification_bias_section}--\ref{LF_section} is the deficiency in faint galaxies relative to the numbers expected based on the CDM model at $5 \lesssim z \lesssim 10$. Here, we analyze the evolution of the UV LF across the redshift range concerned. Figure \ref{multi_fld_cumulative_UV_LF} shows the (unlensed, i.e.~corrected for lensing magnification) cumulative UV LF, $\Phi(<\!M_\mathrm{UV,lim})$, at $z \sim 5$\,--\,10 respectively for the parallel fields and the cluster fields, for which we have different observing depths and hence different predicted normalizations. The data points were determined using the ``$1/V_\mathrm{max}$'' method described in Section \ref{LF_section}, where we corrected for the overcounting of multiply lensed regions (and images) in the cluster fields as before.

Over the redshift range $5 \lesssim z \lesssim 8$, the observed galaxy number density in the parallel fields follows a stable decline towards increasing redshift as predicted by the model LFs for both the standard CDM and $\psi$DM models. Given the relatively shallow observing depths of the parallel fields compared with the cluster fields, we are not able to distinguish between the only slightly differing predictions of these two DM models for the parallel fields (see Figure \ref{LF_mag_bias_comparison}). On the other hand, with the amplified faint-end differences between the UV LFs predicted by different DM models in the highly magnified cluster fields, the cumulative galaxy number density at $z \sim 5$\,--\,7 appears to be well described by the $\psi$DM model (which features a faint-end turnover in the UV LF) with $m_\mathrm{B} = 1.6 \times 10^{-22} \, \mathrm{eV}$.

At yet higher redshifts, however, there was a faster-than-expected plunge with increasing redshift in the number density of galaxies, as reflected also in the non-detection of $M_\mathrm{UV} \lesssim -20$ galaxies in the parallel fields at $z > 8$ (see panel (b) of Figure \ref{UV_LF_inv_V_max}). The same is also seen in the cluster fields beyond $z \sim 7$, where the decline in galaxy number density with redshift is again steeper than expected even for the $\psi$DM model. In fact, the galaxy number density in the cluster fields at $z > 7$ is surprisingly comparable to that in the parallel fields despite the enhanced effective flux limit, with a deficiency that is roughly an order of magnitude below that expected for $\psi$DM model predictions with, say $m_{22} = 1.6$. Note that the data point at $z = 9.5$ in the cluster fields was inferred from just a single galaxy detection, and so the steeper than expected decline in number density likely continues beyond $z > 9$ given the trend at slightly lower redshifts.

Accordingly, our results suggest that there was a sustained deficit of faint (sub-$L_\psi$, see Equation (\ref{LF_parameters})) galaxies at $5 \lesssim z \lesssim 10$ that favours a slow rollover in the UV LF, at odds with the predictions of the standard CDM model but more consistent with the predictions of $\psi$DM model for a (reduced) boson mass $0.8 \lesssim m_{22} \lesssim 1.6$. Moreover, the formation of extremely bright ($M_\mathrm{UV} \lesssim -21$) galaxies was comparatively less efficient at $z \gtrsim 8$ than at later epochs (lower redshifts). The lack of such bright galaxies cannot be explained even by the $\psi$DM-motivated model LFs, which have essentially the same asymptotic bright-end behaviour as the Schechter function in the standard CDM model.

We verified the aforementioned phenomenological picture of early galaxy formation by studying the evolution of the cosmic star formation rate (SFR) density in the HFF fields. The cosmic SFR density was derived from the luminosity density using the standard SFR--UV conversion relation of \citet{1998ApJ...498..106M}, whereby
\begin{equation}
L_\mathrm{UV} = 8.0 \times 10^{27} \Bigg(\frac{\mathrm{SFR}}{M_\odot \, \mathrm{yr}^{-1}}\Bigg) \, \mathrm{erg} \, \mathrm{s}^{-1} \, \mathrm{Hz}^{-1}
\end{equation}
assuming a \citet{1955ApJ...121..161S} initial mass function (IMF) over the stellar mass range 0.1\,--\,125 $M_\odot$, and a steady SFR after an intial burst when the main sequence turnoff mass falls below $\sim$$10 \, M_\odot$ (as commonly adopted in the literature). Rather than following the popular practice of obtaining the best-fit parameters of an assumed parametric form of the UV LF, and then integrating this LF down to a given lower luminosity limit to infer the cosmic SFR density, we derived it by summing up the contributions to the UV luminosity density from every galaxy candidate in our sample. Our approach is therefore immune to assumptions about the form of the analytical function that should be fitted to the measured UV LF, a form which may not truthfully reflect the actual LF.

We also accounted for extinction by dust enshrouding the galaxies, which scatters light at UV light to IR wavelengths, by making use of the infrared excess (IRX) -- UV continuum slope ($\beta$) relationship constructed by \citet{2016ApJ...833...72B} for $z \sim 2$\,--\,10 galaxies as given by $A_\mathrm{UV} = 1.5 (\beta + 2.23)$. This extinction law was only applied to relatively low-$z$ and luminous ($\gtrsim\!L_\star$) galaxies having $\beta > -2.23$ (see Equation (\ref{beta_formula})) such that $A_\mathrm{UV} > 0$. \citet{2016ApJ...833...72B} postulated that the dust extinction for fainter galaxies is indeed much lower and that the stellar mass instead of $\beta$ serves as a better proxy for IRX and hence $A_\mathrm{UV}$. We would, however, like to avoid as much as possible the highly model-dependent estimations of stellar masses, provided that the effect of dust extinction on the overall cosmic SFR density is presumably dominated by the most luminous galaxies, making corrections for dust extinction in low-luminosity galaxies relatively unimportant. In any case, despite the lack of consensus on the high-$z$ IRX--$\beta$ relationship between different studies \citep{2017MNRAS.467.1360B}, we shall stress that the issue of dust correction is of little concern to the purpose of our work since it affects the data and the model predictions in exactly the same manner, and hence does not play any role in terms of distinguishing between different DM models. Furthermore, the predicted UV luminosity densities (or equivalently, cosmic SFR densities) with respect to the CDM and $\psi$DM model LFs can be straightforwardly computed utilizing Equations (\ref{CDM_lensed_cumulative_LF}) and (\ref{psiDM_lensed_cumulative_LF}) respectively, i.e.~$\rho_\mathrm{UV} = L_\star \Phi(\alpha + 1) = L_\star [\mu\Phi_\mathrm{lensed}(\mu,\alpha + 1)]$, where we made the replacement $\alpha \to \alpha + 1$ to take care of the extra factor of $L_\mathrm{UV}$ in the integral.

\begin{figure*}[tp!]
\centering
\includegraphics[width=127.5mm]{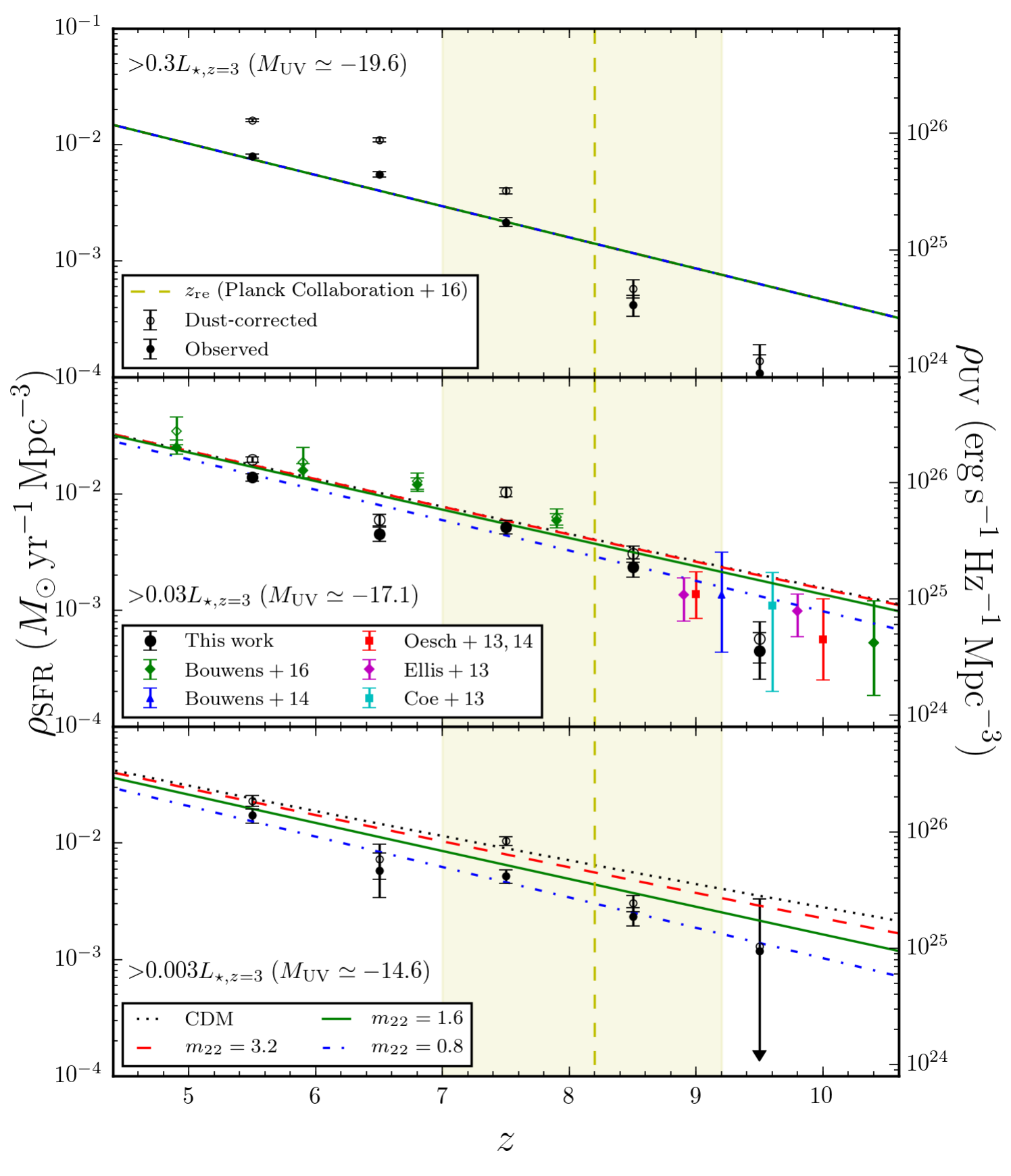}
\caption{\label{multi_fld_SFR_density}Cosmic star formation rate density ({\it left axis}) at $z \sim 5$\,--\,10 inferred from the UV luminosity density ({\it right axis}) via the \citet{1998ApJ...498..106M} conversion relation assuming a \citet{1955ApJ...121..161S} IMF. In all the three panels, we present both the observed luminosity densities ({\it black solid circles}) and the dust-corrected SFR densities ({\it black hollow circles}) employing the IRX-$\beta$ relationship of \citet{2016ApJ...833...72B}, whereas for clarity the various model predictions are shown without being corrected for extinction. In the top panel, we integrated the luminosity density down to $0.3 L_{\star,z=3}$, which is slightly brightwards of the detection limit in the parallel fields at $z \sim 10$ (see Figure \ref{LF_mag_bias_comparison}), thus allowing us to utilize all the data from the twelve HFF fields. In the middle and bottom panels, we lowered the integration limits to $0.03 L_{\star,z=3}$ and $0.003 L_{\star,z=3}$ respectively, where we only made use of the data from the cluster fields. The former limit is similar to most of the values adopted by the literature, so we provide as well the constraints from \citet{2016ApJ...833...72B} ({\it green solid and hollow diamonds for uncorrected and corrected SFR densities respectively}; CANDELS+HUDF09+HUDF12+ERS+BoRG/HIPPIES), \citet{2014ApJ...795..126B} ({\it blue triangle}, CLASH), \citet{2013ApJ...773...75O,2014ApJ...786..108O} ({\it red squares}, HUDF+CANDELS), \citet{2013ApJ...763L...7E} ({\it magenta diamonds}, HUDF12), and \citet{2013ApJ...762...32C} ({\it cyan square}, CLASH). The latter limit corresponds roughly to the minimum effective flux limit in the cluster fields at $z \sim 10$ (see Figure \ref{LF_mag_bias_comparison}), which covers an appreciable portion of galaxies at the faint-end UV LF. Also overlaid in each panel is the redshift of ``instantaneous'' reionization reported by \citet{2016A&A...596A.108P} ({\it yellow dashed line sandwiched between asymmetric uncertainties represented by the yellow shaded regions}; \texttt{lollipop}+PlanckTT) for reference. The model predictions in all the panels are computed from the usual set of CDM and $\psi$DM model LFs as labelled in the bottom panel.}
\end{figure*}

\begin{deluxetable*}{c c c c c c c c c}
\tablecaption{UV luminosity density and cosmic star formation rate density estimates from the HFF fields\label{SFRD_table}}
\tablehead{\colhead{$z$} & \colhead{} & \colhead{\phm{a}No.~of galaxies\tablenotemark{a}} & \colhead{} & \colhead{$\rho_\mathrm{UV}$} & \colhead{} & \multicolumn{3}{c}{\phm{b}$\rho_\mathrm{SFR}$\tablenotemark{b}} \\
\colhead{} & \colhead{} & \colhead{} & \colhead{} & \colhead{($10^{25} \, \mathrm{erg} \, \mathrm{s}^{-1} \, \mathrm{Hz}^{-1} \, \mathrm{Mpc}^{-3}$)} & \colhead{} & \multicolumn{3}{c}{($10^{-3} \, M_\odot \, \mathrm{yr}^{-1} \, \mathrm{Mpc}^{-3}$)} \\
\colhead{} & \colhead{} & \colhead{} & \colhead{} & \colhead{} & \colhead{} & \colhead{Uncorrected} & \colhead{} & \colhead{\phm{c}Dust-corrected\tablenotemark{c}}
}
\startdata
\multicolumn{9}{c}{$>\!0.3 L_{\star,z=3} \ (M_\mathrm{UV} \simeq -19.6)$} \\
\hline
5.5 &  & 64 &  & $6.370 \pm 0.260$ &  & $7.962 \pm 0.324$ &  & $16.187 \pm 0.463\phm{1}$ \\
6.5 &  & 43 &  & $4.455 \pm 0.230$ &  & $5.569 \pm 0.287$ &  & $10.982 \pm 0.404\phm{1}$ \\
7.5 &  & 15 &  & $1.729 \pm 0.151$ &  & $2.161 \pm 0.189$ &  & $4.008 \pm 0.258$ \\
8.5 &  & 4 &  & $0.337 \pm 0.070$ &  & $0.421 \pm 0.088$ &  & $0.583 \pm 0.104$ \\
9.5 &  & 1 &  & $0.087 \pm 0.038$ &  & $0.109 \pm 0.047$ &  & $0.138 \pm 0.053$ \\
\hline
\multicolumn{9}{c}{$>\!0.03 L_{\star,z=3} \ (M_\mathrm{UV} \simeq -17.1)$} \\
\hline
5.5 &  & 125 &  & $11.103 \pm 0.787\phm{1}$ &  & $13.879 \pm 0.984\phm{1}$ &  & $19.720 \pm 1.129\phm{1}$ \\
6.5 &  & 29 &  & $3.629 \pm 0.495$ &  & $4.537 \pm 0.619$ &  & $6.015 \pm 0.687$ \\
7.5 &  & 10 &  & $4.156 \pm 0.554$ &  & $5.195 \pm 0.692$ &  & $10.402 \pm 0.911\phm{1}$ \\
8.5 &  & 7 &  & $1.878 \pm 0.337$ &  & $2.347 \pm 0.421$ &  & $3.065 \pm 0.481$ \\
9.5 &  & 1 &  & $0.360 \pm 0.155$ &  & $0.450 \pm 0.194$ &  & $0.572 \pm 0.219$ \\
\hline
\multicolumn{9}{c}{$>\!0.003 L_{\star,z=3} \ (M_\mathrm{UV} \simeq -14.6)$} \\
\hline
5.5 &  & 150 &  & $13.805 \pm 1.969\phm{1}$ &  & $17.257 \pm 2.462\phm{1}$ &  & $23.098 \pm 2.523\phm{1}$ \\
6.5 &  & 34 &  & $4.663 \pm 1.927$ &  & $5.829 \pm 2.409$ &  & $7.308 \pm 2.427$ \\
7.5 &  & 10 &  & $4.156 \pm 0.554$ &  & $5.195 \pm 0.692$ &  & $10.402 \pm 0.911\phm{1}$ \\
8.5 &  & 7 &  & $1.878 \pm 0.337$ &  & $2.347 \pm 0.421$ &  & $3.065 \pm 0.481$ \\
9.5 &  & 2 &  & $0.944 \pm 1.702$ &  & $^{\phm{\ }}1.180_{-1.180\phm{1}}^{+2.128\phm{1}}\phm{\ }$ &  & $^{\phm{\ }}1.301_{-1.301\phm{1}}^{+2.130\phm{1}}\phm{\ }$ \\
\enddata
\tablenotetext{a}{The composite galaxy sample assembled from all the cluster fields and parallel fields was used for deriving the UV luminosity density contributed by $>\!0.3 L_{\star,z=3}$ galaxies, whereas we only used the cluster field galaxies for the remaining two determinations with fainter luminosity limits.}
\tablenotetext{b}{SFR densities were computed from the UV luminosity densities according to the conversion relation of \citet{1998ApJ...498..106M} and assuming a 0.1\,--\,125 $M_\odot$ \citet{1955ApJ...121..161S} IMF.}
\tablenotetext{c}{Extinction corrections based on the IRX-$\beta$ relationship of \citet{2016ApJ...833...72B}, specified by $A_\mathrm{UV} = 1.5 (\beta + 2.23)$, were applied individually to galaxies with $\beta > -2.23$, which correspond to the predominant sample of relatively low-$z$ and intrinsically luminous galaxies. The remaining ones are expected to have significantly lower dust extinctions that we do not consider in this work.}
\end{deluxetable*}

The estimated cosmic SFR density is plotted against redshift in Figure \ref{multi_fld_SFR_density} (and also tabulated in Table \ref{SFRD_table}), where the three panels (from top to bottom) correspond to three descending lower limits of integration for the UV luminosity. By lowering the integration limit in discrete steps, we can see the effect of including only galaxies at the bright end and then gradually adding those towards the faint end. To begin with, we adopted a lower integration limit of $0.3 L_{\star,z=3}$, i.e.~$M_\mathrm{UV} \simeq -19.6$ (see the top panel of Figure \ref{multi_fld_SFR_density}). The UV luminosity density thus determined comprises solely the population of galaxies at the bright end, for which we have shown earlier to be sparsely populated at $z \gtrsim 8$. We see a similarly steeper than expected decline in the UV luminosity density at essentially the same redshift as found previously, $z \simeq 8$, reinforcing our earlier argument for the inefficient formation of galaxies at the bright-end UV LF prior to the redshift of ``instantaneous'' reionization at $z \simeq 8.2$ \citep{2016A&A...596A.108P}. In spite of this, the data at $z < 8$ are in excellent agreement with the predicted (log-)linear evolution of cosmic SFR density regardless of the choice made for the model LF.

The observed trend of a sharp rise in the cosmic SFR density (with decreasing redshifts) during $z \sim 8$\,--\,10 still holds after lowering the integration limit to $0.03 L_{\star,z=3}$, i.e.~$M_\mathrm{UV} \simeq -17.1$ (see the middle panel of Figure \ref{multi_fld_SFR_density}). Such a luminosity is roughly the detection limit of typical deep field surveys used in other studies. As can be seen, our estimates at $z \gtrsim 8$ are broadly consistent with the results compiled from a number of recent determinations of the high-$z$ cosmic SFR density, including CLASH \citep{2012ApJS..199...25P} lensed $z \sim 9$\,--\,11 detections. Even so, our results seem to yield an arguably steeper slope, $d\rho_\mathrm{SFR}/dz$, that is in moderate tension with claims that the evolution of the cosmic SFR density at very high redshifts is marginally consistent with a simple extrapolation of the UV LF at $z \sim 4$\,--\,8 (e.g.~\citealt{2013ApJ...762...32C,2013ApJ...763L...7E,2014ApJ...795..126B}), but is in line with the contrasting view of \citet{2013ApJ...773...75O,2014ApJ...786..108O} for a more rapid decline in the cosmic SFR density at $z > 8$. \citet{2015ApJ...813...21M} demonstrated that, in the context of CDM, the non-linear evolution of the cosmic SFR density can plausibly be modelled by including multiple phases of star formation during halo assembly, which spanned less than $100 \, \mathrm{Myr}$ at $z \gtrsim 8$, instead of the usual assumption of a constant SFR dominated by young stellar populations at relatively low redshifts. Therefore, it would be illuminating to incorporate this factor into the $\psi$DM model in future studies to see whether the accelerated SFR can also be reproduced in such context.

Disregarding the rapid decline in the cosmic SFR density beyond $z \sim 8$, the overall normalization of the data points at $z \lesssim 8$ is evidently best-matched by the predictions of the $\psi$DM model for $m_{22} = 0.8$ (blue dash-dot line).\footnote{The little kink at $z = 6.5$ is resulted from the non-detection of $M_\mathrm{UV} < -21$ galaxies in the cluster fields (see panel (a) of Figure \ref{UV_LF_inv_V_max}), whereas six of them could indeed be identified in the parallel fields (see Table \ref{par_inv_V_max_LF}). Since the $z = 6.5$ data point for the integration limit at $0.3 L_{\star,z=3}$ (see the top panel of Figure \ref{multi_fld_SFR_density}), which combines data from both the cluster and parallel fields to gain more statistics within a larger search volume, shows no significant discrepancy compared to the model predictions, it is likely that the kink here is merely a consequence of cosmic variance where the number density of these bright-end galaxies in the cluster fields underrepresents the average cosmic value.} Given that the bright-end galaxy population is well described by a standard Schechter function (as illustrated in the top panel of Figure \ref{multi_fld_SFR_density}), a net downward offset in the normalization between the sequentially adopted integration limits of $M_\mathrm{UV} \simeq -19.6$ and $-17.1$ implies a sub-Schechter UV LF (i.e.~a shallower faint-end slope), where the extent of the offset reflects the degree of suppression of the UV LF within this magnitude range.

To fully utilize the magnification-boosted observing depth in the HFF cluster fields, we further extended our analysis of the cosmic SFR density down to a lower integration limit of $0.003 L_{\star,z=3}$, i.e.~$M_\mathrm{UV} \simeq -14.6$ (see the bottom panel of Figure \ref{multi_fld_SFR_density}), which almost coincides with the maximal effective detection limit at $z \sim 10$. In this way, we can probe much deeper towards the faint end of the UV LF, where the differences between various model predictions are larger owing to the extended coverage of the faint-end turnover featured in the $\psi$DM model LFs. Consequently, the persistent lack of galaxies relative to the CDM prediction becomes even clearer across the entire redshift range, further strengthening all the previous arguments for the existence of a slow rollover in the UV LF.

Lastly, we comment on the much larger uncertainty associated with the data point at $z = 9.5$ in the bottom panel of Figure \ref{multi_fld_SFR_density} compared with the corresponding data points in top and middle panels. In the bottom panel, by integrating to a lower luminosity limit, we added one extra galaxy with $M_\mathrm{UV} \simeq -16$, bringing the total number of galaxies in this bin from one to two. This additional faint galaxy is only detectable due to its relatively high magnification, so the inclusion of its contribution to the data point does not affect our earlier conclusion that the bright-end population was underpopulated at $z \gtrsim 8$. Because of its relatively high magnification, however, the effective volume searched for this galaxy is relatively small, leading to the larger uncertainty in the cosmic SFR density in this redshift bin.

\section{Robustness of Results Against Alternative Lens Modelling Approaches}\label{lens_model_dependence}

Accurate magnification estimates of individual lensed galaxies are crucial to the determination of the faint-end LF slope. The inherent difficulty in modelling the deflection field near the critical curves unavoidably leads to relatively large uncertainties in the predicted magnification factors of highly magnified galaxies, which mostly correspond to those residing at the faint-end of the UV LF. In the preceding anaylses, we estimated the magnification factors of our galaxy candidates using the free-form/semi-parametric WSLAP+ lens models. In this section, we examine the robustness of our results against alternative lens modelling approaches. Specifically, we made use of the publicly available CATS (Clusters As TelescopeS) lens models for the six HFF clusters \citep{2009ApJ...707L.163S,2012MNRAS.426.3369J,2015MNRAS.452.1437J,2015MNRAS.446.4132J,2014MNRAS.444..268R,2010MNRAS.402L..44R}, constructed with the parametric LENSTOOL modelling algorithm \citep{2009MNRAS.395.1319J},\footnote{The overall mass distributions of the galaxy clusters were modelled by two major components, including the contributions from one or several cluster-scale halo(s) as well as those from individual cluster members, all of which were parameterized as Pseudo Isothermal Elliptical Mass Distributions (PIEMDs).} to repeat the tests performed in the previous sections. We restricted the identification of galaxy candidates in the cluster fields within the regions spanned mutually by the fields of view of the WSLAP+ and CATS models, from which we recovered $>\!90\%$ (240 out of 260) of the $z \geq 4.75$ galaxy candidates selected solely from the WSLAP+ model coverages.

\begin{figure}[tp!]
\centering
\includegraphics[width=85mm]{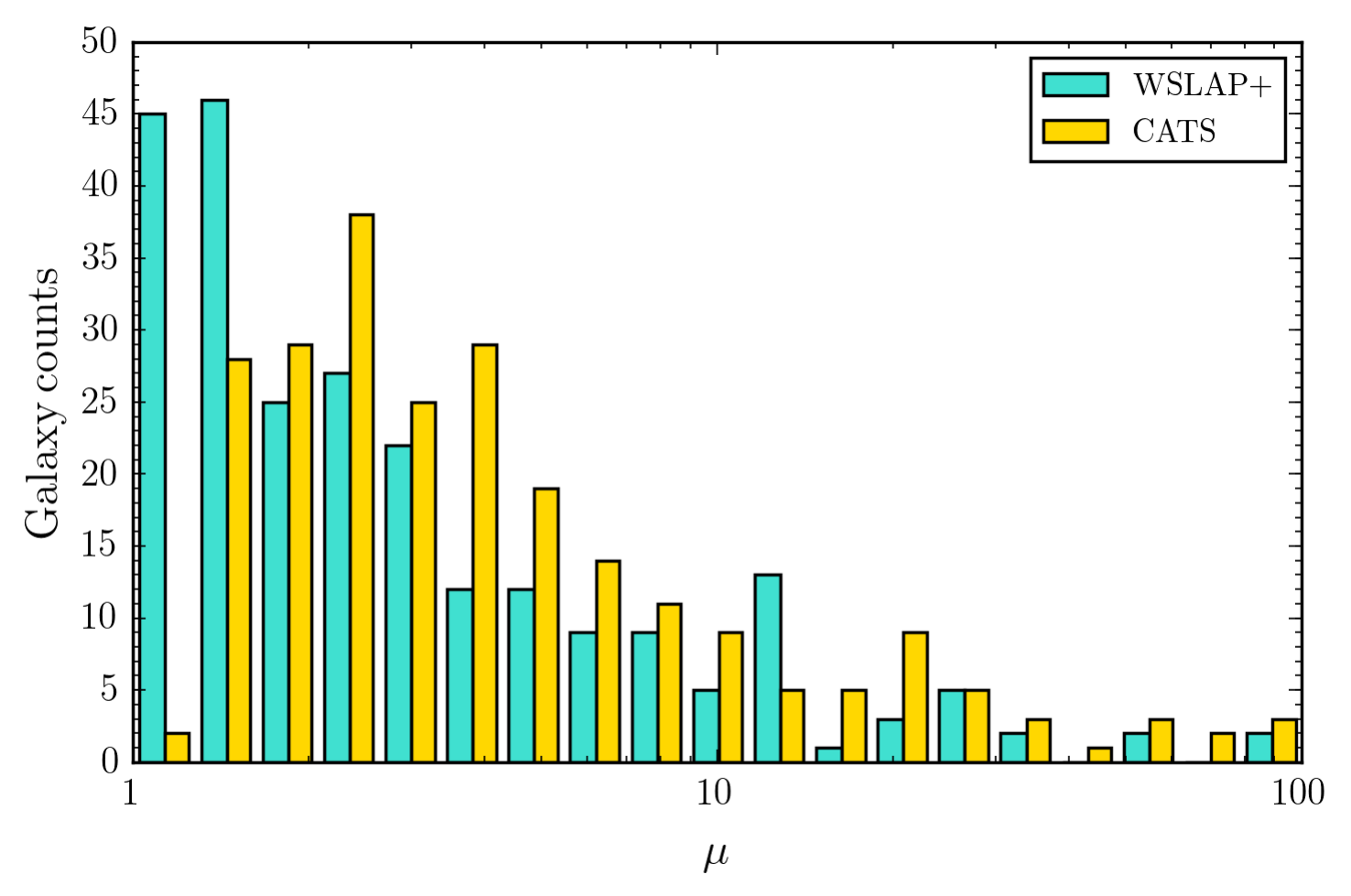}
\caption{\label{multi_clt_magnification_distribution_CATS_comparison}Magnification factor distributions of all the mutually selected galaxy candidates in the six HFF cluster fields, which were estimated respectively using the free-form/semi-parametric WSLAP+ lens models ({\it turquoise}) and the parametric CATS/LENSTOOL lens models ({\it gold}).}
\end{figure}

Prior to presenting the CATS-based version of the results, it is worth comparing the model-predicted magnifications of the galaxy candidates estimated using the WSLAP+ and CATS models respectively. Figure \ref{multi_clt_magnification_distribution_CATS_comparison} shows the distributions of the magnification factors estimated from the two sets of lens models for the galaxy candidates in the cluster fields. There is a slight offset between the peaks of the two distributions, with the CATS-predicted magnification factors being generally higher than the WSLAP+ equivalents, particularly in the low-magnification regime. The underlying cause of this discrepancy is most likely rooted in the inherently different approaches adopted by the two algorithms for modeling the cluster-scale halo. For example, in the cluster outskirts where a sizeable fraction of galaxy candidates was identified, the cluster-scale halo was essentially modelled non-parametrically in WSLAP+ but parameterized as a PIEMD in LENSTOOL (or similar analytical profiles in other parametric approaches), so the radial mass density profile drops away and levels off more quickly in the previous case than in the latter, resulting in generally lower magnifications predicted by the WSLAP+ models than the CATS models for background galaxies lying at large clustercentric radii. A direct consequence of the higher magnifications predicted by the CATS models is the fainter UV luminosities derived for the galaxies than estimated using the WSLAP+ models, implying a larger population of faint galaxies that could possibly steepen the inferred faint-end slope of the UV LF.

\begin{figure*}[tp!]
\centering
\gridline{\fig{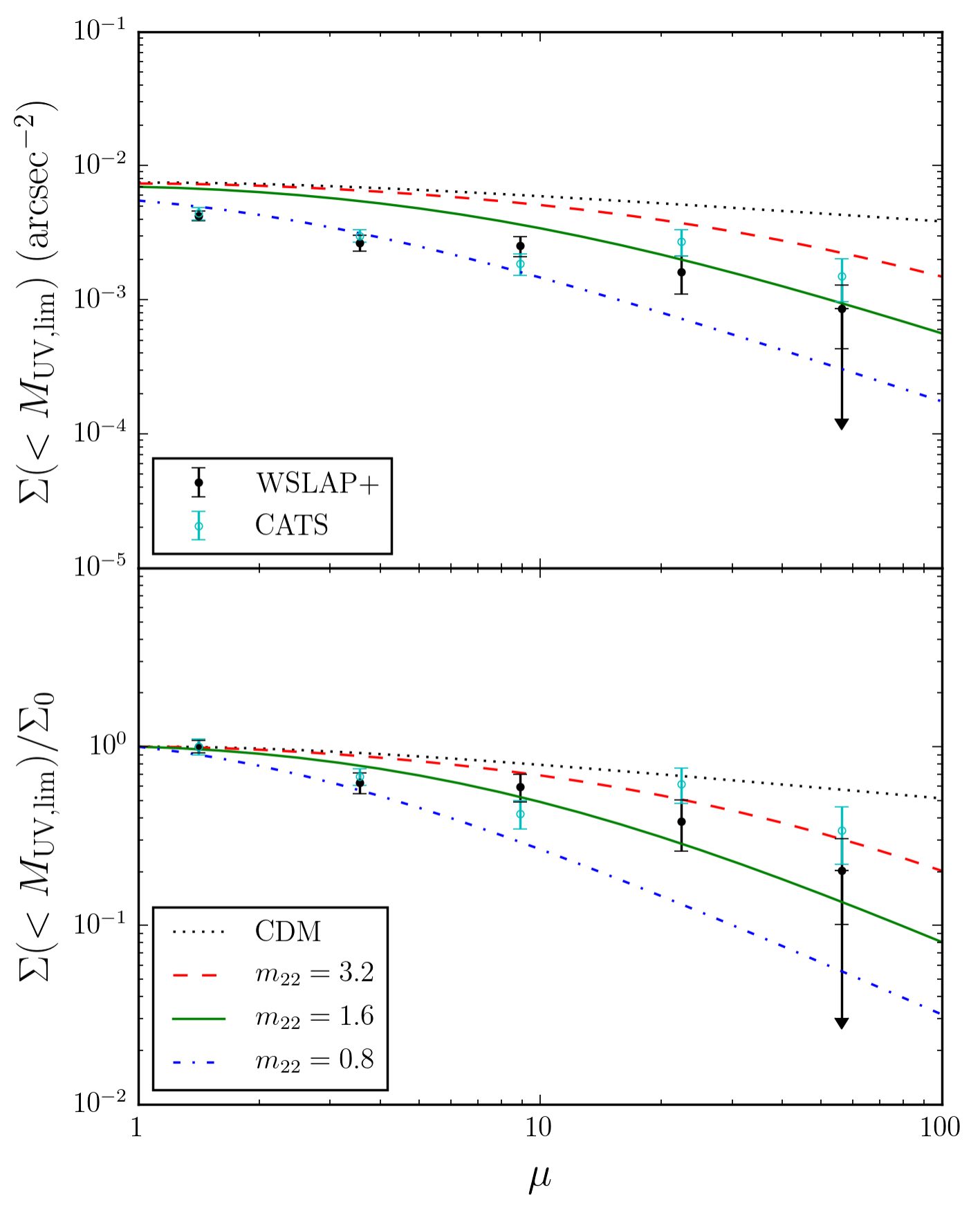}{85mm}{(a) Magnification bias.} \fig{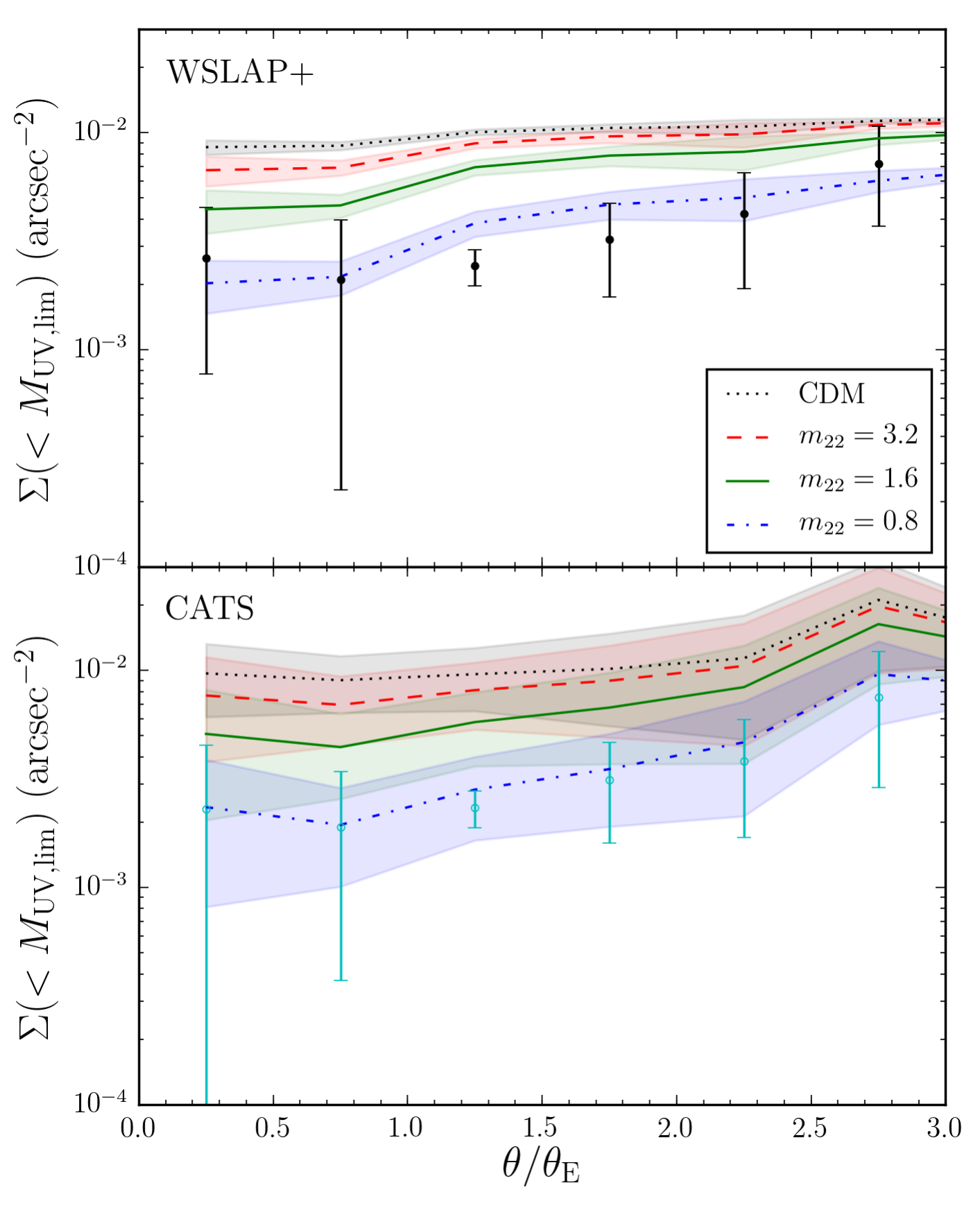}{85mm}{(b) Clustercentric radial density profile.}}
\gridline{\fig{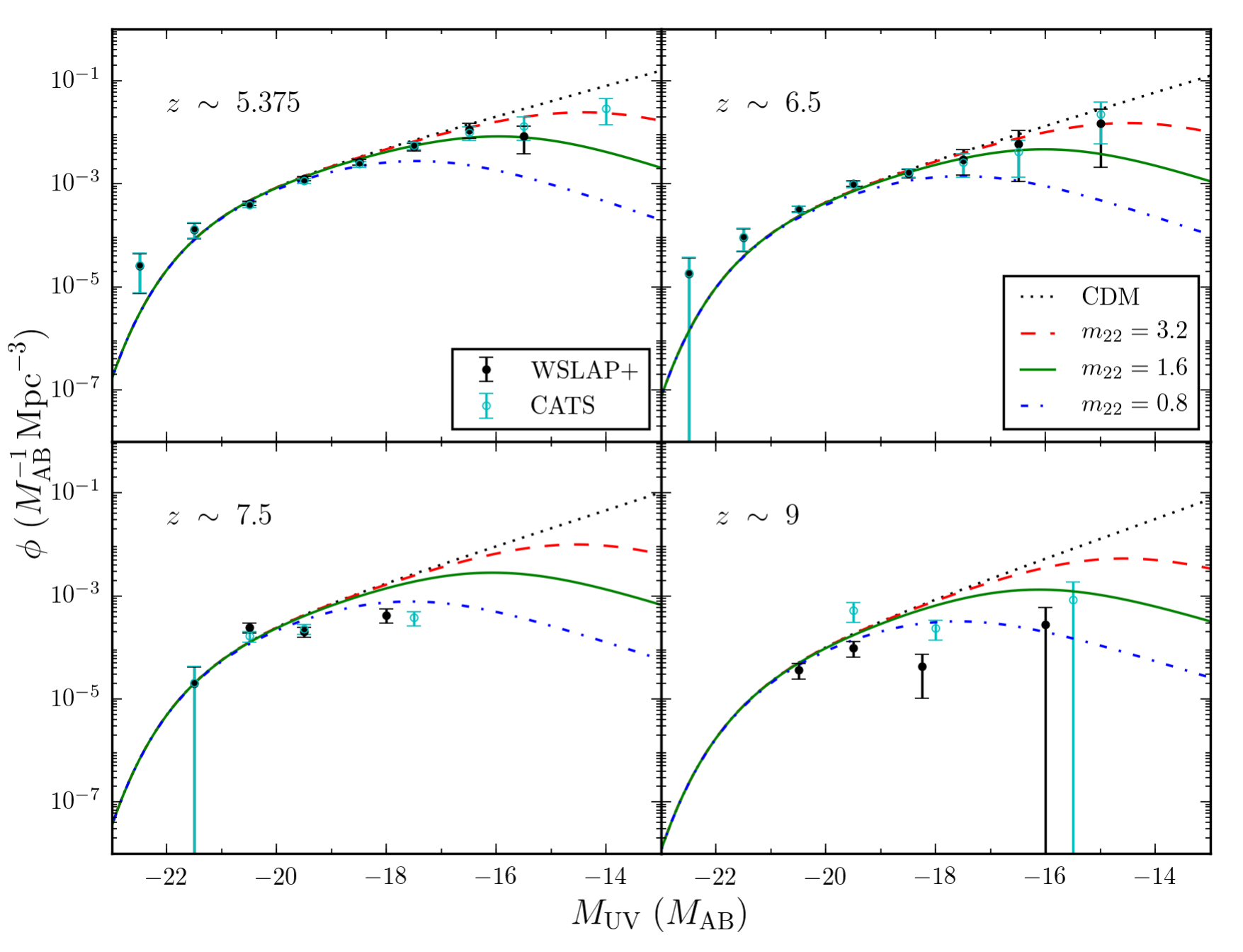}{127.5mm}{(c) SWML UV LF.}}
\caption{\label{results_CATS_comparison}Comparisons of the (a) magnification biases, (b) clustercentric radial density profiles, and (c) stepwise maximum likelihood UV LFs determined respectively using the galaxy magnifications predicted by the WSLAP+ models and the CATS/LENSTOOL models. The different coloured curves are predictions made from the usual reference set of CDM and $\psi$DM model LFs used throughout the paper.}
\end{figure*}

We thus consider explicitly whether the central conclusions reached in the previous sections using the WSLAP+ lens models continue to hold or are negated when the CATS lens models are used instead. As mentioned above, since the CATS models prefer a somewhat higher mean magnification of the galaxy candidates, we would expect the outcome of switching to these parametric lens models to be most noticeably reflected by a flattening of the slope of the derived magnification bias owing to the increase in the inferred number of highly magnified galaxies. In Figure \ref{results_CATS_comparison}(a), we show the results of the magnification bias tests performed in Section \ref{magnification_bias_section} but now using the CATS models, and for ease of comparison also the previous results obtained using the WSLAP+ models. Despite the enhanced galaxy number density at high magnifications resulting in a shallower slope for the magnification bias using the CATS models compared with that using the WSLAP+ models, the data points still deviate from the prediction assuming a standard CDM Schechter LF at a significance level of $\simeq$$5.34\sigma$.\footnote{We doubled the magnification bin widths in this reconstruction of the magnification bias compared with that presented in Section \ref{magnification_bias_section} so as to smooth out the relatively noisy fluctuations in the high-$\mu$ end when using the galaxy magnifications predicted by the CATS models. For reference, if we adopt the same doubled bin widths, the statistical significance of the deviation of the magnification bias from the CDM prediction using the WSLAP+ magnifications is at a $\simeq$$5.02\sigma$ level.} Instead, as before, the data can be better described by a UV LF exhibiting a faint-end turnover with a characteristic $\psi$DM boson mass of $\sim$$10^{-22} \, \mathrm{eV}$. Our conclusion of the existence of a faint-end turnover in the UV LF therefore appears to be robust against the particular choice of lens models for the HFF clusters.

Analogously to Section \ref{radial_test_section}, a different perspective of the magnification bias is presented in Figure \ref{results_CATS_comparison}(b), which shows how the projected galaxy number density changes as a function of clustercentric radial distance (scaled with the Einstein radius) for the WSLAP+ and CATS models respectively. The clustercentric radial density profiles compiled respectively from the WSLAP+-selected and CATS-selected galaxy candidates are plotted, along with the predictions from various CDM and $\psi$DM model LFs, where we determined the expected (weighted) averages and intercluster cosmic variances using the corresponding lens models. Apart from the larger prediction uncertainties using the CATS models, the data are clearly in better agreement with the $\psi$DM prediction for a boson mass of $m_\mathrm{B} = 0.8 \times 10^{-22} \, \mathrm{eV}$, just as what we found using the WSLAP+ models.

Finally, as a last consistency check, we compare in Figure \ref{results_CATS_comparison}(c) the stepwise maximum likelihood UV LFs reconstructed from the HFF galaxies with their UV luminosities inferred respectively using the magnification estimates of the WSLAP+ and CATS models (also tabulated in Table \ref{CATS_UV_LF_SWML_table}). Except for the redshift interval centered at $z \sim 5.375$ , in which the higher magnifications predicted by the CATS models lead to a handful of galaxies having substantially fainter UV absolute magnitudes compared with those using the WSLAP+ models, there is a reassuring agreement between both reconstructions in all the redshift intervals. Nonetheless, the subtle yet recognizable modification in the deduced luminosity distribution of the galaxies, brought about by the particular choice of lens models used, illuminates the most important message of our work: given the depth currently reached by even the deepest images of the most powerfully lensed fields, it is perilous to use the standard LF determination of lensed galaxies (where a correction for their different magnifications is of course necessary) to differentiate between the predictions of different DM models for the UV LF. Our message echoes that of \citet{2017ApJ...843..129B}, who illustrated the impact of the uncertainty between the magnifications predicted by different lens models on the inferred faint-end slope of the UV LF. In their work, \citet{2017ApJ...843..129B} used a suite of publicly available lens models, based on both parametric and non-parametric approaches, to study the same problem in great detail under an independent formalism. The difficulty in accurately determining the faint-end LF slope using lensed galaxies lends emphasis to our approach utlizing the magnification bias, which as we have demonstrated provides a more reliable probe of the faint-end slope of the UV LF.


\section{Conclusion}\label{conclusion}

Armed with the magnification boost provided by the most powerful gravitational lenses in the universe, we gained unprecedented depth to probe the faint end of the UV luminosity function (LF) at $z \sim 5$\,--\,10. Given the inherent difficulties in accurately conducting photometric measurements of comparatively faint high-$z$ galaxies amongst crowded bright cluster members, as well as thoroughly identifying existing multiply lensed images of source galaxies, however, it is technically challenging in practice to precisely determine the faint-end slope of the UV LF, which is crucial for discriminating between galaxy-formation predictions of the Wave Dark Matter ($\psi$DM) and the standard Cold Dark Matter (CDM) models.

To overcome the aforementioned problems, we devised a test to constrain the shape of  the faint-end LF utilizing the phenomenon of magnification bias in strongly lensed fields (e.g.~\citealt{1995ApJ...438...49B}), where we analyzed the dependence of the cumulative UV LF on the image-plane magnification factor. A great strength of this test is that it requires no information about the UV luminosity (or apparent magnitude) distribution of the galaxy sample, making it essentially immune to photometric uncertainties.  More importantly, the magnification bias is locally determined by the magnification factor in the image plane, in contrast to conventional LF reconstructions that is performed globally in the source plane, implying that we do not have to correct for the overcounting of multiply lensed regions (not only images) in the former case as required in the latter. Furthermore, we demonstrated in Figure \ref{results_CATS_comparison}(a), with the use of magnification estimates from completely independent parametric and semi-parametric lens models respectively, that the magnification bias test is not highly sensitive to the magnification uncertainties of individual galaxies, provided that our major goal is to distinguish whether or not there exists a faint-end turnover in the UV LF. Utilizing data from the six cluster fields taken in the {\it Hubble} Frontier Fields (HFF) program, our results show that (see Figure \ref{magnification_bias}):
\begin{itemize}
	\item{The observed magnification bias deviates from the model prediction using a standard CDM Schechter LF at $>$$6\sigma$ significance level (slope-only; $>$$20\sigma$ if considered together with the normalization).}
	\item{The observed magnification bias favours a turnover in the faint-end UV LF that, in the context of the $\psi$DM model, corresponds to a DM boson mass of $0.8 \times 10^{-22} \, \mathrm{eV} \lesssim m_\mathrm{B} \lesssim 3.2 \times 10^{-22} \, \mathrm{eV}$.}
\end{itemize}
The strongly negative magnification bias arises because the number of galaxies from the faint-end LF magnified above the detection threshold by lensing fails to compensate for the reduction in source-plane area probed. The same effect is also reflected by the enhanced deficit in the projected galaxy number density with decreasing clustercentric radius (see Figure \ref{multi_clt_n(theta)}), which can be understood from the fact that the cluster core around the Einstein radius is in general more highly magnified than in the outskirts.

The clear implication of a faint-end rollover in the UV LF from the negative magnification bias was checked for self-consistency with a direct determination of the UV LF. After carefully correcting for all the multiply lensed images identified in the cluster fields, we adopted the stepwise maximum likelihood method \citep{1988MNRAS.232..431E} to reconstruct the UV LF. This approach is only weakly dependent on the field-to-field cosmic variance in the normalization of the UV LF \citep{2015ApJ...803...34B}, and avoids the highly convoluted volume estimates in strongly lensed fields owing to multiple distorted appearances of certain source-plane regions in the image plane. We found that (see Figure \ref{UV_LF_multi_works}):
\begin{itemize}
	\item{The reconstructed UV LF agrees well with other previous blank-field studies to have a Schechter-like bright end at $z \sim 5$\,--\,8 as predicted by the standard CDM model or the $\psi$DM model.}
	\item{The reconstructed UV LF disfavours, but does not completely rule out, a steep power-law buildup at the faint end as motivated by the lack of galaxy detections at $M_\mathrm{UV} \gtrsim -15$ for $z \sim 5$\,--\,7, and at $M_\mathrm{UV} \gtrsim -17$ for $z \sim 7$\,--\,10.}
	\item{The reconstructed UV LF features a deficit of $M_\mathrm{UV} \lesssim -21$ galaxies with respect to the model predictions at $z \sim 8$\,--\,10, the absence of which is consistent with other blank-field studies.}
\end{itemize}

The non-detection of $\gtrsim\!L_{\star,z=3}$ galaxies at $z \gtrsim 8$ was further examined by studying the evolution of the cosmic SFR density (or UV luminosity density), which imposes a heavier weighting on the bright-end UV LF, across the redshift range $5 < z < 10$. Through successively lowering the integration limit for computing the UV luminosity density, we deduced that (see Figure \ref{multi_fld_SFR_density}):
\begin{itemize}
	\item{The cosmic SFR density increased with time at a much more rapid rate than predicted even by the $\psi$DM model LF during $z \gtrsim 8$.}
	\item{The cosmic SFR density kept on increasing at later epochs, but at a slower rate that is in agreement with the (log-)linear evolution predicted by the $\psi$DM model LF for a DM boson mass of $m_\mathrm{B} \simeq 0.8 \times 10^{-22} \, \mathrm{eV}$; the transition redshift of $z \simeq 8$ is consistent with the redshift of instantaneous reionization at $8.2_{-1.2}^{+1.0}$ as recently preferred by \citet{2016A&A...596A.108P}.}
\end{itemize}
The significant deficit of galaxies, especially the luminous ones, at $z \gtrsim 8$ can possibly be reconciled with the predictions by considering previously neglected factors while modelling the UV LF, such as multiple phases of star formation during halo assembly \citep{2015ApJ...813...21M} or other yet poorly understood baryonic physics including, but not limited to, supernova or radiative feedback mechanisms. In addition, the fewer than expected galaxies at this early epoch may pose a challenge to fulfill alone the total UV luminosity budget required to trigger cosmic reionization.

Moreover, in the context of $\psi$DM, S16 inferred the expected UV LF from the simulated $\psi$DM halo MF based on the conditional LF formalism \citep{2005ApJ...627L..89C}, resting on specific assumptions concerning the (halo) mass -- (galaxy) luminosity relation ($M_\mathrm{h}-L_\mathrm{UV}$ relation) and its redshift dependence. While this approach had been shown to successfully reproduce model LFs from simulated CDM halo MFs that consistently match with the observed high-$z$ (bright-end) UV LF (e.g.~\citealt{2015ApJ...803...34B}), the same assumed $M_\mathrm{h}-L_\mathrm{UV}$ relation presumably does not apply equally well to the $\psi$DM model, particularly at very high redshifts ($z \gtrsim 8$) where DM halo formation differs more drastically between the standard CDM and the $\psi$DM scenarios than at relatively low redshifts. It is hereby premature to precisely constrain the $\psi$DM particle mass from the HFF results solely using the $\psi$DM model LF given explicitly by Equation (\ref{psiDM_LF}), before future studies are done to obtain a more appropriate $M_\mathrm{h}-L_\mathrm{UV}$ relation in the $\psi$DM model, which may also shed some light on the non-linear evolution of the cosmic SFR density.

In summary, what we can firmly conclude from the HFF data is the likely existence of a faint-end turnover in the $z \sim 5$\,--\,10 UV LF, inferred unambiguously from the strongly negative magnification bias. We emphasize that such a conclusion is not altered by the drop in the normalization of the UV LF at $z \gtrsim 8$, owing to the fact that the projected galaxy number density we measured in the magnification bias test is dominated by galaxies located at $z \sim 4.75$\,--\,7 as evident in Figure \ref{multi_clt_z_binned_magnification_distribution}.

We now look ahead to upcoming observations (either of blank or lensed fields) using the {\it James Webb Space Telescope} ({\it JWST}\,), which will be equipped with an enhanced sensitivity to reach a limiting magnitude as faint as $m_\mathrm{lim} \sim 31$\,--\,32 \citep{2006SSRv..123..485G,2015ApJ...813...54T}, and thus giving us the opportunity to precisely constrain the shape of the faint-end turnover in the UV LF. Through a simple extrapolation of the observed rapidly declining trend in galaxy number density at $z \gtrsim 8$, we anticipate the detection of near to nothing -- perhaps just a handful of galaxies -- at $z > 10$ in deep NIR imaging. Consequently, an observational verification of this prediction will provide a strong support for alternatives to the standard CDM model, such as the $\psi$DM model in which halo and galaxy formation are suppressed and delayed. Otherwise, the $\psi$DM particle mass of $\sim$$10^{-22} \, \mathrm{eV}$ will need to be revised dramatically upwards so as to contradict with that inferred from the kpc-scale cores of local dSph galaxies \citep{2014NatPh..10..496S}.


\acknowledgements

We would like to thank Dan Coe for providing most of the HFF photometric catalogs that we used in our work. TB thanks the Visiting Research Professors Scheme at the University of Hong Kong for generous support. JL acknowledges a seed fund for basic research from the University of Hong Kong to initiate this work, and support from the Research Grants Council of Hong Kong through grant 17319316 for the conduct and completion of this work. JMD acknowledges the support of project AYA2015-64508-P (MINECO/FEDER, UE). Part of this work (the A370 catalogs) is based on data and catalog products from HFF-DeepSpace, funded by the National Science Foundation and Space Telescope Science Institute (operated by the Association of Universities for Research in Astronomy, Inc., under NASA contract NAS5-26555). This research made use of SAOImage DS9, an astronomical imaging and data visualization application \citep{2000ascl.soft03002S}, and Astropy, a community-developed core Python package for Astronomy \citep{2013A&A...558A..33A}. Other heavily dependent Python packages involved in this project for scientific computing and data visualization purposes include NumPy \citep{5725236}, SciPy \citep{scipy}, and Matplotlib \citep{4160265}.


\appendix
\renewcommand*{\thetable}{\Alph{section}\arabic{table}}


\onecolumngrid
\section{Information of selected galaxy candidates}\label{galaxy_info}
\setcounter{table}{0}

We provide below a sample table of the detailed information of representative galaxy candidates in the twelve HFF target fields analyzed in our work (see Table \ref{gal_info_sample_table}). The full data set is available in the electronic edition of this article as a machine-readable table. The description of each column (from left to right) of the table is laid out as follows. First column: target clusters. Second column: field types (cluster/parallel). Third column: galaxy IDs (arranged in ascending order of photometric redshifts for the respective target fields). Fourth and fifth columns: right ascension and declination (J2000.0) coordinates. Sixth column: Bayesian photometric redshifts with 95\% confidence intervals. Seventh column: apparent magnitudes with associated uncertainties given by $1\sigma$ upper limits in measured flux. Eighth column: HFF filters corresponding to the measured apparent magnitudes shown in the entries to the left. Ninth column: magnification factors. For the cluster fields, the magnification values were computed from the free-form lens models described in Section \ref{lensing_effects} using Equation (\ref{magnification_formula}), assuming the quoted $z_\mathrm{photo}$ estimates. The magnification uncertainties were assumed to be 10\% if $\mu_\mathrm{lens} < 10$ as are usually obtained in the low-$\mu$ regime. In contrast, when the galaxies are fairly close to the model critical curves, i.e.~$\mu_\mathrm{lens} \gtrsim 10$, the typical error in the model deflection field is $\delta\alpha \sim 1\arcsec$, and we approximated the magnification factor to fall off as $\mu_\mathrm{lens} \sim \mu_0 (1\arcsec/\theta)$ near a critical curve where $\mu_0 \sim 100$, thus we could roughly estimate the magnification uncertainties to be $\delta\mu_\mathrm{lens} \sim \mu_0(1\arcsec/\theta)^2 \sim 0.01\mu_\mathrm{lens}^2$. On the other hand, we assumed a fiducial 5\% magnification throughout all the parallel fields (i.e.~$\mu_\mathrm{fid} = 1.05$), considering that they would not be completely free of any lensing effect from their cluster counterparts. Tenth column: UV (1500\,\AA) absolute magnitudes determined using Equation (\ref{M_UV_formula}), with uncertainties estimated by taking into account the uncertainties in the photo-$z$'s (sixth column) (both explicitly through Equation (\ref{M_UV_formula}) and implicitly through the magnification factor $\mu(\boldsymbol{\theta},z)$, the UV continuum slope $\beta(z,M_\mathrm{UV})$, and the luminosity distance $D_\mathrm{L}(z)$), the uncertainties in the apparent magnitudes (seventh column), and the systematic uncertainties in the magnification factors (ninth column). Note that for more highly magnified (and hence intrinsically fainter) galaxies, the errors in their estimated UV absolute magnitudes are increasingly dominated by the magnification uncertainties, and so they are also more likely to be underestimated than overestimated.

\startlongtable
\begin{deluxetable*}{c c c c c c c c c c}
\tablecaption{Sample table of the detailed information of individual galaxy candidates\label{gal_info_sample_table}}
\tablehead{
\colhead{Target} & \colhead{Field} & \colhead{ID} & \colhead{RA} & \colhead{Dec} & \colhead{\phm{a}$z_\mathrm{photo}$\tablenotemark{a}} & \colhead{$m_\mathrm{filter}$} & \colhead{Filter} & \colhead{$\mu_\mathrm{lens/fid} $} & \colhead{$M_\mathrm{UV}$} \\
\colhead{cluster} & \colhead{type} & \colhead{} & \colhead{(J2000.0)} & \colhead{(J2000.0)} & \colhead{(95\% C.I.)} & \colhead{($m_\mathrm{AB}$)} & \colhead{} & \colhead{} & \colhead{($M_\mathrm{AB}$)}
}
\startdata
A2744 & cluster & 31 & 3.5907599 & -30.3955780 & $5.610_{-0.650}^{+0.650}$ & $24.71\pm0.07$ & F105W & $25.48\pm6.49$ & $-18.37_{-0.26}^{+0.34}$ \\
A2744 & parallel & 23 & 3.4655130 & -30.3902908 & $5.488_{-0.317}^{+0.159}$ & $26.77\pm0.04$ & F105W & 1.05 & $-19.72_{-0.08}^{+0.09}$ \\
MACS0416 & cluster & 37 & 64.0478253 & -24.0827722 & $6.402_{-0.250}^{+0.270}$ & $27.55\pm0.06$ & F105W & $1.24\pm0.12$ & $-19.01_{-0.13}^{+0.15}$ \\
MACS0416 & parallel & 71 & 64.1213651 & -24.1258530 & $7.624_{-0.473}^{+0.174}$ & $27.24\pm0.06$ & F125W & 1.05 & $-19.78_{-0.08}^{+0.09}$ \\
MACS0717 & cluster & 29 & 109.4087796 & 37.7483846 & $5.926_{-0.169}^{+0.137}$ & $27.39\pm0.06$ & F105W & $11.54\pm1.33$ & $-16.63_{-0.14}^{+0.16}$ \\
MACS0717 & parallel & 91 & 109.3437987 & 37.8292365 & $8.180_{-0.430}^{+0.253}$ & $27.89\pm0.08$ & F140W & 1.05 & $-19.23_{-0.10}^{+0.11}$ \\
MACS1149 & cluster & 1 & 177.4180016 & 22.4135989 & $4.755_{-0.223}^{+0.271}$ & $28.37\pm0.06$ & F814W & $1.14\pm0.11$ & $-17.79_{-0.14}^{+0.16}$ \\
MACS1149 & parallel & 79 & 177.4097104 & 22.2886229 & $6.225_{-0.323}^{+0.321}$ & $27.20\pm0.05$ & F105W & 1.05 & $-19.50_{-0.09}^{+0.10}$ \\
AS1063 & cluster & 19 & 342.1592837 & -44.5439621 & $6.281_{-0.176}^{+0.187}$ & $27.50\pm0.07$ & F105W & $4.42\pm0.44$ & $-17.63_{-0.12}^{+0.14}$ \\
AS1063 & parallel & 43 & 342.2999926 & -44.5594339 & $6.215_{-0.388}^{+0.369}$ & $28.28\pm0.07$ & F105W & 1.05 & $-18.40_{-0.11}^{+0.12}$ \\
A370 & cluster & 11 & 39.9752050 & -1.5688089 & $5.119_{-0.306}^{+0.306}$ & $25.86\pm0.06$ & F105W & $7.87\pm0.79$ & $-18.35_{-0.16}^{+0.19}$ \\
A370 & parallel & 41 & 40.0514046 & -1.6192631 & $6.485_{-0.374}^{+0.374}$ & $26.33\pm0.46$ & F105W & 1.05 & $-20.45_{-0.39}^{+0.62}$ \\
\enddata
\tablecomments{The relevant information of twelve selected galaxy candidates is shown here as an illustrative sample of the full data set, which can be accessed in machine-readable format in the electronic edition of this article.}
\tablenotetext{a}{For the A370 cluster and parallel fields, the uncertainties in photometric redshifts are assumed to be $0.05 (1 + z_\mathrm{photo})$, i.e.~5\% errors.}
\end{deluxetable*}


\onecolumngrid
\section{Information of multiply lensed galaxies}\label{multiple_image_info}
\setcounter{table}{0}

We tabulate in the following the information of the multiply lensed images that we identified from our pool of selected galaxy candidates presented in Table \ref{gal_info_sample_table} (with its full version available in the electronic edition of this article) for each cluster field. At the same time, we also record the parameter choices that we chose to represent the multiply lensed source galaxies when reconstructing the UV LF. A useful rule of thumb for us to decide which parameters should be used is to average the photometric redshift estimates among members belonging to a given multiple image system (since lensing magnification does not change the observed colour of a galaxy and hence should not affect photo-$z$ estimation), and to assign the source galaxy with the inferred UV absolute magnitude of the least magnified member (because more highly magnified galaxies are prone to higher magnification and thus UV absolute magnitude uncertainties). These criteria were applied to the majority of the identified multiple image systems, with a few exceptional cases where we specify under the respective tables with explanation (e.g.~we adopted the spectroscopic redshift instead of the photometric redshift of a given multiply lensed galaxy whenever available, so as to arrive at the most accurate UV luminosity estimate possible). Note that although the odd number theorem in strong gravitational lensing (e.g.~\citealt{2010arXiv1010.3416P}) prescribes that multiply lensed images normally form in odd numbers, it is often not the case as can be seen in the identifications listed below, caused by two major underlying reasons. First, many of the ``missing'' counterimages are located inside the exclusion regions that we defined earlier (several are obscured by the diffuse light from foreground cluster members, for example), within which we omitted the inclusion of all galaxy candidates to ensure data completeness in our analysis. Second, a few of these images, if not buried inside the exclusion regions, are likely to be too faint to be detected owing to their relatively low magnifications, unlike their highly magnified counterparts that are sufficiently bright to be identified.

\startlongtable
\begin{deluxetable*}{c c c c c c c c c}
\tablecaption{Detailed information of multiply lensed galaxy candidates in the A2744 cluster field\label{a2744_multi_im_table}}
\tablehead{
\colhead{System} & \colhead{Galaxy ID} & \colhead{RA} & \colhead{Dec} & \colhead{$z_\mathrm{photo}$} & \colhead{$m_\mathrm{filter}$} & \colhead{Filter} & \colhead{$\mu_\mathrm{lens} $} & \colhead{$M_\mathrm{UV}$} \\
\colhead{} & \colhead{} & \colhead{(J2000.0)} & \colhead{(J2000.0)} & \colhead{($\pm1\sigma$)} & \colhead{($m_\mathrm{AB}$)} & \colhead{} & \colhead{} & \colhead{($M_\mathrm{AB}$)}
}
\startdata
1 & 31 & 3.5907599 & -30.3955780 & $5.610\pm0.650$ & $24.71\pm0.07$ & F105W & $25.48\pm6.49$ & $-18.37_{-0.26}^{+0.34}$ \\
1 & 32 & 3.5761424 & -30.4044930 & $5.610\pm0.650$ & $26.57\pm0.17$ & F105W & $2.14\pm0.21$ & $-19.19_{-0.24}^{+0.31}$ \\
2 & 56 & 3.5978104 & -30.3959820 & $6.650\pm0.750$ & $28.45\pm0.10$ & F105W & $1.85\pm0.19$ & $-17.69_{-0.18}^{+0.22}$ \\
2 & 57 & 3.5804218 & -30.4050720 & $6.650\pm0.750$ & $28.22\pm0.10$ & F105W & $2.28\pm0.23$ & $-17.70_{-0.18}^{+0.21}$ \\
3 & 66 & 3.5925126 & -30.4014840 & $9.830\pm0.330$ & $27.93\pm0.36$ & F160W & $13.42\pm1.80$ & $-16.63_{-0.33}^{+0.48}$ \\
3 & 67 & 3.5950299 & -30.4007520 & $9.830\pm0.330$ & $28.31\pm0.44$ & F160W & $19.10\pm3.65$ & $-15.85_{-0.40}^{+0.65}$ \\
\enddata
\tablecomments{The information regarding these three multiply lensed galaxies was taken from \citet{2014ApJ...797...98L}, where they had already assumed and assigned the same photo-$z$ (and symmetric errors) for members belonging to a given multiple image system.}
\end{deluxetable*}

\startlongtable
\begin{deluxetable*}{c c c c}
\tablecaption{Parameter choices for the multiply lensed source galaxies in the A2744 cluster field}
\tablehead{
\colhead{System} & \colhead{$z$} & \colhead{$\mu$} & \colhead{$M_\mathrm{UV}$} \\
\colhead{} & \colhead{($\pm1\sigma$)} & \colhead{} & \colhead{($M_\mathrm{AB}$)}
}
\startdata
1 & $5.610\pm0.650$ & $2.14\pm0.21$ & $-19.19_{-0.24}^{+0.31}$ \\
\phm{$^\mathrm{a}$}2\tablenotemark{a} & $6.650\pm0.750$ & $2.28\pm0.23$ & $-17.70_{-0.18}^{+0.21}$ \\
\phm{$^\mathrm{b}$}3\tablenotemark{b} & $9.830\pm0.330$ & $19.10\pm3.65$ & $-15.85_{-0.40}^{+0.65}$ \\
\enddata
\tablenotetext{a}{The estimated parameters for galaxy 57 were used to represent this source galaxy, despite being more highly magnified, due to the cleaner sky background than that of its counterimage (galaxy 56) which is located close to a bright cluster member.}
\tablenotetext{b}{We represented this source galaxy with the estimated parameters of the more highly magnified galaxy 67, since its counterimage (galaxy 66) unfortunately coincides with a red stellar spike that inevitably adds noise to the measured apparent magnitude.}
\end{deluxetable*}

\startlongtable
\begin{deluxetable*}{c c c c c c c c c}
\tablecaption{Detailed information of multiply lensed galaxy candidates in the MACS0416 cluster field\label{macs0416_multi_im_table}}
\tablehead{
\colhead{System} & \colhead{Galaxy ID} & \colhead{RA} & \colhead{Dec} & \colhead{$z_\mathrm{photo}$} & \colhead{$m_\mathrm{filter}$} & \colhead{Filter} & \colhead{$\mu_\mathrm{lens} $} & \colhead{$M_\mathrm{UV}$} \\
\colhead{} & \colhead{} & \colhead{(J2000.0)} & \colhead{(J2000.0)} & \colhead{(95\% C.I.)} & \colhead{($m_\mathrm{AB}$)} & \colhead{} & \colhead{} & \colhead{($M_\mathrm{AB}$)}
}
\startdata
1 & 14 & 64.0350788 & -24.0855172 & $5.449_{-0.317}^{+0.123}$ & $27.13\pm0.04$ & F105W & $2.43\pm0.24$ & $-18.45_{-0.12}^{+0.14}$ \\
1 & 15 & 64.0229895 & -24.0772689 & $5.449_{-0.307}^{+0.107}$ & $25.94\pm0.03$ & F105W & $3.74\pm0.37$ & $-19.17_{-0.12}^{+0.13}$ \\
2 & 10 & 64.0232972 & -24.0750555 & $5.282_{-0.746}^{+0.232}$ & $28.38\pm0.10$ & F105W & $2.47\pm0.25$ & $-17.16_{-0.17}^{+0.20}$ \\
2 & 31 & 64.0333114 & -24.0842855 & $5.950_{-4.842}^{+0.292}$ & $27.77\pm0.09$ & F105W & $11.29\pm1.27$ & $-16.28_{-0.22}^{+0.28}$ \\
3 & 26 & 64.0471382 & -24.0611402 & $5.755_{-0.301}^{+0.106}$ & $27.28\pm0.04$ & F105W & $11.94\pm1.43$ & $-16.67_{-0.15}^{+0.17}$ \\
3 & 33 & 64.0510729 & -24.0665094 & $5.971_{-4.870}^{+0.265}$ & $28.25\pm0.10$ & F105W & $2.59\pm0.26$ & $-17.39_{-0.40}^{+0.63}$ \\
3 & 35 & 64.0492248 & -24.0633480 & $6.070_{-0.168}^{+0.172}$ & $27.99\pm0.06$ & F105W & $49.46\pm24.47$ & $-14.47_{-0.45}^{+0.78}$ \\
4 & 8 & 64.0508199 & -24.0664977 & $5.232_{-4.729}^{+0.217}$ & $25.30\pm0.03$ & F105W & $2.72\pm0.27$ & $-20.07_{-0.58}^{+1.34}$ \\
4 & 30 & 64.0481750 & -24.0624053 & $5.821_{-0.184}^{+0.136}$ & $27.70\pm0.06$ & F105W & $27.70\pm7.67$ & $-15.35_{-0.28}^{+0.37}$ \\
4 & 36 & 64.0435601 & -24.0589974 & $6.268_{-0.188}^{+0.186}$ & $27.26\pm0.05$ & F105W & $3.03\pm0.30$ & $-18.29_{-0.12}^{+0.13}$ \\
\enddata
\end{deluxetable*}

\startlongtable
\begin{deluxetable*}{c c c c}
\tablecaption{Parameter choices for the multiply lensed source galaxies in the MACS0416 cluster field}
\tablehead{
\colhead{System} & \colhead{$z$} & \colhead{$\mu$} & \colhead{$M_\mathrm{UV}$} \\
\colhead{} & \colhead{(95\% C.I.)} & \colhead{} & \colhead{($M_\mathrm{AB}$)}
}
\startdata
1 & \phm{$^\mathrm{a}$}5.365\tablenotemark{a} & $2.43\pm0.24$ & $-18.45_{-0.12}^{+0.14}$ \\
2 & $5.616_{-2.794}^{+0.262}$ & $2.47\pm0.25$ & $-17.16_{-0.17}^{+0.20}$ \\
3 & $5.932_{-1.780}^{+0.181}$ & $2.59\pm0.26$ & $-17.39_{-0.40}^{+0.63}$ \\
4 & \phm{$^\mathrm{a}$}6.145\tablenotemark{a} & $2.72\pm0.27$ & $-20.07_{-0.58}^{+1.34}$ \\
\enddata
\tablenotetext{a}{MUSE spectroscopic redshifts of the same multiple image systems reported by \citet{2017AandA...600A..90C}.}
\end{deluxetable*}

\startlongtable
\begin{deluxetable*}{c c c c c c c c c}
\tablecaption{Detailed information of multiply lensed galaxy candidates in the MACS0717 cluster field\label{macs0717_multi_im_table}}
\tablehead{
\colhead{System} & \colhead{Galaxy ID} & \colhead{RA} & \colhead{Dec} & \colhead{$z_\mathrm{photo}$} & \colhead{$m_\mathrm{filter}$} & \colhead{Filter} & \colhead{$\mu_\mathrm{lens} $} & \colhead{$M_\mathrm{UV}$} \\
\colhead{} & \colhead{} & \colhead{(J2000.0)} & \colhead{(J2000.0)} & \colhead{(95\% C.I.)} & \colhead{($m_\mathrm{AB}$)} & \colhead{} & \colhead{} & \colhead{($M_\mathrm{AB}$)}
}
\startdata
1 & 8 & 109.3723583 & 37.7399219 & $5.186_{-0.106}^{+0.113}$ & $26.66\pm0.04$ & F105W & $1.70\pm0.17$ & $-19.22_{-0.11}^{+0.13}$ \\
1 & 14 & 109.3914508 & 37.7670463 & $5.414_{-0.156}^{+0.094}$ & $27.36\pm0.16$ & F105W & $2.74\pm0.27$ & $-18.09_{-0.18}^{+0.21}$ \\
2 & 18 & 109.3804006 & 37.7494593 & $5.510_{-0.142}^{+0.135}$ & $27.42\pm0.04$ & F105W & $3.31\pm0.33$ & $-17.86_{-0.12}^{+0.13}$ \\
2 & 21 & 109.3755173 & 37.7412054 & $5.663_{-0.115}^{+0.116}$ & $27.28\pm0.05$ & F105W & $2.86\pm0.29$ & $-18.19_{-0.11}^{+0.13}$ \\
3 & 22 & 109.3769838 & 37.7364526 & $5.674_{-0.164}^{+0.263}$ & $26.82\pm0.05$ & F105W & $1.61\pm0.16$ & $-19.26_{-0.12}^{+0.14}$ \\
3 & 24 & 109.3862180 & 37.7519303 & $5.750_{-0.181}^{+0.172}$ & $26.34\pm0.03$ & F105W & $3.65\pm0.37$ & $-18.88_{-0.12}^{+0.13}$ \\
3 & 27 & 109.3988103 & 37.7650724 & $5.872_{-0.136}^{+0.128}$ & $27.56\pm0.06$ & F105W & $4.61\pm0.46$ & $-16.95_{-0.12}^{+0.13}$ \\
4 & 32 & 109.4136639 & 37.7346388 & $6.143_{-0.101}^{+0.096}$ & $26.68\pm0.03$ & F105W & $2.06\pm0.21$ & $-19.27_{-0.11}^{+0.12}$ \\
4 & 33 & 109.4128619 & 37.7338114 & $6.176_{-0.243}^{+0.108}$ & $26.68\pm0.05$ & F105W & $2.96\pm0.30$ & $-18.88_{-0.12}^{+0.13}$ \\
\enddata
\end{deluxetable*}

\startlongtable
\begin{deluxetable*}{c c c c}
\tablecaption{Parameter choices for the multiply lensed source galaxies in the MACS0717 cluster field}
\tablehead{
\colhead{System} & \colhead{$z$} & \colhead{$\mu$} & \colhead{$M_\mathrm{UV}$} \\
\colhead{} & \colhead{(95\% C.I.)} & \colhead{} & \colhead{($M_\mathrm{AB}$)}
}
\startdata
1 & $5.300_{-0.131}^{+0.104}$ & $1.70\pm0.17$ & $-19.22_{-0.11}^{+0.13}$ \\
2 & $5.587_{-0.129}^{+0.126}$ & $2.86\pm0.29$ & $-18.19_{-0.11}^{+0.13}$ \\
3 & $5.765_{-0.160}^{+0.188}$ & $1.61\pm0.16$ & $-19.26_{-0.12}^{+0.14}$ \\
4 & $6.160_{-0.172}^{+0.102}$ & $2.06\pm0.21$ & $-19.27_{-0.11}^{+0.12}$ \\
\enddata
\end{deluxetable*}

\startlongtable
\begin{deluxetable*}{c c c c c c c c c}
\tablecaption{Detailed information of multiply lensed galaxy candidates in the MACS1149 cluster field\label{macs1149_multi_im_table}}
\tablehead{
\colhead{System} & \colhead{Galaxy ID} & \colhead{RA} & \colhead{Dec} & \colhead{$z_\mathrm{photo}$} & \colhead{$m_\mathrm{filter}$} & \colhead{Filter} & \colhead{$\mu_\mathrm{lens} $} & \colhead{$M_\mathrm{UV}$} \\
\colhead{} & \colhead{} & \colhead{(J2000.0)} & \colhead{(J2000.0)} & \colhead{(95\% C.I.)} & \colhead{($m_\mathrm{AB}$)} & \colhead{} & \colhead{} & \colhead{($M_\mathrm{AB}$)}
}
\startdata
1 & 11 & 177.4180579 & 22.3975864 & $5.020_{-0.232}^{+0.135}$ & $27.35\pm0.05$ & F105W & $1.22\pm0.12$ & $-18.84_{-0.13}^{+0.14}$ \\
1 & 28 & 177.4188306 & 22.4004707 & $5.471_{-4.662}^{+0.252}$ & $26.79\pm0.06$ & F105W & $1.52\pm0.15$ & $-19.30_{-0.54}^{+1.12}$ \\
2 & 31 & 177.3869602 & 22.4013863 & $5.569_{-5.052}^{+0.308}$ & $27.56\pm0.08$ & F105W & $3.03\pm0.30$ & $-17.82_{-0.44}^{+0.74}$ \\
2 & 37 & 177.4016203 & 22.4102605 & $5.786_{-5.083}^{+0.329}$ & $28.36\pm0.10$ & F105W & $2.95\pm0.30$ & $-17.11_{-0.39}^{+0.62}$ \\
2 & 41 & 177.3877076 & 22.4058340 & $5.918_{-0.354}^{+0.233}$ & $28.32\pm0.08$ & F105W & $8.25\pm0.82$ & $-16.06_{-0.13}^{+0.15}$ \\
3 & 42 & 177.4191160 & 22.3987516 & $5.928_{-0.475}^{+0.305}$ & $28.20\pm0.10$ & F105W & $1.24\pm0.12$ & $-18.24_{-0.16}^{+0.19}$ \\
3 & 46 & 177.4198695 & 22.4009412 & $6.013_{-4.831}^{+0.361}$ & $28.21\pm0.12$ & F105W & $1.25\pm0.12$ & $-18.24_{-0.49}^{+0.93}$ \\
4 & 48 & 177.4066732 & 22.3843234 & $6.089_{-0.424}^{+0.386}$ & $28.04\pm0.08$ & F105W & $2.67\pm0.27$ & $-17.60_{-0.14}^{+0.17}$ \\
4 & 56 & 177.4120760 & 22.3890548 & $6.719_{-0.277}^{+0.293}$ & $27.80\pm0.06$ & F105W & $8.72\pm0.87$ & $-16.65_{-0.14}^{+0.16}$ \\
\enddata
\end{deluxetable*}

\startlongtable
\begin{deluxetable*}{c c c c}
\tablecaption{Parameter choices for the multiply lensed source galaxies in the MACS1149 cluster field}
\tablehead{
\colhead{System} & \colhead{$z$} & \colhead{$\mu$} & \colhead{$M_\mathrm{UV}$} \\
\colhead{} & \colhead{(95\% C.I.)} & \colhead{} & \colhead{($M_\mathrm{AB}$)}
}
\startdata
1 & $5.246_{-2.447}^{+0.194}$ & $1.22\pm0.12$ & $-18.84_{-0.13}^{+0.14}$ \\
2 & $5.758_{-3.496}^{+0.290}$ & $2.95\pm0.30$ & $-17.11_{-0.39}^{+0.62}$ \\
3 & $5.971_{-2.653}^{+0.333}$ & $1.24\pm0.12$ & $-18.24_{-0.16}^{+0.19}$ \\
4 & $6.404_{-0.351}^{+0.340}$ & $2.67\pm0.27$ & $-17.60_{-0.14}^{+0.17}$ \\
\enddata
\end{deluxetable*}

\startlongtable
\begin{deluxetable*}{c c c c c c c c c}
\tablecaption{Detailed information of multiply lensed galaxy candidates in the AS1063 cluster field\label{as1063_multi_im_table}}
\tablehead{
\colhead{System} & \colhead{Galaxy ID} & \colhead{RA} & \colhead{Dec} & \colhead{$z_\mathrm{photo}$} & \colhead{$m_\mathrm{filter}$} & \colhead{Filter} & \colhead{$\mu_\mathrm{lens} $} & \colhead{$M_\mathrm{UV}$} \\
\colhead{} & \colhead{} & \colhead{(J2000.0)} & \colhead{(J2000.0)} & \colhead{(95\% C.I.)} & \colhead{($m_\mathrm{AB}$)} & \colhead{} & \colhead{} & \colhead{($M_\mathrm{AB}$)}
}
\startdata
1 & 13 & 342.1761301 & -44.5426618 & $5.815_{-0.120}^{+0.104}$ & $26.92\pm0.04$ & F105W & $9.54\pm0.95$ & $-17.28_{-0.11}^{+0.12}$ \\
1 & 14 & 342.1643644 & -44.5302361 & $5.846_{-0.323}^{+0.253}$ & $27.43\pm0.05$ & F105W & $5.74\pm0.57$ & $-17.33_{-0.12}^{+0.14}$ \\
2 & 16 & 342.1712986 & -44.5198062 & $5.923_{-0.066}^{+0.053}$ & $25.64\pm0.02$ & F105W & $2.08\pm0.21$ & $-20.23_{-0.11}^{+0.12}$ \\
2 & 17 & 342.1908925 & -44.5374622 & $6.024_{-0.034}^{+0.035}$ & $24.78\pm0.01$ & F105W & $6.57\pm0.66$ & $-19.88_{-0.10}^{+0.12}$ \\
\enddata
\end{deluxetable*}

\startlongtable
\begin{deluxetable*}{c c c c}
\tablecaption{Parameter choices for the multiply lensed source galaxies in the AS1063 cluster field}
\tablehead{
\colhead{System} & \colhead{$z$} & \colhead{$\mu$} & \colhead{$M_\mathrm{UV}$} \\
\colhead{} & \colhead{(95\% C.I.)} & \colhead{} & \colhead{($M_\mathrm{AB}$)}
}
\startdata
1 & $5.831_{-0.222}^{+0.179}$ & $5.74\pm0.57$ & $-17.33_{-0.12}^{+0.14}$ \\
2 & \phm{$^\mathrm{\,a}$}$5.923_{-0.066}^{+0.053}$\tablenotemark{\,a} & $2.08\pm0.21$ & $-20.23_{-0.11}^{+0.12}$ \\
\enddata
\tablenotetext{a}{The photo-$z$ of galaxy 16 was adopted since it suffers from comparatively less severe contamination from intracluster light.}
\end{deluxetable*}

\startlongtable
\begin{deluxetable*}{c c c c c c c c c}
\tablecaption{Detailed information of multiply lensed galaxy candidates in the A370 cluster field\label{a370_multi_im_table}}
\tablehead{
\colhead{System} & \colhead{Galaxy ID} & \colhead{RA} & \colhead{Dec} & \colhead{\phm{a}$z_\mathrm{photo}$\tablenotemark{a}} & \colhead{$m_\mathrm{filter}$} & \colhead{Filter} & \colhead{$\mu_\mathrm{lens} $} & \colhead{$M_\mathrm{UV}$} \\
\colhead{} & \colhead{} & \colhead{(J2000.0)} & \colhead{(J2000.0)} & \colhead{} & \colhead{($m_\mathrm{AB}$)} & \colhead{} & \colhead{} & \colhead{($M_\mathrm{AB}$)}
}
\startdata
1 & 11 & 39.9752050 & -1.5688089 & $5.119\pm0.306$ & $25.86\pm0.06$ & F105W & $7.87\pm0.79$ & $-18.35_{-0.16}^{+0.19}$ \\
1 & 30 & 39.9795227 & -1.5717750 & $5.970\pm0.349$ & $25.66\pm0.04$ & F105W & $9.85\pm0.99$ & $-18.54_{-0.12}^{+0.14}$ \\
2 & 22 & 39.9652834 & -1.5878085 & $5.750\pm0.338$ & $27.71\pm0.27$ & F105W & $11.84\pm1.40$ & $-16.25_{-0.27}^{+0.37}$ \\
2 & 23 & 39.9636212 & -1.5868808 & $5.750\pm0.338$ & $26.82\pm0.15$ & F105W & $11.00\pm1.21$ & $-17.21_{-0.18}^{+0.22}$ \\
3 & 25 & 39.9692510 & -1.5664433 & $5.830\pm0.341$ & $26.92\pm0.14$ & F105W & $3.60\pm0.36$ & $-18.33_{-0.19}^{+0.23}$ \\
3 & 31 & 39.9858098 & -1.5713028 & $6.029\pm0.351$ & $25.96\pm0.11$ & F105W & $2.21\pm0.22$ & $-19.89_{-0.17}^{+0.19}$ \\
\enddata
\tablenotetext{a}{Photometric redshift errors are assumed to scale as $0.05 (1 + z_\mathrm{photo})$.}
\end{deluxetable*}

\startlongtable
\begin{deluxetable*}{c c c c}
\tablecaption{Parameter choices for the multiply lensed source galaxies in the A370 cluster field}
\tablehead{
\colhead{System} & \colhead{$z$} & \colhead{$\mu$} & \colhead{$M_\mathrm{UV}$} \\
\colhead{} & \colhead{} & \colhead{} & \colhead{($M_\mathrm{AB}$)}
}
\startdata
1 & $5.119\pm0.306$ & $7.87\pm0.79$ & $-18.35_{-0.16}^{+0.19}$ \\
2 & \phm{$^\mathrm{\,a}$}5.750\tablenotemark{\,a} & $11.84\pm1.40$ & $-16.25_{-0.27}^{+0.37}$ \\
3 & $6.029\pm0.351$ & $2.21\pm0.22$ & $-19.89_{-0.17}^{+0.19}$ \\
\enddata
\tablecomments{The photo-$z$'s of galaxies 11 and 31 were adopted for systems 1 and 3 respectively because they are located at less contaminated sky regions than their counterimages.}
\tablenotetext{a}{MUSE spectroscopic redshift of the same multiple image system reported by \citet{2017MNRAS.469.3946L}.}
\end{deluxetable*}


\onecolumngrid
\section{Supplementary tables of various UV luminosity function estimates}\label{UV_LF_supplementary_tables}
\setcounter{table}{0}

In this appendix, we tabulate the stepwise UV LFs determined respectively for the six HFF cluster fields, the six HFF parallel fields, and all the twelve HFF fields, using the ``$1/V_\mathrm{max}$'' method. We also tabulate the stepwise maximum likelihood UV LF reconstructed for all the target fields where we used the CATS lens models to compute the magnification estimates of individual galaxies.

\startlongtable
\begin{deluxetable*}{c c c c c c c c c c c c c c c}
\tablecaption{Stepwise UV LF in the six HFF cluster fields reconstructed using the ``$1/V_\mathrm{max}$'' method\label{clt_inv_V_max_LF}}
\tablehead{\colhead{$M_\mathrm{UV}$} & \colhead{} & \colhead{No.~of galaxies} & \colhead{} & \colhead{$\phi$} & \colhead{} & \colhead{} & \colhead{} & \colhead{} & \colhead{} & \colhead{$M_\mathrm{UV}$} & \colhead{} & \colhead{No.~of galaxies} & \colhead{} & \colhead{$\phi$} \\
\colhead{} & \colhead{} & \colhead{} & \colhead{} & \colhead{($M_\mathrm{AB}^{-1} \, \mathrm{Mpc}^{-3}$)} & \colhead{} & \colhead{} & \colhead{} & \colhead{} & \colhead{} & \colhead{} & \colhead{}  & \colhead{} & \colhead{}& \colhead{($M_\mathrm{AB}^{-1} \, \mathrm{Mpc}^{-3}$)}
}
\startdata
\multicolumn{5}{c}{$z \sim 5.375$} & & & & & & \multicolumn{5}{c}{$z \sim 6.5$} \\
\hline
-22.5 & & 0 & & $<\!0.000193$ & & & & & & -22.5 & & 0 & & $<\!0.000228$ \\
-21.5 & & 1 & & $\leq\!0.000193$ & & & & & & -21.5 & & 0 & & $<\!0.000228$ \\
-20.5 & & 6 & & $0.001156 \pm 0.000472$ & & & & & & -20.5 & & 2 & & $0.000456 \pm 0.000322$ \\
-19.5 & & 25 & & $0.004816 \pm 0.000963$ & & & & & & -19.5 & & 8 & & $0.001824 \pm 0.000645$ \\
-18.5 & & 68 & & $0.013100 \pm 0.001589$ & & & & & & -18.5 & & 12 & & $0.002736 \pm 0.000790$ \\
-17.5 & & 57 & & $0.023416 \pm 0.003572$ & & & & & & -17.5 & & 7 & & $0.004098 \pm 0.0001575$ \\
-16.5 & & 21 & & $0.035983 \pm 0.008534$ & & & & & & -16.5 & & 3 & & $0.010057 \pm 0.005910$ \\
-15.5 & & 5 & & $0.049303 \pm 0.022721$ & & & & & & -15.5 & & 1 & & $\leq\!0.020752$ \\
-14.5 & & 0 & & $<\!0.023796$ & & & & & & -14.5 & & 1 & & $\leq\!0.026675$ \\
-13.5 & & 0 & & $<\!0.052387$ & & & & & & -13.5 & & 0 & & $<\!0.067898$ \\
\hline
\multicolumn{5}{c}{$z \sim 7.5$} & & & & & & \multicolumn{5}{c}{$z \sim 9$} \\
\hline
-22.5 & & 0 & & $<0.000258$ & & & & & & -22.5 & & 0 & & $<\!0.000290$ \\
-21.5 & & 1 & & $\leq\!0.000258$ & & & & & & -21.5 & & 0 & & $<\!0.000290$ \\
-20.5 & & 4 & & $0.001033 \pm 0.000516$ & & & & & & -20.5 & & 3 & & $0.000870 \pm 0.000502$ \\
-19.5 & & 0 & & $<0.000258$ & & & & & & -19.5 & & 4 & & $0.001159 \pm 0.000580$ \\
-18.5 & & 3 & & $0.001065 \pm 0.000659$ & & & & & & -18.5 & & 1 & & $\leq\!0.000290$ \\
-17.5 & & 2 & & $0.002679 \pm 0.002024$ & & & & & & -17.5 & & 0 & & $<\!0.001677$ \\
-16.5 & & 0 & & $<\!0.004577$ & & & & & & -16.5 & & 0 & & $<\!0.006701$ \\
-15.5 & & 0 & & $<\!0.018130$ & & & & & & -15.5 & & 1 & & $\leq\!0.016371$ \\
-14.5 & & 0 & & $<\!0.048022$ & & & & & & -14.5 & & 0 & & $<\!0.060766$ \\
-13.5 & & 0 & & $<\!0.087502$ & & & & & & \nodata & & \nodata & & \nodata \\
\enddata
\end{deluxetable*}

\startlongtable
\begin{deluxetable*}{c c c c c c c c c c c c c c c}
\tablecaption{Stepwise UV LF in the six HFF parallel fields reconstructed using the ``$1/V_\mathrm{max}$'' method\label{par_inv_V_max_LF}}
\tablehead{\colhead{$M_\mathrm{UV}$} & \colhead{} & \colhead{No.~of galaxies} & \colhead{} & \colhead{$\phi$} & \colhead{} & \colhead{} & \colhead{} & \colhead{} & \colhead{} & \colhead{$M_\mathrm{UV}$} & \colhead{} & \colhead{No.~of galaxies} & \colhead{} & \colhead{$\phi$} \\
\colhead{} & \colhead{} & \colhead{} & \colhead{} & \colhead{($M_\mathrm{AB}^{-1} \, \mathrm{Mpc}^{-3}$)} & \colhead{} & \colhead{} & \colhead{} & \colhead{} & \colhead{} & \colhead{} & \colhead{}  & \colhead{} & \colhead{}& \colhead{($M_\mathrm{AB}^{-1} \, \mathrm{Mpc}^{-3}$)}
}
\startdata
\multicolumn{5}{c}{$z \sim 5.375$} & & & & & & \multicolumn{5}{c}{$z \sim 6.5$} \\
\hline
-22.5 & & 2 & & $0.000028 \pm 0.000020$ & & & & & & -22.5 & & 1 & & $\leq\!0.000020$ \\
-21.5 & & 9 & & $0.000125 \pm 0.000042$ & & & & & & -21.5 & & 5 & & $0.000099 \pm 0.000044$ \\
-20.5 & & 26 & & $0.000362 \pm 0.000071$ & & & & & & -20.5 & & 16 & & $0.000316 \pm 0.000079$ \\
-19.5 & & 73 & & $0.001017 \pm 0.000119$ & & & & & & -19.5 & & 48 & & $0.000947 \pm 0.000137$ \\
-18.5 & & 134 & & $0.001950 \pm 0.000170$ & & & & & & -18.5 & & 53 & & $0.001356 \pm 0.000208$ \\
-17.5 & & 20 & & $0.003730 \pm 0.001831$ & & & & & & \nodata & & \nodata & & \nodata \\
\hline
\multicolumn{5}{c}{$z \sim 7.5$} & & & & & & \multicolumn{5}{c}{$z \sim 9$} \\
\hline
-22.5 & & 0 & & $<\!0.000022$ & & & & & & -22.5 & & 0 & & $<\!0.000013$ \\
-21.5 & & 0 & & $<\!0.000022$ & & & & & & -21.5 & & 0 & & $<\!0.000013$ \\
-20.5 & & 8 & & $0.000176 \pm 0.000062$ & & & & & & -20.5 & & 0 & & $<\!0.000013$ \\
-19.5 & & 10 & & $0.000220 \pm 0.000069$ & & & & & & -19.5 & & 4 & & $0.000051 \pm 0.000025$ \\
-18.5 & & 8 & & $0.000242 \pm 0.000105$ & & & & & & -18.5 & & 1 & & $\leq\!0.000029$ \\
\enddata
\end{deluxetable*}

\startlongtable
\begin{deluxetable*}{c c c c c c c c c c c c c c c}
\tablecaption{Stepwise UV LF in the twelve HFF fields reconstructed using the ``$1/V_\mathrm{max}$'' method\label{UV_LF_inv_V_max_table}}
\tablehead{\colhead{$M_\mathrm{UV}$} & \colhead{} & \colhead{No.~of galaxies} & \colhead{} & \colhead{$\phi$} & \colhead{} & \colhead{} & \colhead{} & \colhead{} & \colhead{} & \colhead{$M_\mathrm{UV}$} & \colhead{} & \colhead{No.~of galaxies} & \colhead{} & \colhead{$\phi$} \\
\colhead{($M_\mathrm{AB}$)} & \colhead{} & \colhead{} & \colhead{} & \colhead{($M_\mathrm{AB}^{-1} \, \mathrm{Mpc}^{-3}$)} & \colhead{} & \colhead{} & \colhead{} & \colhead{} & \colhead{} & \colhead{($M_\mathrm{AB}$)} & \colhead{}  & \colhead{} & \colhead{}& \colhead{($M_\mathrm{AB}^{-1} \, \mathrm{Mpc}^{-3}$)}
}
\startdata
\multicolumn{5}{c}{$z \sim 5.375$} & & & & & & \multicolumn{5}{c}{$z \sim 6.5$} \\
\hline
-22.50 & & 2 & & $0.000026 \pm 0.000018$ & & & & & & -22.50 & & 1 & & $\leq\!0.000018$ \\
-21.50 & & 10 & & $0.000130 \pm 0.000041$ & & & & & & -21.50 & & 5 & & $0.000091 \pm 0.000041$ \\
-20.50 & & 32 & & $0.000416 \pm 0.000074$ & & & & & & -20.50 & & 18 & & $0.000327 \pm 0.000077$ \\
-19.50 & & 98 & & $0.001274 \pm 0.000129$ & & & & & & -19.50 & & 56 & & $0.001017 \pm 0.000136$ \\
-18.50 & & 202 & & $0.002724 \pm 0.000193$ & & & & & & -18.50 & & 65 & & $0.001491 \pm 0.000197$ \\
-17.50 & & 78 & & $0.022080 \pm 0.003509$ & & & & & & -17.50 & & 7 & & $0.004098 \pm 0.001575$ \\
-16.50 & & 21 & & $0.035983 \pm 0.008534$ & & & & & & -16.50 & & 3 & & $0.010057 \pm 0.005910$ \\
-15.50 & & 5 & & $0.049303 \pm 0.022721$ & & & & & & -15.50 & & 1 & & $\leq\!0.020752$ \\
-14.50 & & 0 & & $<\!0.023796$ & & & & & & -14.50 & & 1 & & $\leq\!0.026675$ \\
-13.50 & & 0 & & $<\!0.052387$ & & & & & & -13.50 & & 0 & & $<\!0.067898$ \\
\hline
\multicolumn{5}{c}{$z \sim 7.5$} & & & & & & \multicolumn{5}{c}{$z \sim 9$} \\
\hline
-22.50 & & 0 & & $<\!0.000020$ & & & & & & -22.50 & & 0 & & $<\!0.000012$ \\
-21.50 & & 1 & & $\leq\!0.000020$ & & & & & & -21.50 & & 0 & & $<\!0.000012$ \\
-20.50 & & 12 & & $0.000243 \pm 0.000070$ & & & & & & -20.50 & & 3 & & $0.000037 \pm 0.000021$ \\
-19.50 & & 10 & & $0.000202 \pm 0.000064$ & & & & & & -19.50 & & 8 & & $0.000098 \pm 0.000035$ \\
-18.50 & & 11 & & $0.000801 \pm 0.000556$ & & & & & & -18.50 & & 2 & & $0.000085 \pm 0.000064$ \\
-17.50 & & 2 & & $0.002679 \pm 0.002024$ & & & & & & -17.50 & & 0 & & $<\!0.001677$ \\
-16.50 & & 0 & & $<\!0.004577$ & & & & & & -16.50 & & 0 & & $<\!0.006701$ \\
-15.50 & & 0 & & $<\!0.018130$ & & & & & & -15.50 & & 1 & & $\leq\!0.016371$ \\
-14.50 & & 0 & & $<\!0.048022$ & & & & & & -14.50 & & 0 & & $<\!0.060766$ \\
-13.50 & & 0 & & $<\!0.087502$ & & & & & & \nodata & & \nodata & & \nodata \\
\enddata
\end{deluxetable*}

\startlongtable
\begin{deluxetable*}{c c c c c c c c c c c c c c c}
\tablecaption{Stepwise maximum likelihood UV LF in the twelve HFF fields using magnification estimates from the CATS lens models\label{CATS_UV_LF_SWML_table}}
\tablehead{\colhead{$M_\mathrm{UV}$} & \colhead{} & \colhead{No.~of galaxies} & \colhead{} & \colhead{$\phi$} & \colhead{} & \colhead{} & \colhead{} & \colhead{} & \colhead{} & \colhead{$M_\mathrm{UV}$} & \colhead{} & \colhead{No.~of galaxies} & \colhead{} & \colhead{$\phi$} \\
\colhead{($M_\mathrm{AB}$)} & \colhead{} & \colhead{} & \colhead{} & \colhead{($M_\mathrm{AB}^{-1} \, \mathrm{Mpc}^{-3}$)} & \colhead{} & \colhead{} & \colhead{} & \colhead{} & \colhead{} & \colhead{($M_\mathrm{AB}$)} & \colhead{}  & \colhead{} & \colhead{}& \colhead{($M_\mathrm{AB}^{-1} \, \mathrm{Mpc}^{-3}$)}
}
\startdata
\multicolumn{5}{c}{$z \sim 5.375$} & & & & & & \multicolumn{5}{c}{$z \sim 6.5$} \\
\hline
-22.50 & & 2 & & $0.000027 \pm 0.000019$ & & & & & & -22.50 & & 1 & & $0.000019 \pm 0.000019$ \\
-21.50 & & 10 & & $0.000134 \pm 0.000045$ & & & & & & -21.50 & & 5 & & $0.000094 \pm 0.000044$ \\
-20.50 & & 29 & & $0.000389 \pm 0.000042$ & & & & & & -20.50 & & 17 & & $0.000319 \pm 0.000045$ \\
-19.50 & & 86 & & $0.001152 \pm 0.000124$ & & & & & & -19.50 & & 52 & & $0.000976 \pm 0.000136$ \\
-18.50 & & 178 & & $0.002427 \pm 0.000319$ & & & & & & -18.50 & & 66 & & $0.001662 \pm 0.000312$ \\
-17.50 & & 89 & & $0.005533 \pm 0.000930$ & & & & & & -17.50 & & 8 & & $0.002614 \pm 0.001251$ \\
-16.50 & & 27 & & $0.009842 \pm 0.002754$ & & & & & & -16.50 & & 4 & & $0.004187 \pm 0.002840$ \\
-15.50 & & 8 & & $0.013352 \pm 0.006348$ & & & & & & -15.00 & & 3 & & $0.022666 \pm 0.016708$ \\
-14.00 & & 5 & & $0.029494 \pm 0.015356$ & & & & & & -13.50 & & 0 & & $<\!0.052415$ \\
\hline
\multicolumn{5}{c}{$z \sim 7.5$} & & & & & & \multicolumn{5}{c}{$z \sim 9$} \\
\hline
-22.50 & & 0 & & $<\!0.000021$ & & & & & & -21.50 & & 0 & & $<\!0.000012$ \\
-21.50 & & 1 & & $0.000021_{-0.000021}^{+0.000022}$ & & & & & & -19.50 & & 9 & & $0.000523 \pm 0.000219$ \\
-20.50 & & 8 & & $0.000167 \pm 0.000039$ & & & & & & -18.00 & & 3 & & $0.000236 \pm 0.000099$ \\
-19.50 & & 11 & & $0.000229 \pm 0.000053$ & & & & & & -15.50 & & 1 & & $0.000857_{-0.000857}^{+0.001011}$ \\
-17.50 & & 12 & & $0.000377 \pm 0.000116$ & & & & & & \nodata & & \nodata & & \nodata \\
\enddata
\tablecomments{Upper limits for null-detection bins were estimated using the ``$1/V_\mathrm{max}$'' method assuming $\phi_k < 1 / V_\mathrm{max}(M_k)$.}
\end{deluxetable*}


\twocolumngrid
\bibliographystyle{aasjournal}
\bibliography{references}


\end{document}